\documentclass[aps,pre,floats,superscriptaddress,nofootinbib,floatfix,twocolumn,longbibliography,showkeys]{revtex4-1}
\usepackage{amssymb,amsmath,amsfonts,amsthm}
\usepackage{graphicx}
\usepackage{newtxtext,newtxmath}
\usepackage{dcolumn}
\usepackage{bm}
\usepackage{comment}
\usepackage{csquotes}
\usepackage{mathtools}
\usepackage{gensymb}
\usepackage{hyperref}
\usepackage{makecell}
\usepackage{changepage,orcidlink}
\usepackage{multirow}
\usepackage[table]{xcolor}
\usepackage[mathscr]{euscript}
\usepackage{ulem}
\usepackage{comment}
\graphicspath{{./fig/}}

\hypersetup{
	colorlinks=true,
	linkcolor=blue,
	filecolor=blue,      
	urlcolor=blue,
	citecolor=blue,
}

\begin{document}

\title{
Chemotaxis-induced linear instabilities and pattern formation in a reaction-diffusion model}
\author{Mintu Karmakar\,\orcidlink{0000-0003-1037-2580}}
\email{mkarmakar094@ucas.ac.cn, mkarmakar094@gmail.com}
\affiliation{Wenzhou Institute of the University of Chinese Academy of Sciences, Wenzhou, Zhejiang 325011, China}
\affiliation{School of Physical Sciences, University of Chinese Academy of Sciences, Beijing 100049, China}
\affiliation{Departament de F\'isica de la Mat\`eria Condensada, Universitat de Barcelona, Mart\'i i Franqu\`es 1, E08028 Barcelona, Spain}
\author{Abhik Basu}\email{abhik.123@gmail.com, abhik.basu@saha.ac.in}
\affiliation{Theoretical Physics Division, Saha Institute of
Nuclear Physics, 1/AF Bidhannagar, Calcutta 700064, West Bengal, India}
\date{\today}

\begin{abstract}
We study linear instabilities and patterns in a spatially extended version of the Selkov model for glycolysis that includes the diffusion of individual species and the chemotactic interactions between the two species. We show that this model has two types of bifurcations. The first is a zero-wavevector, finite-frequency Hopf bifurcation. This leads to a growing oscillatory state that is spatially homogeneous.  The second is a Turing instability with saddle-node bifurcation. This leads to a growing inhomogeneous state with a steady pattern. The pattern is characterized by a finite selected wavevector, $k_c$. The value of $k_c$ depends sensitively on the specific nature of the underlying chemotaxis. Linear stability analysis reveals an intriguing nonmonotonic dependence of $k_c$ on the chemotaxis parameters, with the possibility of re-entrant transitions between uniform and patterned states. For sufficiently strong chemotaxis with both species either mutually attracting or repelling each other, an unusual instability emerges where the preferred wavevector diverges as the finite instability threshold on the strength of the chemotaxis parameters is approached. Our results from the linear stability analysis are complemented by extensive direct numerical simulations (DNS) of the governing nonlinear partial differential equations in two dimensions.  Our DNS studies reveal a complex dependence of the patterns on the chemotaxis parameters. The DNS studies show signatures of nonmonotonic behavior and re-entrant transitions. These also confirm the instabilities predicted by the linear stability analysis. In addition, the studies display spatially uniform but oscillatory states for specific model parameters.
The DNS studies also show that the steady patterns in this model can be spots and also, unexpectedly, stripes. The stable striped patterns found in the DNS studies can be reconciled by the stability properties of the linear amplitude equations for the striped patterns. The DNS studies further unveil transitions between spots and stripes, which are tuned by the chemotaxis parameters and also the diffusivity ratios. We argue that the spot-stripe transitions should be due to nonlinear effects. The phenomenological implications of our results are discussed.
\end{abstract}
\maketitle

\section{Introduction}

For more than a century, various types of patterns, such as convection rolls, Taylor columns, viscous fingering, droplet formation due to surface tension, etc., have been observed and extensively studied in the field of fluid mechanics~\cite{thermohaline_convec, bin_liq_convec, palme, kuz, strogatz, binfluid_inst, soret, silber, moore, holli, futt, gall}. In 1952, Turing presented a somewhat different method for pattern generation in which two hostile species interact and undergo diffusive motion~\cite{turing}. A critical ratio of the two diffusivities causes the Turing pattern. Numerous studies on reaction diffusion systems have been conducted since Turing's initial work. A generic two-species reaction diffusion system can be constructed by adding the diffusive terms to the dynamical system with two distinct diffusivities. It is theoretically possible for the dynamical system instabilities to compete with the pattern-forming instability, and thereby create a rich system.  The system variables in a reaction-diffusion model are fields (i.e., functions of time and space)~\cite{ricard, predator, *predator1, ref1, ref2, post_turing, post_turing_a, post_turing_b, post_turing_c, post_turing_2, post_turing_2a, post_turing_2b, post_turing_2c}. 

The stable fixed point of the dynamical system defines the state of the physical system, and an instability of the fixed point causes the physical state to become unstable. When a system parameter is changed, a stable fixed point becomes unstable, which is a bifurcation that typically results in a new stable state for the system. The control parameter is the parameter in question. Different types of bifurcations may occur if there are multiple control parameters. Each bifurcation will have a surface defined in the parameter space, and these surfaces may intersect. When there are two control parameters, the surfaces will be curves that can end at a specific parameter value, which is known as a codimension-two point. A recent study illustrates this in the context of the diffusive Selkov model, i.e., the Selkov model with diffusion of the two species~\cite{ab_jkb}, highlighting the existence of two distinct instabilities in this model and their meeting at a codimension two point in the parameter space. Yet another well-known example is the Brusselator model~\cite{bruss, bruss1, bruss2, bruss3, bruss4}, showing a large number of interesting possibilities. From a general point of view, principal theoretical issues in pattern formation include how simple dynamical rules generate complex periodic structures or patterned steady states, transitions between different patterned states and the associated turning parameters, robustness of the patterns against noise, etc. See Refs.~\cite{rmp, book} for extensive discussions and reviews on pattern formation in various model systems.

Chemotaxis is a fundamental strategy in many active matter and biological systems; see Ref.~\cite{lowen} for a recent review. In living systems, it allows individual cells to move both individually and collectively. Cells use this technique to identify signaling chemicals and move along concentration gradients. In many living systems, it is an essential feature. Examples include bacterial colonies~\cite{bact_colony}, tissue structure~\cite{murray, cancer}, and even the immune system~\cite{immune}. Beyond natural biological systems, chemotaxis-like mechanisms are essential in synthetic~\cite{synthetic} and reconstituted~\cite{recons} systems. For instance, the directed movement of colloidal particles in response to (self-generated) solute concentration gradients is driven by various phoretic processes~\cite{synthetic, synthetic1, artificial, ramin1, ramin2}. Additionally, cross-diffusion of enzymes with their substrates and products enables the enzymes' catalysis-driven fluxes to cause their aggregation~\cite{cross_ensyme}. Lastly, chemotaxis-like cross-diffusion occurs with respect to substrate and product densities when enzyme-enriched condensates interact with the substrates and products. It has been demonstrated that these condensates exhibit self-propulsion via organizing the substrate distribution and catalysis-driven fluxes~\cite{frey_enzyme, berg}. The Keller–Segel model, which described the accumulation of Dictyostelium discoideum, was the first theoretical description of chemotactic aggregation~\cite{keller_segel}. A wide range of spatiotemporal patterns, such as aggregate coarsening, stable aggregate sizes, and spatiotemporally chaotic aggregate dynamics, have been revealed by the analysis of numerous variations of this model since then. Specifically, interesting patterns are produced by the interaction of chemotaxis with cell growth and death~\cite{chemo1, chemo2}. More complex signaling dynamics~\cite{erwin_chemo_2} and a variety of cell and chemoattractant species~\cite{wolansky, liu, frey_chemo_3} must be taken into account in order to characterize complicated biological chemotaxis systems. In yet another study, Ref~\cite{cross_diff} illustrates how cross diffusion can induce patterns in a discrete two-component predator-prey model. Given that chemotaxis by itself is a number-conserving process, how that conspires with other pattern forming elements in the systems (which may or may not be overall number-conserving) to produce steady patterns and the latter's dependence on the specific nature of chemotaxis are some of the generic questions that are currently being explored.

Given that the addition of diffusion in dynamical systems, i.e., reaction-diffusion models, can exhibit patterns, and chemotactic interactions in multi-species systems can also induce pattern formations, it is imperative to study the competition between chemotaxis and reaction-diffusion instabilities in models with specific structures. To that end, in this article we present a variation and extension of the previous studies. This has several novel features, which lead to some new physical phenomena. In this work, we theoretically explore this competition and its consequences on pattern formation. As a concrete example, we propose and use a spatially extended version of the Selkov model for glycolysis, which combines self-diffusion of the two species and chemotactic interactions between them. The latter also contains the physics of cross diffusion of the two species. We systematically study the model dynamical equations analytically within linear stability analyses, together with extensive numerical solutions of the model equations in the appropriate regions of the parameter space. Our most surprising result is that the chemotactic interactions crucially affect the emerging patterns. This is clearly observed in the sensitive dependence on the selected wavevector $k_c$ on the chemotaxis parameters, including a nonmonotonic dependence of $k_c$ on the chemotaxis parameters. To complement our analytical, linear stability analysis results, we have performed extensive direct numerical simulations (DNS) of the model equations in two dimensions (2D). Our DNS studies capture the complex dependence of the steady patterns on the chemotaxis parameters. In particular, for the purposes of quantifying the patterns observed in the DNS studies, we have defined quantities, which can serve as markers for the pattern amplitudes and display strong variations with the chemotaxis parameters. These variations have a clear correspondence with the variation of $k_c$ with the same parameters in the linear stability analysis.  Our DNS studies also reveal transitions between distinct types of patterns - spots and stripes - tuned by the chemotaxis parameters in the model.  The remainder of the article is organized as follows. In Section~\ref{princi}, we set up our model equations and provide a summary of our principal results. Then in Section~\ref{linear}, we study the linear stability analysis of the model and provide a phase diagram. We further calculate the selected wavevector $k_c$. We further extract its sensitivity (including nonmonotonic dependence in some of the cases considered here) on the chemotaxis parameters and their consequences on the patterns that may be observed. Next in Section~\ref{dns}, we discuss the numerical results on the patterns and oscillations obtained from the DNS of the model equations. Then in Section~\ref{ampli}, we set up the linear amplitude equations for striped patterns. A short summary of our main results is available in Section~\ref{table-results} in tabular form. In Section~\ref{summ}, we summarize and discuss our results.
In the Appendices, we provide additional numerical results and technical details. Appendix~\ref{sec:numerical_methods} describes the numerical methods used in the DNS studies of the model. Appendices~\ref{geom} and \ref{ini_cond} examine, respectively, the influence of domain geometry by comparing square and rectangular systems, and the dependence of the observed patterns on the initial conditions. In Appendix~\ref{nonlin-ampli}, we use a perturbative analysis to show how nonlinear chemotactic effects can modify the linear stability of stripe patterns, thus facilitating stripe-to-spot and spot-to-stripe transitions. Appendix~\ref{noises} discusses the effects of noise on steady-state patterns, while Appendix~\ref{diff-eff} examines how variations in diffusivities affect pattern selection. Finally, Appendix~\ref{selected-kc} presents additional results on the dependence of the selected wavevector on the chemotaxis parameters, and Appendix~\ref{osci-pattern} provides further DNS results on oscillatory patterns.





\section{Model equations and principal results}\label{princi}

In this work, we study two-component reaction-diffusion models and investigate how chemotactic interactions between the two species can affect the emerging patterns. A generic two-component reaction-diffusion model for variables $\phi$ and $\psi$ consists of the dynamical equations
\begin{eqnarray}
    \frac{\partial \phi}{\partial t} &=&f_\phi(\phi,\psi) + D_1\nabla^2 \phi,\label{modbasica}\\
 \frac{\partial \psi}{\partial t} &=& f_\psi(\phi,\psi) + D_2\nabla^2 \psi,\label{modbasicb}
\end{eqnarray}
 where $f_\phi(\phi,\psi),\,f_\psi(\phi,\psi)$ are {\it nonconserving} reaction functions that model local (onsite) reactions between the two species, and $D_1, D_2$ are the two diffusivities. In general, Eqs.~(\ref{modbasica}) and (\ref{modbasicb}) admit two fixed points $\phi=\phi_0,\,\psi=\psi_0$, where $\phi_0,\,\psi_0$ are constants corresponding to a spatially uniform steady state. The stability of this uniform state may be destroyed by tuning the parameters that define functions $f_\phi,\,f_\psi$, or the diffusivities $D_1, D_2$, or by further introducing inter-species chemotactic interactions (see below), which are abundant in nature. Our focus here is the latter. This may be systematically explored either by studying the linear instabilities around the above fixed points or by numerically solving (\ref{modbasica}) and (\ref{modbasicb}). The details of the outcome should depend on the choice of reaction functions $f_\phi,\,f_\psi$. At the fixed points corresponding to the uniform states,
 \begin{eqnarray}
     f_\phi(\phi_0,\psi_0)=0=f_\psi(\phi_0,\psi_0).
 \end{eqnarray}
 Expanding around the uniform state with $\rho_1=\phi_0+\phi$, $\rho_2=\psi_0+\psi$, we obtain the dynamical equations for $\phi$ and $\psi$. We wish to study how the uniform states defined by $\phi=0,\psi=0$ become unstable, giving rise to patterns. A brief account of the present study, highlighting the generic aspects, is available in the associated short paper (ASP)~\cite{asp}. 

 To proceed systematically and for the purposes of explicit results, we work with the Selkov model for glycolysis~\cite{strogatz}.
 
 We first write down the diffusive Selkov model equations~\cite{ab_jkb}. The Selkov model for glycolysis was introduced to model glycolytic oscillations and has two species. In the Selkov model,
 \begin{eqnarray}
     F_\phi(\phi,\psi)&=& -\phi + a\psi + \phi^2\psi,\\
     F_\psi(\phi,\psi)&=& b - a\psi -\phi^2\psi,
 \end{eqnarray}
 where $a,b$ parametrize the reaction functions. With these,
 the model equations read
\begin{eqnarray}
 \frac{\partial \phi}{\partial t} &=&-\phi + a\psi + \phi^2\psi + D_1\nabla^2 \phi,\label{moda}\\
 \frac{\partial \psi}{\partial t} &=&  b - a\psi -\phi^2 \psi + D_2\nabla^2 \psi,\label{modb}
\end{eqnarray}
{where $\phi,\psi$ are dimensionless concentrations of ADP (adenosine diphosphate)  and
F6P (fructose-6-phosphate), respectively~\cite{strogatz}; see Ref.~\cite{selkov1968} for the model equations in the original form.} The
terms $D_1\nabla^2\phi$ and $D_2\nabla^2 \psi$ in (\ref{mod1}) and (\ref{mod2}) respectively
 represent the diffusion of the two species in space. The conventional Selkov model does
not include any diffusion~\cite{selkov1968}. All the parameters $a,b, D_1, D_2$ are positive. Parameters $a$ and $b$ control the reactions between the two species.
Notice that without diffusion, (\ref{moda}) and (\ref{modb}) are just two coupled ordinary 
differential equations (ODEs) that define a dynamical system, whereas with 
diffusion, they become partial differential equations (PDEs). Reaction-diffusion equations (\ref{moda}) and (\ref{modb}) have no conservation laws and can show patterns, and hence provide a platform to study how the pattern forming instabilities of (\ref{moda}) and (\ref{modb}) are affected by the mutual chemotactic interactions, which are conserving (see below). 

We now extend Eqs.~(\ref{moda}) and (\ref{modb}) by incorporating the physics of chemotaxis. In chemotaxis between two species, each one responds to the spatial variations in the density of the other by having a nonzero current parallel or antiparallel to and proportional to the gradient of the concentration of the other species. The currents of the two species ${\bf j}_1$ and ${\bf j}_2$ due to chemotaxis are 
\begin{eqnarray}
    &&{\bf j}_1  = -\xi_1 \phi {\boldsymbol\nabla}\psi,\label{curr1}\\
    &&{\bf j}_2  = -\xi_2 \psi {\boldsymbol\nabla}\phi,\label{curr2}
\end{eqnarray}
see also Ref.~\cite{chemo_aip} for a partly related model study. The chemotaxis coefficients $\xi_1,\xi_2$ can be individually positive or negative. When both $\xi_1,\xi_2>0(<)$, the currents ${\bf j}_1,{\bf j}_2$ are (anti) parallel to the concentration gradient of the other species, whereas, these parameters can be of mutually opposite signs too: if $\xi_1>0$ and $\xi_2<0$, ${\bf j}_1$ is along ${\boldsymbol\nabla} \psi$ and ${\bf j}_2$ is antiparallel to ${\boldsymbol\nabla} \phi$, and vice versa if $\xi_1,\xi_2$ reverse their signs. 
We now include these contributions in (\ref{moda}) and (\ref{modb}) above to obtain
\begin{eqnarray}
 \frac{\partial \phi}{\partial t} &=&-\phi + a\psi + \phi^2\psi + D_1\nabla^2 \phi +\xi_1 {\boldsymbol\nabla}\cdot (\phi{\boldsymbol\nabla}\psi),\label{mod1}\\
 \frac{\partial \psi}{\partial t} &=&  b - a\psi -\phi^2 \psi + D_2\nabla^2 \psi+\xi_2 {\boldsymbol\nabla}\cdot (\psi{\boldsymbol\nabla}\phi).\label{mod2}
\end{eqnarray}
Physically, when both $\xi_1,\xi_2$ are positive (negative), i.e., with $\xi_1\xi_2>0$, they attract (repel) each other; we call ``reciprocal chemotaxis''. In contrast, when one of them is positive and the other negative, i.e., with $\xi_1\xi_2<0$, one is attracted to the other, whereas the second one is repelled, i.e., chasing phenomenon, or a pursuer-evader relationship. We call it ``nonreciprocal chemotaxis''. Notice that due to the reactions, Eqs.~(\ref{mod1}) and (\ref{mod2}) {\it do not} conserve numbers. However, the chemotaxis processes conserve the number of individual species.  
If we set $\xi_1=0=\xi_2$ and $D_1=0=D_2$ in Eqs.~(\ref{mod1}) and (\ref{mod2}), the original Selkov model is obtained~\cite{strogatz}. In what follows, we study Eqs.~(\ref{mod1}) and (\ref{mod2}) in two dimensions (2D) with periodic boundary conditions.

By considering small fluctuations about the fixed points of (\ref{mod1}) and (\ref{mod2}), we show that such a uniform state corresponding to these fixed points can be unstable in two distinct ways. First of all, in some regions of the parameter space spanned by $a$, $b$, the model admits a spatially uniform oscillatory instability with a frequency 
\begin{equation}
    \omega_0 = \pm\sqrt{a+b^2}.
\end{equation}
This implies a steady oscillation that occurs in the entire system; the system remains spatially homogeneous everywhere. The phase boundary in the $a-b$ plane that demarcates a steady homogeneous phase and a phase with oscillatory instability is
\begin{equation}
    b^2 = \frac{1}{2}\bigg(1-2a \pm\sqrt{1-8a}\bigg),\label{hopf-line},
\end{equation}
which is the condition for Hopf bifurcation in this model, and is unsurprisingly identical to the one found in Refs.~\cite{strogatz,ab_jkb}. Secondly, saddle-node or Turing instability may occur in other regions of the $a-b$ plane, which are characterized by finite wavevector instabilities, resulting in steady patterns in the long-time limit. The preferred or selected wavevector $k_c$ is given by
\begin{eqnarray}
    k_c^2=-\frac{D_1(a+b^2)-D_2\frac{b^2-a}{b^2+a}+\xi_2b-\frac{2\xi_1b^3}{a+b^2}}{2\bigg[D_1D_2-\frac{\xi_1\xi_2b^2}{a+b^2}\bigg]}.\label{qc}
\end{eqnarray}
The corresponding phase boundary separating a patterned state from the uniform state in the $a-b$ plane is given by
\begin{eqnarray}
   && D_1(a+b^2)-D_2\frac{b^2-a}{b^2+a}+\xi_2b-\frac{2\xi_1b^3}{a+b^2}\nonumber \\&& = -2\bigg(D_1D_2-\frac{\xi_1\xi_2b^2}{a+b^2}\bigg)^{1/2}\sqrt{a+b^2}.\label{pattern-boundary}
\end{eqnarray}
Thus, the reaction-diffusion system given by (\ref{mod1}) and (\ref{mod2}) admits both type III${_o}$ (periodic in time but uniform in space) and type I${_s}$ (stationary in time but periodic in space) linear instabilities in the classification of Ref.~\cite{rmp}. Our principal focus here is on the type I${_s}$, which can potentially give rise to stationary patterns, as we find in our DNS studies.

Equation~(\ref{qc}) reveals a complex dependence of $k_c^2$ on the chemotaxis parameters $\xi_1,\xi_2$. Considering $\xi_1=\xi_2$ for simplicity (fully reciprocal chemotaxis), it can be shown that $k_c^2$ has a nonmonotonic dependence on $\xi>0$ when $b^2>a$, or if $\xi<0$, a similar nonmonotonic dependence of $k_c^2$ on $\xi$ is observed for $b^2<a$.  On the other hand, if $\xi_1=-\xi_2=\xi$ (extreme nonreciprocal chemotaxis), $k_c^2$ depends nonmonotonically on $\xi>0$ in a manner {\it different} from the fully reciprocal case. Such nonmonotonic dependence can potentially give rise to intriguing {\it re-entrant} behavior in the system, tuned by $\xi$. Furthermore, Eq.~(\ref{qc})   can be used to show that for large enough $D_1, D_2$, $k_c$ vanishes, the pattern disappears, and a uniform state ensues. 


Focusing on the type I${_s}$-type instabilities, 
our DNS studies of \eqref{mod1} and \eqref{mod2} support the above results. Intriguingly, both spots and striped patterns are found. The stability of the striped patterns can be shown by using the amplitude equations correspond to the model equations (\ref{mod1}) and (\ref{mod2}). Furthermore, our DNS studies are able to detect the nonmonotonic dependence of the pattern on $\xi_1,\xi_2$ in some of the cases studied, as predicted by the linear stability analysis. Our DNS studies also reveal that with increasing noise, the patterns get less and less sharp, eventually disappearing for sufficiently high noise.

In the following, we systematically obtain the above results.

\section{Linear stability analysis}\label{linear}

At the fixed points of the model equations (\ref{mod1}) and (\ref{mod2}), the densities $\phi$ and $\psi$ are constants, which are given by
\begin{equation}
 \phi^*=b,\;\;\;\psi^*=\frac{b}{a+b^2};\label{fps}
\end{equation}
a superscript * denotes a fixed point value. 
We now linearize Eqs.~(\ref{mod1}) and (\ref{mod2}) around the fixed points (\ref{fps}). We find
\begin{eqnarray}
 \frac{\partial u}{\partial t} &=& \frac{b^2-a}{b^2+a} u + av + b^2 v + D_1\nabla^2 u+\xi_1 b\nabla^2 v,\label{modlin1} \\
 \frac{\partial v}{\partial t} &=& -(b^2+a)v - \frac{2ub^2}{b^2+a} + D_2\nabla^2 v\nonumber \\&&+\xi_2\frac{b}{a+b^2}\nabla^2 u,\label{modlin2}
\end{eqnarray}
where $u=\phi-\phi^*,\,v=\psi - \psi^*$. Since (\ref{modlin1}) and (\ref{modlin2}) are PDEs, they actually correspond to an {\it infinite} number of modes. The conventional methods for characterizing the instability of a uniform state to a patterned state or a spatially uniform but oscillatory state are used in partial differential equations (\ref{modlin1}) and (\ref{modlin2}). The general criterion for separating the various wavenumber modes is just the existence of the Fourier transforms of the functions $u({\bf x}, t)$  and $v({\bf x}, t)$, as the instability emerges from a linearized system. Furthermore, these modes can be conveniently labeled by using the Fourier wavevector $k$, assuming periodic boundary conditions. Notice that chemotactic interactions give rise to {\it cross diffusivities} $\xi_1 b$ and $\xi_2\frac{b}{a+b^2}$  respectively, in the linearized equations (\ref{modlin1}) and (\ref{modlin2}). The cross diffusivities can be positive or negative.

Assuming time-dependence of the form $\exp(\Lambda t)$, where $\Lambda$ is a growth rate, i.e., setting $u({\bf x}, t), v({\bf x}, t)\sim \exp(\Lambda t)$, we get for the linear stability matrix $J$ for the pair of equations (\ref{modlin1}) and (\ref{modlin2})  in the Fourier space
 \begin{equation}
 J(k^2) = \left(\begin{array}{cc}
                 \frac{b^2 -a}{b^2+a^2} - D_1k^2 & a+b^2-\xi_1k^2b\\
                 -\frac{2b^2}{a+b^2} -\xi_2 k^2 \frac{b}{a+b^2}& -a -b^2 - D_2k^2
                \end{array}
\right).\label{j-matrix}
\end{equation}
The corresponding eigenvalues, labeled by wavevector $k$, are given by
\begin{eqnarray}
 \Lambda_\pm (k^2) &=&\frac{1}{2}\bigg[{\tt Tr}\pm \sqrt{{\tt Tr}^2 - 4\,{\tt Det}}\bigg],\label{eigen1}
\end{eqnarray}
where ${\tt Tr}$ and ${\tt Det}$, respectively, are the trace and determinant of the matrix $J(k^2)$ in (\ref{j-matrix}) above, 
and both ${\tt Tr}$ and ${\tt Det}$ are functions of   $k^2=||{\bf k}||^2$ (implying the Euclidean norm). Here, periodic boundary condition in 2D implies $k_x=2\pi n_x/L_x,\,k_y=2\pi n_y/L_y,\,n_x,n_y=0,\pm 1,\pm 2...$, $k^2=k_x^2 + k_y^2$ where $L_x,\,L_y$ are the linear system sizes along the $x$- and $y$-directions. In the infinite system size limit $L_x,L_y\rightarrow \infty$, and $k_x,k_y$ become quasi-continuous. Unsurprisingly, each of these wavevector modes is independent of each other for the linear equations (\ref{modlin1}) and (\ref{modlin2}). This clearly allows us to examine the linear instability of the individual wavevector modes, and thereby identify the {\it marginal} mode(s) (i.e., the ones which neither grow nor decay) at the threshold of the linear instabilities. The remaining modes should decay at the instability threshold.

For the matrix (\ref{j-matrix}),
\begin{eqnarray}
  &&{\tt Tr}(k^2)=-1+\frac{2b^2}{a+b^2}-D_1k^2-a-b^2-D_2 k^2,\label{trace}\\  
  && {\tt Det}(k^2)=-\bigg(-a-b-D_2k^2 + 2b^2 + \frac{2b^2D_2k^2}{a+b^2}\nonumber \\&&-(a+b^2)D_1k^2 -D_1D_2k^4\bigg)+ 2b^2+\xi_2bk^2-\frac{2\xi_1b^3k^2}{a+b^2}\nonumber \\&&-\frac{\xi_1\xi_2 b^2k^4}{a+b^2}.\label{det}
\end{eqnarray}
Notice that the chemotaxis terms do not enter ${\tt Tr}(k^2)$, but affects ${\tt Det}(k^2)$.
Growth rate $\Lambda(k^2)$ satisfies a quadratic equation of the form
\begin{eqnarray}
    \Lambda^2 - \Lambda B(k^2) + C(k^2)=0,
\end{eqnarray}
where $B(k^2)\equiv {\tt Tr}(k^2)$ and $C(k^2)\equiv {\tt Det}(k^2)$. The general solution to $\Lambda(k^2)$ is
\begin{eqnarray}
    \Lambda(k^2) = \frac{1}{2}\bigg[-B\pm\sqrt{B^2-4C}\bigg].\label{eigen2}
\end{eqnarray}

We look for the instabilities now. If we set $k=0$, diffusion and chemotaxis disappear. We get
\begin{eqnarray}
    {\tt Tr}(k=0)&=& -\frac{a-b^2}{a+b^2}-(a+b^2),\\
    {\tt Det}(k=0)&=&a+b^2>0,
\end{eqnarray}
which are unsurprisingly identical to those found in Ref.~\cite{strogatz,ab_jkb}. If we further impose ${\tt Tr}(k=0)=0$, which gives the curve
\begin{equation}
    b^2=\frac{1}{2}\bigg[1-2a\pm\sqrt{1-8a}\bigg]\equiv b^2_\pm.\label{hopf-curve}
\end{equation}
The upper branch of (\ref{hopf-curve}) given by $b=b_+$ intersects the $b$-axis at $(0,1)$, whereas the lower branch $b=b_-$ passes through the origin. 
On the curve (\ref{hopf-curve}), 
\begin{equation}
    \Lambda(k=0)=\pm i\sqrt{{\tt Tr}(k=0)},
\end{equation}
giving purely oscillatory time-dependence for $u,v$. If ${\tt Tr}(k^2)>(<)0$, $\Lambda(k^2=0)$ has a growing (decaying) real part, corresponding to the instability (stability) of the uniform state. Thus, ${\tt Tr}(k^2)=0$ gives the {\it instability threshold} of the zero wavevector Hopf bifurcation. The finite frequency of oscillation $\omega_0$ is given by
\begin{equation}
    \omega_0= \pm \sqrt{{\tt Det}(k^2=0)}=\pm \sqrt{a+b^2}.\label{hopf-freq}
\end{equation}
See also Refs.~\cite{strogatz,ab_jkb}. In the present model, when ${\tt Tr}(k^2=0)<0$, ${\tt Tr}(k^2)<0$ even in the presence of chemotaxis. Hence, chemotaxis does not induce any Hopf bifurcation. We now study the Turing instabilities, i.e., the linear instabilities at a finite wavevector $k$, and explore how they are affected by chemotaxis. From (\ref{eigen2}), the growth rate $\Lambda(k^2)$ can be

(i) real and stable (for negative real parts, i.e., when $B(k^2)>0$ and $B^2(k^2)>4C(k^2)>0$,

(ii) Stable with complex conjugate values ($B(k^2)<0$ and $B^2(k^2)<4C(k^2)$.

(iii) Unstable with one solution for $\Lambda(k^2)$ having a positive real part: $C(k^2)<0$.

Therefore, the threshold of a finite wavevector instability is $C(k_c^2)=0$ at some $k_c$, such that
\begin{equation}
    \Lambda(k_c^2)=\frac{1}{2}\bigg[-B+\sqrt{B^2-4C}\bigg]\equiv \Lambda_+(k_c^2)
\end{equation}
has a positive real part, and hence instability ensues.
 To determine the wavevector $k_c$, we argue that it is the wavevector $k$ for which $C(k^2)$ is a minimum~\cite{ab_jkb}. Hence, we set
\begin{equation}
    \frac{\partial C(k^2)}{\partial k^2}|_{k^2=k_c^2} =0,\;
    \frac{\partial}{\partial k^2}\frac{\partial C(k^2)}{\partial k^2}\bigg|_{k^2=k_c^2}>0,
\end{equation}
together with $C(k_c^2)=0$ at the instability threshold.

\begin{figure}[htb]
\includegraphics[width=4cm]{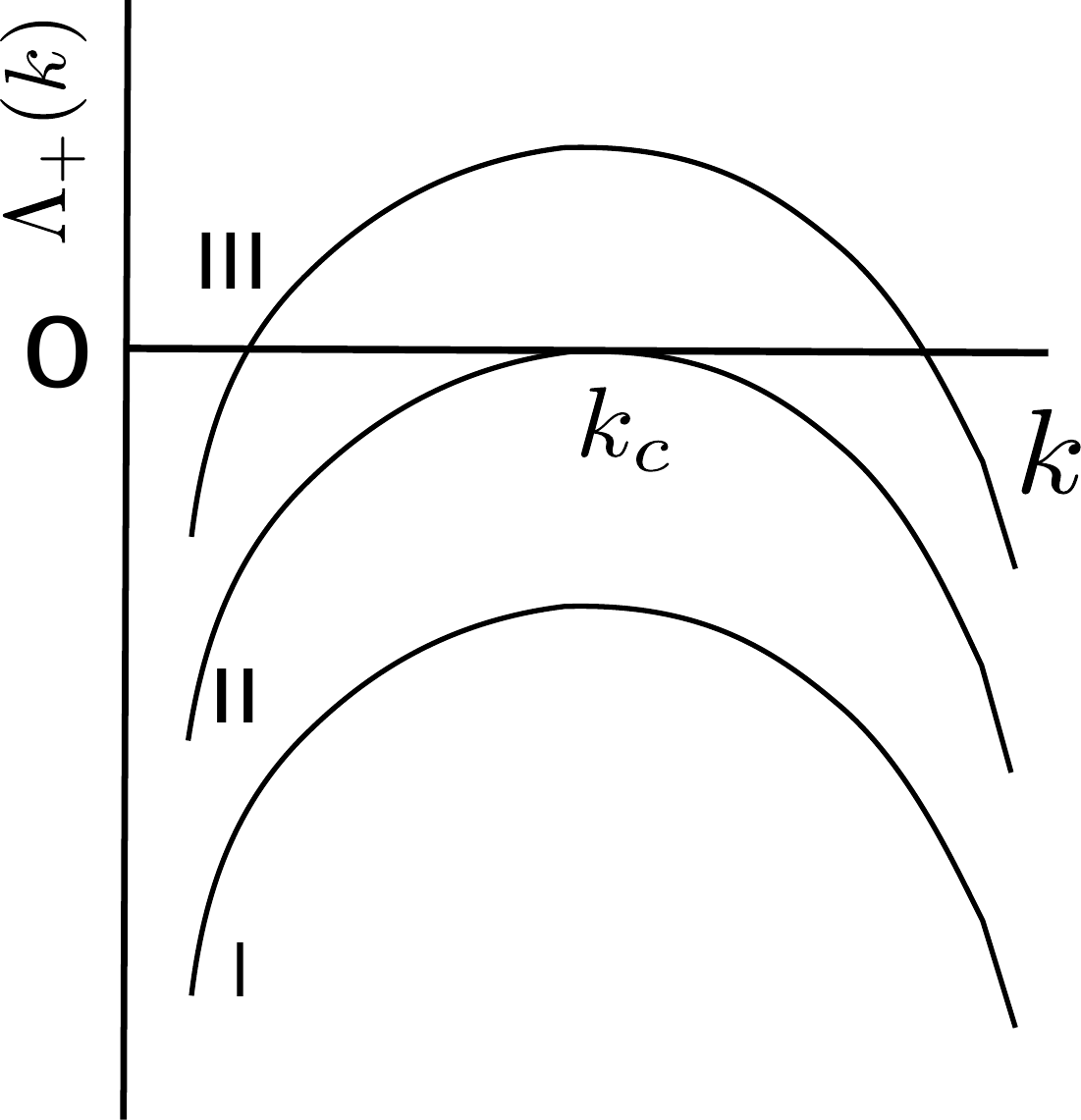}\hfill \includegraphics[width=4cm]{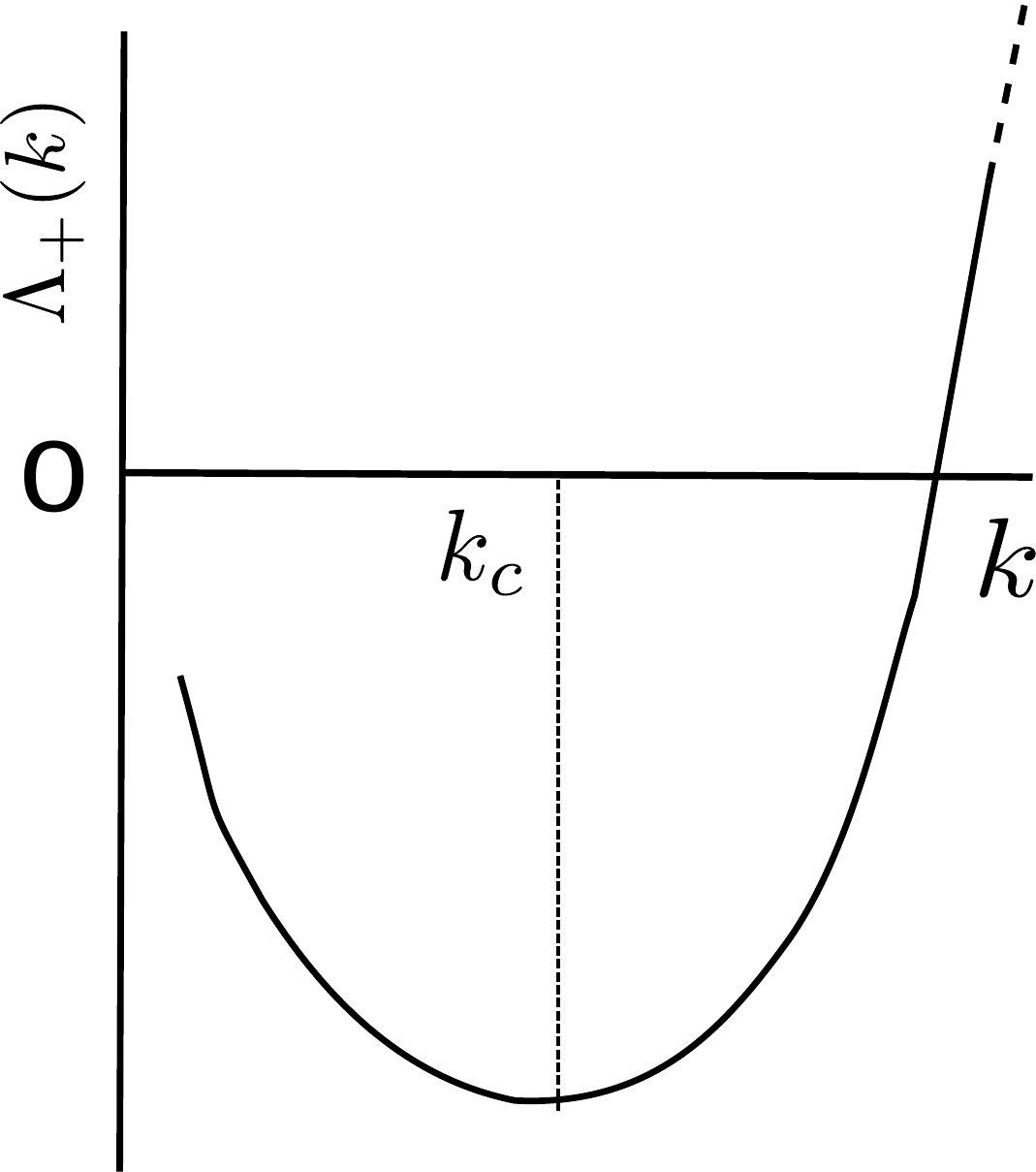}
\caption{Variation of $\Lambda_+(k)$ as a function of $k$. (i) Reciprocal chemotaxis: $-\infty<\xi_1\xi_2<D_1D_2 (a_c+b_c^2)/b_c^2$ for which a pattern with a preferred wavevector $k_c$ is possible. Curves I, II \& III correspond to no linear instability (uniform phase), threshold of instability with $k=k_c$ becoming marginal, and a band of unstable wavevectors with the mode $k=k_c$ having the highest growth rate (pattern) in the linear stability analysis. (ii) Reciprocal chemotaxis: $\xi_1\xi_2 > D_1D_2 (a_c+b_c^2)/b_c^2$ for which there is instability for sufficiently large $k$ for any region of the parameter space and $k\rightarrow \infty$ having the highest growth rate, signaling a chemotaxis induced instability. For nonreciprocal chemotaxis, only the case shown in (i) is possible, since $\tilde D>0$ always. See text.}\label{lambda-plot}
\end{figure}

Simplifying, we find
\begin{equation}
    C(k_c^2)=\tilde Dk_c^4 + \Gamma_1 k_c^2 + \Gamma_2,
\end{equation}
where 
\begin{eqnarray}
 &&   \tilde D\equiv D_1D_2 - \frac{\xi_1\xi_2b^2}{a+b^2}, \label{tilded}\\
 &&\Gamma_1\equiv D_1(a+b^2)-D_2\frac{b^2-a}{a+b^2}+\xi_2b - \frac{2\xi_1b^3}{a+b^2},\label{gamma1}\\
 && \Gamma_2 \equiv a+b^2.
\end{eqnarray}
Using the above conditions, we find
\begin{equation}
    k_c^2 = - \frac{\Gamma_1}{2\sqrt {\tilde D}}.
\end{equation}
Using (\ref{tilded}) and (\ref{gamma1}), we get (\ref{qc}).
We must have
\begin{equation}
    \tilde D>0,
\end{equation}
giving $D_1 D_2>\xi_1\xi_2b^2/(a+b^2)$ and $\Gamma_1<0$, since $k_c^2>0$. However, we notice that $\tilde D$ is {\it not} positive definite. It can be turned negative by making the product $\xi_1\xi_2$ sufficiently large and positive. In that case
\begin{equation}
    \frac{\partial}{\partial k^2}\frac{\partial C(k^2)}{\partial k^2}|_{k^2=k_c^2}<0,
\end{equation}
corresponding to a maximum of $C(k_c^2)$, which vanishes. This means $C(k_c^2)$ is now a {\it maximum}. Since $C(k_c)^2=0$, this further means that $\Lambda(k^2)$ has a positive real part for {\it all} $k^2\neq k_c^2$, indicating a {\it novel} instability induced by chemotaxis. Notice that this is possible only for reciprocal chemotaxis, i.e., when $\xi_1\xi_2>0$. For nonreciprocal chemotaxis, $\xi_1\xi_2<0$, and hence $\tilde D$ is positive definite, ruling out the above chemotaxis-induced instability. We do not discuss this instability further in the present work. See Fig.~\ref{lambda-plot} for plots showing the dependence of $\Lambda_+(k)$ on $k$ for $\tilde D>0$ and $\tilde D<0$.

Focusing on the case with $\tilde D>0$, for which there is a preferred wavevector $k_c$ that has the highest growth rate, we now calculate the phase boundary between the patterned and uniform states in the $a-b$ plane. At the threshold of the pattern formation
\begin{equation}
    C(k_c^2)=\Gamma_2 - \frac{\Gamma_1^2}{4\tilde D}=0,
\end{equation}
giving $\Gamma_1=-2\sqrt{\tilde D}\sqrt {a+b^2}$ at the onset of the pattern. This then gives
\begin{equation}
    D_1(a+b^2)-D_2\frac{b^2-a}{a+b^2} +\xi_2b-\frac{2\xi_1b^3}{a+b^2}= -2\sqrt {\tilde D}\sqrt{a+b^2} \label{pattern-boundary}
\end{equation}
as the phase boundary between the patterned and uniform states in the $a-b$ plane, which depends explicitly on $\xi_1,\xi_2$. Clearly, the boundary (\ref{pattern-boundary}) can exist only if $D_1D_2-\xi_1\xi_2b^2/(a+b^2)>0$. We will see that this is the condition for a pattern to exist in the linear stability analysis.

We now discuss the shape of the phase boundary (\ref{pattern-boundary}) in detail. We first set $a=0$, giving
\begin{equation}
    D_1b^2 -D_2+\xi_2b-2\xi_1b=-2\sqrt{D_1D_2-\xi_1\xi_2}b.\label{a=0}
\end{equation}
This gives the intersection $(0,b)$ between the curve \eqref{pattern-boundary} and the $b$-axis. We must have $D_1D_2>\xi_1\xi_2$. In addition to $b=0$, there is a nonzero solution to \eqref{a=0} at $b=\tilde b$. Solving (\ref{a=0}) and considering only the positive solution, we get
\begin{eqnarray}
    \tilde b&=&\frac{1}{2D_1}\bigg[-2\sqrt{D_1D_2-\xi_1\xi_2}-\xi_2+2\xi_1\nonumber \\ &&+\{(-2\sqrt{ D_1 D_2}-\xi_2+2\xi_1)^2+4D_1D_2\}^{1/2}\bigg].\label{tilde-b}
\end{eqnarray}
Thus, the point of intersection between the phase boundary (\ref{pattern-boundary}) and the $b$-axis $(0,\tilde b)$ depends sensitively on the chemotaxis parameters $\xi_1,\xi_2$. Consider reciprocal chemotaxis with $\xi_1\xi_2 \rightarrow D_1D_2(a+b^2)/b^2$ from below. Then
\begin{equation}
    \tilde b\approx \frac{1}{2D_1}\bigg[-\xi_2+2\xi_1+\{(-\xi_2+2\xi_1)^2 + 4D_1D_2\}^{1/2}\bigg].
\end{equation}
Now assume that one among $|\xi_1|,|\xi_2|$ is very large: $|\xi_1|,|\xi_2|\gg D_1, D_2$. In particular, consider  $\xi_1\gg D_1, D_2$. Then if $\xi_1>0$, we get
\begin{equation}
    \tilde b= \frac{4\xi_1}{D_1},
\end{equation}
which can be very large. However, if $\xi_<0$, $\tilde b~  {\cal O} (1/\xi_1)$, which can be very small, coming close to the origin! Similar conclusions can be reached if $|\xi_2|$ is large and is either negative or positive. 


For $\xi_1=0=\xi_2$, (\ref{tilde-b}) reduces to $\tilde b=(-1+\sqrt 2)\sqrt{D_2/D_1}$, which is unsurprisingly in agreement with Ref.~\cite{ab_jkb}. With chemotaxis on, $\tilde b$ can be continuously tuned by $\xi_1,\xi_2$, and can be more or less than its value without chemotaxis. Lastly, noting that the upper branch of the Hopf line (\ref{hopf-curve}) meets the $b$-axis at $b=1$, $\tilde b$ can be found above or below this intersection point.

If we now set $b=0$ in (\ref{pattern-boundary}), we get
\begin{equation}
    (D_1+D_2)a = -2\sqrt{ D_1D_2}\sqrt a, \label{b-zero}
\end{equation}
which has {\it no} solution for $a>0$. Therefore, the curve (\ref{pattern-boundary}) cannot intersect the $a$-axis for any $a>0$. However, $a=0$ satisfies (\ref{b-zero}). Therefore, the pattern boundary should pass through the origin, similar to the Hopf line~(\ref{hopf-curve}) and the pattern boundary {\it without} chemotaxis, as elucidated in Ref.~\cite{ab_jkb}. We now calculate how the pattern boundary (\ref{pattern-boundary}) approaches the origin. This will tell us
whether (\ref{pattern-boundary}) lies above or below the lower branch of the Hopf curve $b=b_-$ near the origin. Close to the origin, in analogy with the nonchemotactic ($\xi_1=0=\xi_2$) limit result~\cite{ab_jkb}, we write
\begin{eqnarray}
    b=\sqrt a + \Gamma a^\mu,\,\mu> 0.\label{b-form}
\end{eqnarray}
Substitute this to the lhs of (\ref{pattern-boundary}) to get
\begin{eqnarray}
    &&{\rm lhs}= D_1a + D_1 (\sqrt a +\Gamma a^\mu)^2 - D_2 \frac{ (\sqrt a +\Gamma a^\mu)^2 -a}{a+(\sqrt a+\Gamma a^\mu)^2} \nonumber \\&&+ \xi_2  (\sqrt a +\Gamma a^\mu)- \frac{2\xi_1  (\sqrt a +\Gamma a^\mu)^3}{a+ (\sqrt a +\Gamma a^\mu)^2}. \label{lhs}
    \end{eqnarray}
    Similarly, considering the rhs of (\ref{pattern-boundary}), we get
    \begin{equation}
        {\rm rhs}=-2\sqrt{\tilde D}\bigg(a+(\sqrt a+\Gamma a^\mu)^2\bigg)^{1/2}.\label{rhs}
    \end{equation}
Now, assuming $\mu<1/2$, in the limit of $a\rightarrow 0$, we get
    \begin{eqnarray}
    {\rm lhs}\approx D_1 \Gamma^2 a^{2\mu-D_2 + \xi_2\Gamma a^\mu-2\xi_1 a^\mu }\approx -D_2,
\end{eqnarray}
whereas
\begin{equation}
    {\rm rhs}\approx -2\sqrt{\tilde D}\,\Gamma a^\mu.\label{b-expansion}
\end{equation}
Clearly, the assumption of $\mu<1/2$ does not yield a consistent solution. We now consider the opposite case with $\mu>1/2$. In this case from (\ref{lhs}), we find
\begin{eqnarray}
    &&{\rm lhs} \approx 2D_1 a - D_2 \frac{2\Gamma \sqrt a\,a^\mu}{2a} +\xi_2\sqrt a - \xi_1 \frac{2a^{3/2}}{2a} \nonumber \\&&\approx 2aD_1 -D_2\Gamma a^{\mu-1/2}+ (\xi_2-\xi_1)\sqrt a,\label{lhs1}
\end{eqnarray}
and
\begin{equation}
    {\rm rhs} \approx -2\sqrt{\tilde D}\sqrt{2a}. \label{rhs1}
\end{equation}
Now, using (\ref{b-expansion}),
\begin{equation}
    \tilde D=\bigg( D_1D_2-\frac{\xi_1\xi_2b^2}{a+b^2} \bigg)^{1/2}=D_1D_2-\frac{\xi_1\xi_2}{2} \label{tilde-d-near-origin}
\end{equation}
for $a\rightarrow 0$. 
Comparing (\ref{lhs1}) with (\ref{rhs1}), we find
\begin{equation}
    \mu=1,\;\;\Gamma =\bigg[2\sqrt{2\bigg( D_1D_2-\frac{\xi_1\xi_2}{2} \bigg)}+\xi_2-\xi_1\bigg]\frac{1}{D_2}.\label{small-ab-boundary}
\end{equation}
For $\xi_1=0=\xi_2$, the form of (\ref{small-ab-boundary}) is consistent with the corresponding result in Ref.~\cite{ab_jkb}.
Since $\xi_2-\xi_1$ can be positive or negative of arbitrary magnitude, (\ref{pattern-boundary}) can be {\it above} or {\it below} the Hopf line (\ref{hopf-curve}) near the origin: the location of  (\ref{pattern-boundary}) vis-\'a-vis (\ref{hopf-curve}) can be tuned by the chemotaxis coefficients. Consider for instance, the reciprocal case $\xi_1\xi_2>0$ with $\xi_1\xi_2\rightarrow 2D_1D_2$ from below. In this case,  
\begin{equation}
    b\approx \sqrt a + \frac{\xi_2-\xi_1}{D_2}a, \label{small-ab-boundary-reci}
\end{equation}
which can be made {\it above} or {\it below} the Hopf line~(\ref{hopf-curve}) near the origin by tuning $\xi_1,\xi_2$. This coincides with the Hopf boundary near the origin for $\xi_1=\xi_2$. In general, depending upon the relative magnitude of $\xi_1$ and $\xi_2$, (\ref{small-ab-boundary}) may lie above or below the Hopf line~(\ref{hopf-line}).  In the nonreciprocal case with $\xi_1<0,\,\xi_2>0$, the Hopf curve near the origin is {\it always below} the pattern line, since $\Gamma$ is positive definite for $\xi_1<0,\,\xi_2>0$.  On the other hand, for the nonreciprocal case with $\xi_1>0,\,\xi_2<0$, $\Gamma$ can be positive or negative, depending upon the magnitude of $\xi_1,\xi_2$.

We now consider the intersection of the Hopf line~(\ref{hopf-curve}) and pattern boundary~(\ref{pattern-boundary}). From (\ref{hopf-curve})~\cite{ab_jkb}
\begin{equation*}
    \frac{b^2-a}{a+b^2}=a+b^2.
\end{equation*}
Using this in the pattern boundary~(\ref{pattern-boundary}), we get
\begin{eqnarray}
    (D_1-D_2)(a+b^2)+\xi_2 b - \frac{2\xi_1b^3}{a+b^2}= - 2\sqrt{D_1D_2}\sqrt{a+b^2}.
\end{eqnarray}
Furthermore,
\begin{equation}
    \frac{2\xi_1b^2}{a+b^2}=\xi_1\frac{b^2+a+b^2-a}{a+b^2}=\xi_1 + \xi_1 (a+b^2).
\end{equation}
Using the above, the pattern boundary may be written as
\begin{equation}
    (a+b^2)(D_1-D_2)+b[\xi_2 -\xi_1 (a+b^2)]=-2\sqrt{D_1D_2}\sqrt{a+b^2},
\end{equation}
which may be considered as a quadratic equation in $\sqrt P,\;P\equiv a+b^2$. Since $P>0$, both $(D_1-D_2-b\xi_1$ and $(\xi_2-\xi_1)$ cannot be positive for a solution of $P$ to exist; at least one of them should be negative. Solving, we get

\begin{figure}[htb]
    \includegraphics[width=\columnwidth]{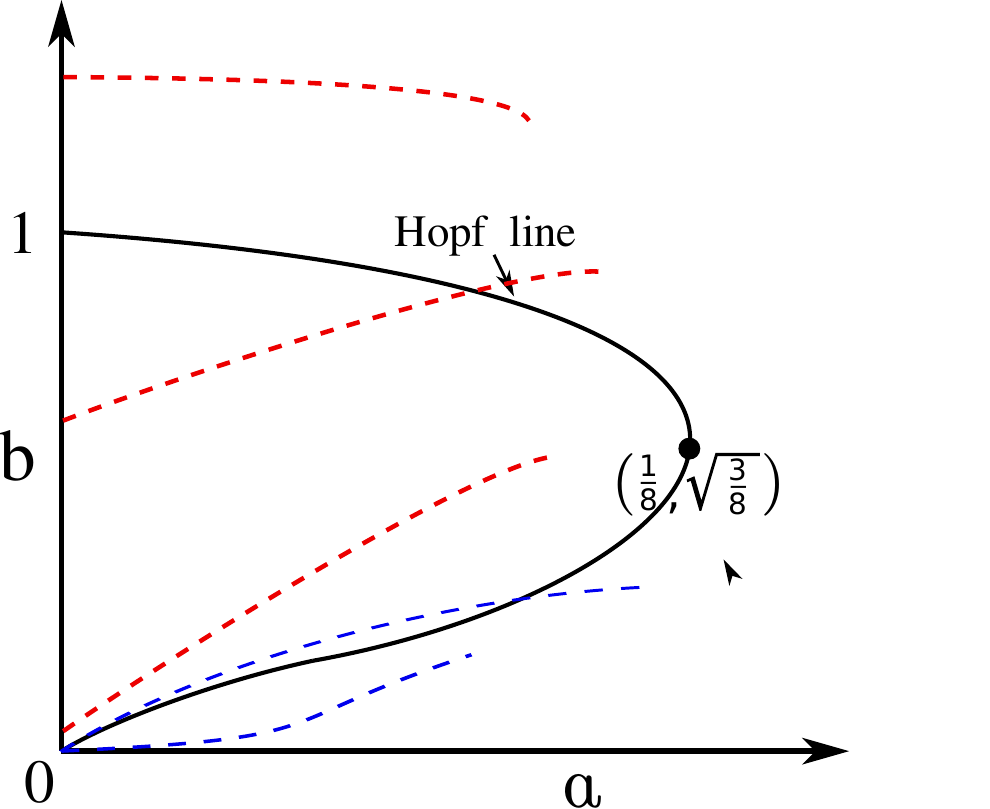}
    \caption{Schematic phase boundaries in the $a-b$ plane. The black continuous line is the Hopf boundary given by~\eqref{hopf-line}. The blue and red broken lines are the lower and upper branches of the pattern boundaries for various $D,\xi_1,\xi_2$. The blue broken lines can be above or below the Hopf boundary; the red broken lines can meet the $b$-axis close to or away from the origin (see text). }\label{phase-diag}
    \end{figure}

\begin{widetext}
\begin{equation}
P\equiv a+b^2    = \frac{1}{D_1-D_2-b\xi_1}\bigg[D_1D_2\pm\bigg\{D_1D_2-b(\xi_2-\xi_1)(D_1-D_2-b\xi_1)\bigg\}\pm 2\sqrt{D_1D_2}\bigg\{D_1D_2-b(\xi_2-\xi_1)(D_1-D_2-b\xi_1)\bigg\}^{1/2}\bigg]\equiv M (\text{say}).
\end{equation}

\end{widetext}
Notice that $M$, parametrized by the model parameters, is a function of $b$ only.
Then using the Hopf line equation~(\ref{hopf-curve}), we  get
\begin{equation}
b^2-a=(a+b^2)^2=M^2,    
\end{equation}
which in turn gives
\begin{equation}
    b^2 = (M+M^2)/2. \label{bsol}
\end{equation}
Note that (\ref{bsol}) is an {\it implicit} equation for $b$, which can be solved explicitly, e.g., graphical methods. The solution for $a$ can be found:
\begin{equation}
    a=(M-M^2)/2>0.
\end{equation}
For a physically acceptable solution, we must have
\begin{eqnarray}
&& P\geq 0,\\
    &&M\geq M^2,\nonumber\\
    && \Delta_1\equiv D_1D_2 - b(\xi_2-\xi_1) (D_1-D_2-b\xi_1)\geq 0.
\end{eqnarray}
The number of intersection points between the Hopf line~\eqref{hopf-curve} and the pattern boundary~\eqref{pattern-boundary} can be zero or more. Indeed, given (\ref{tilde-b}) and (\ref{small-ab-boundary}) above, the pattern boundary is entirely enclosed within the Hopf line (or vise-versa) if $\tilde b<1$ and $\Gamma>0$ ($\tilde b>1$ and $\Gamma<0$), together with $\Delta_1<0$. Else if $\tilde b<1$ and $\Gamma>0$ (or $\tilde b>1$ and $\Gamma<0$) together with $\Delta_1>0$, there should be even number of  intersection points. On the other hand, if $\tilde b<1$ and $\Gamma<0$, or $\tilde b>1$ and $\Gamma>0$, there should be {\it odd} numbers of intersection points between the two lines. A schematic phase diagram is shown in Fig.~\ref{phase-diag}. Detailed enumeration of the phase diagram is beyond the scope of the present work.

For simplicity and in order to reduce the number of parameters, in the DNS studies (see next Section), we have set  $\xi_2=0,\xi_1\neq 0$, $\xi_1=0,\xi_2\neq 0$, $\xi_1=\xi_2=\xi$ (fully reciprocal chemotaxis), or $\xi_1=-\xi_2=\xi)$ (extreme nonreciprocal chemotaxis), where $\xi$ itself can be positive and negative. Then, the selected wavevector $k_c^2$ has a complex dependence on the single chemotaxis parameter $\xi$. 

In the extreme limit, if $\xi_2=0$, then 
\begin{equation}
    k_c^2=\frac{1}{2D_1D_2}\bigg[-D_1(a+b^2)+D_2\frac{b^2-a}{a+b^2}+\frac{2\xi_1b^3}{a+b^2}\bigg],\label{extreme-xi1}
\end{equation}
and if $\xi_1=0$
\begin{equation}
    k_c^2=\frac{1}{2D_1D_2}\bigg[-D_1(a+b^2)+D_2\frac{b^2-a}{a+b^2}-\xi_2 b\bigg].\label{extreme-xi2}
\end{equation}
From (\ref{extreme-xi1}), we find that $k_c^2$ increases monotonically with $\xi_1>0$, or $k_c^2$ decreases monotonically as $|\xi_1|$ increases for $\xi_1<0$. In contrast, from (\ref{extreme-xi2}) we find that $k_c^2$ decreases monotonically as $\xi_2>0$ increases, or $k_c^2$ increases monotonically if $|\xi_2|$ increases for $\xi_2<0$.

\begin{figure}[htb]
\includegraphics[width=4cm]{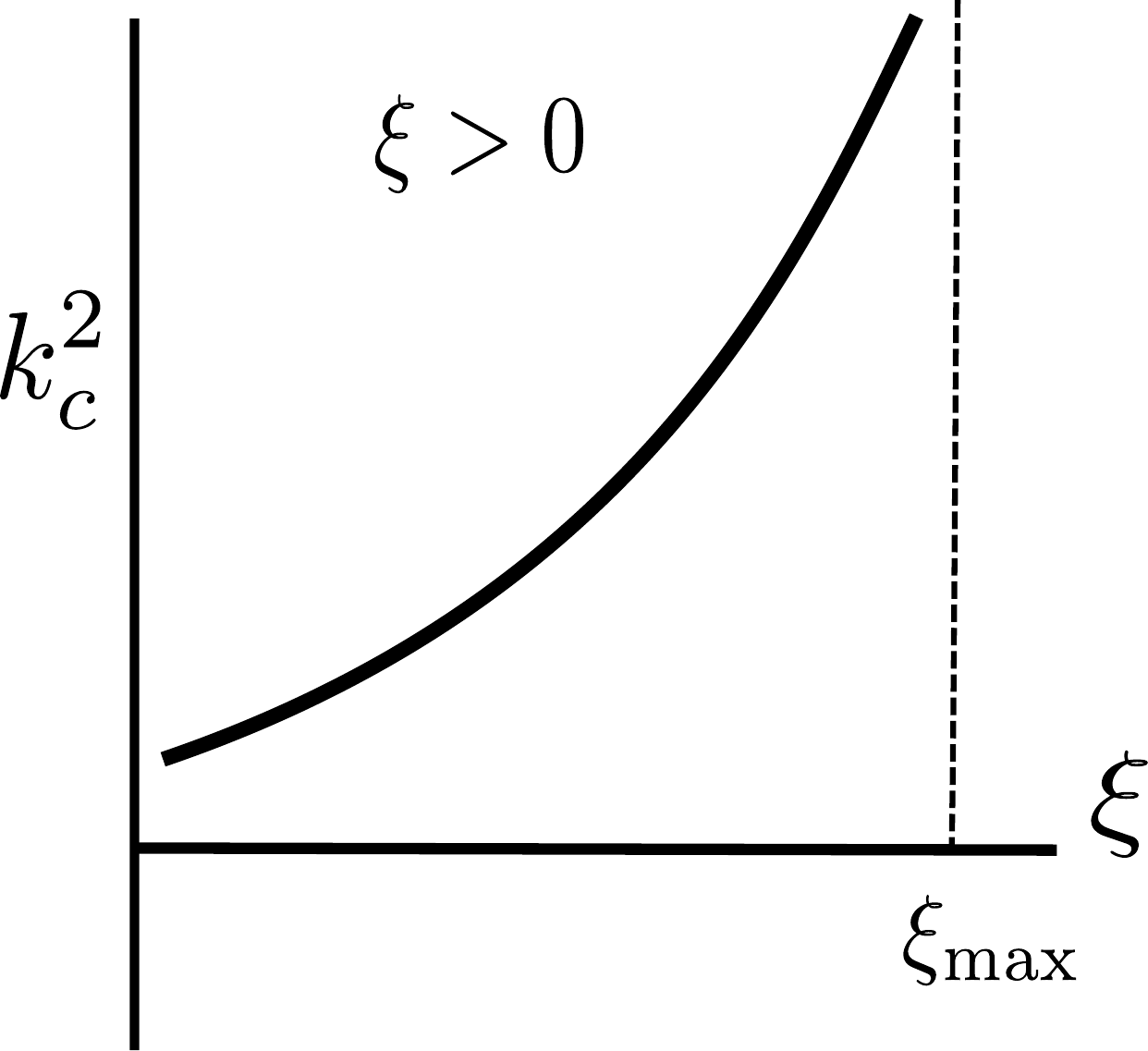}\hfill \includegraphics[width=4cm]{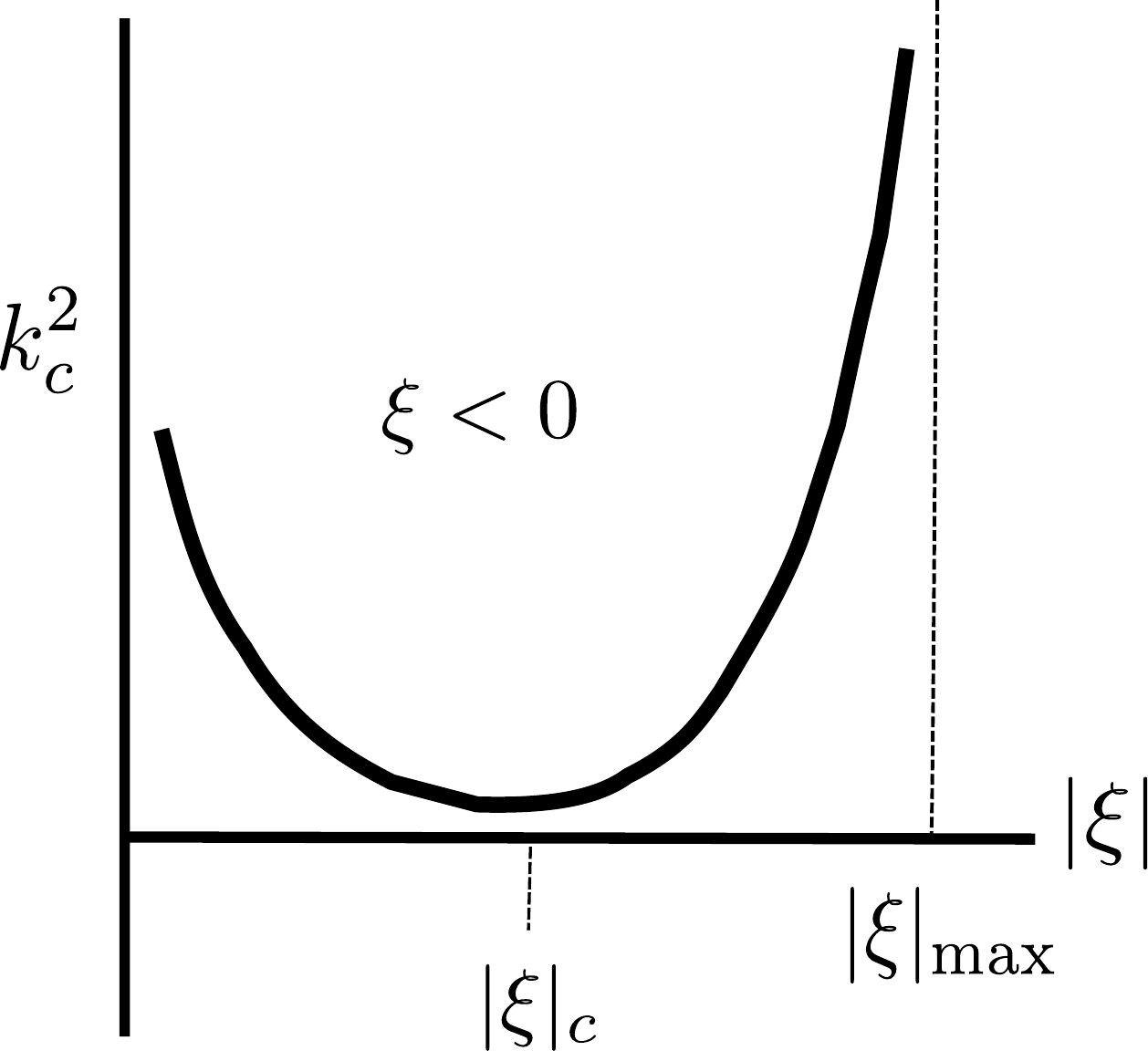}\\
\includegraphics[width=4cm]{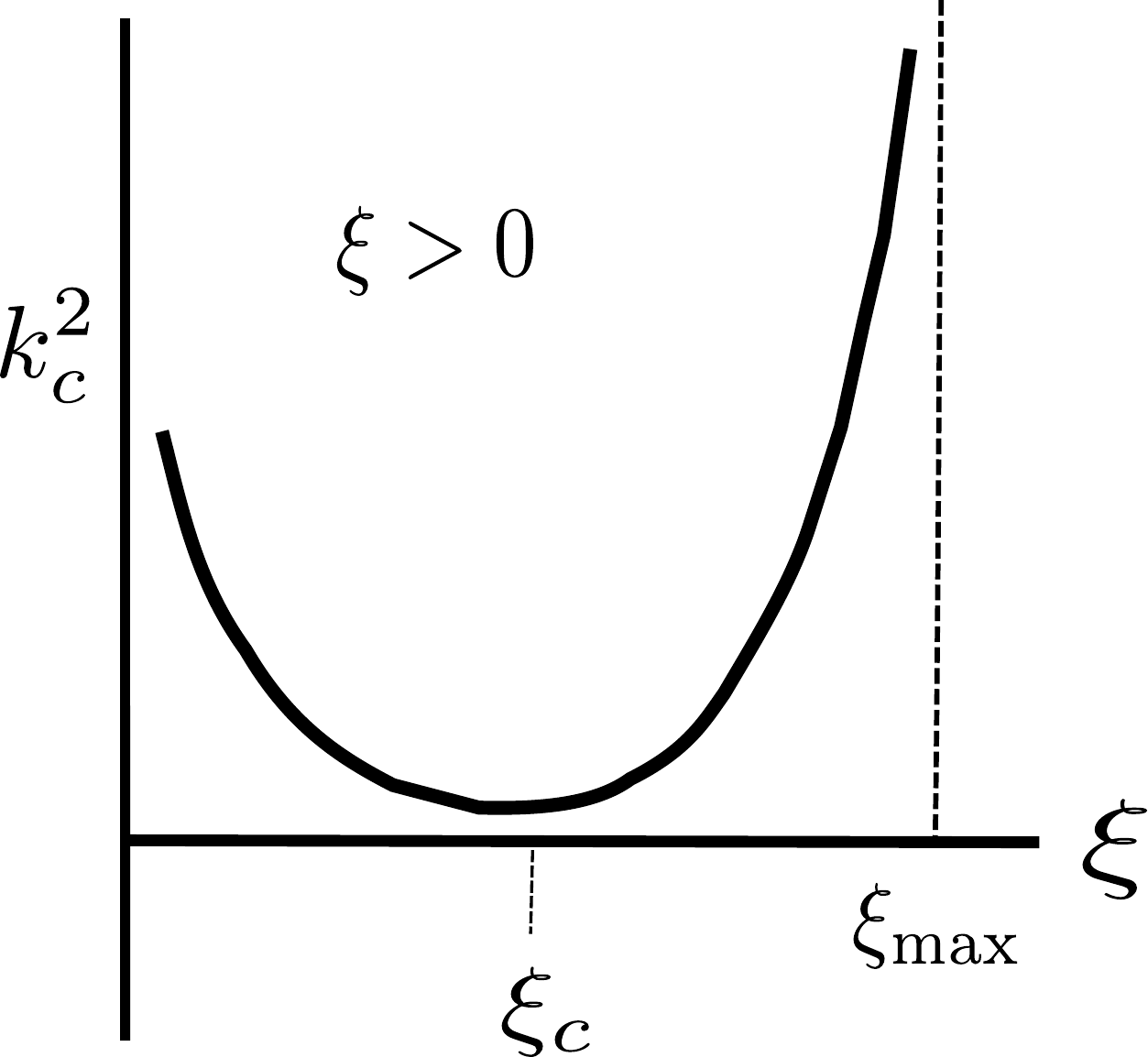}\hfill \includegraphics[width=4cm]{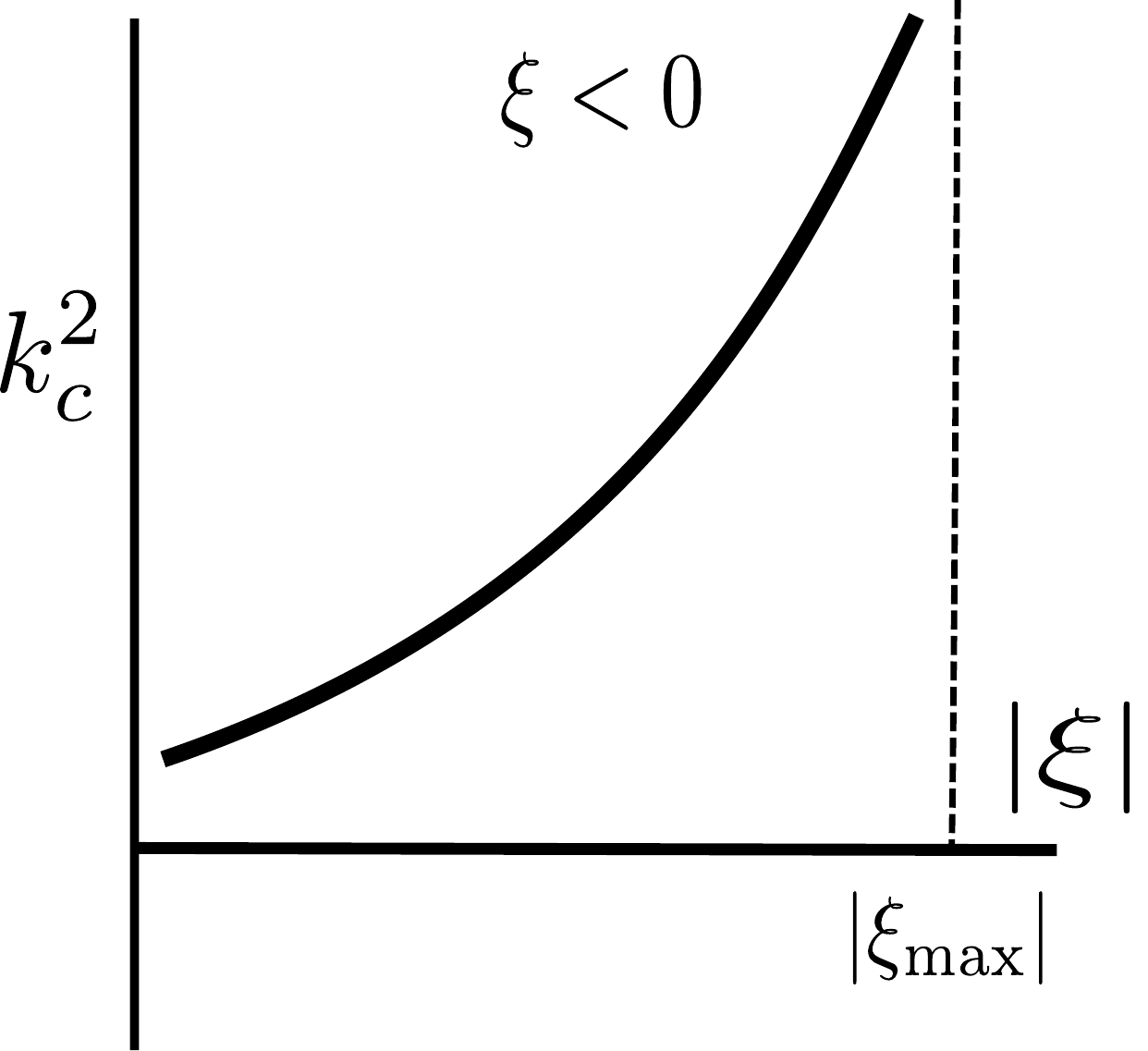}
\caption{Schematic variation of $k_c^2$ with $\xi$ in the fully reciprocal chemotaxis case : (top, left) (assuming $b^2>a$) $\xi>0$ with $k_c^2$ rising monotonically with $\xi$, diverging as $\xi\rightarrow \xi_\text{max}=D_1D_2(a+b^2)/b^2$, (top, right) (assuming $b^2>a$) $\xi<0$ with $k_c^2$ depending nonmonotonically on $\xi$, having a minimum at $\xi=\xi_\text{reci}$ and eventually diverging as $\xi\rightarrow \xi_\text{max}=\bigg[D_1D_2(a+b^2)/b^2\bigg]^{1/2}$, (bottom, left) (assuming $b^2<a$) $\xi>0$ with $k_c^2$ depending nonmonotonically on $\xi$, having a minimum at $\xi=\xi_\text{reci}$ and eventually diverging as $\xi\rightarrow \xi_\text{max}$, (bottom, right) (assuming $b^2<a$) $\xi<0$ with $k_c^2$ rising monotonically with $\xi$, diverging as $\xi\rightarrow \xi_\text{max}=D_1D_2(a+b^2)/b^2$.  }
\label{kc-xi-plot-reci}
\end{figure}

When both $\xi_1,\xi_2$ are nonzero, the dependence of $k_c^2$ on them is more complex. Considering $\xi_1=\xi=\xi_2$,
\begin{equation}
    k_c^2=\frac{-D_1(a+b^2)+D_2\frac{b^2-a}{a+b^2}+\xi b \frac{b^2-a}{a+b^2}}{2[D_1D_2-\frac{\xi^2b^2}{a+b^2}]},\label{kc-xi-DNS-reci}
\end{equation}
in the fully reciprocal chemotaxis case, and with $\xi_1=\xi=-\xi_2$
\begin{equation}
    k_c^2=\frac{-D_1(a+b^2)+D_2\frac{b^2-a}{a+b^2}+\xi b \frac{a+3b^2}{a+b^2}}{2[D_1D_2+\frac{\xi^2b^2}{a+b^2}]}\label{kc-xi-DNS-nonreci}
\end{equation}
in the extreme nonreciprocal chemotaxis case. Equations~(\ref{kc-xi-DNS-reci}) and (\ref{kc-xi-DNS-nonreci}) reveal how the pattern may change as the chemotaxis interaction parameters are varied. Consider first Eq.~(\ref{kc-xi-DNS-reci}) for the reciprocal case. For $\xi>0$ if $b^2<a$, the numerator of (\ref{kc-xi-DNS-reci}) that depends {\it linearly} on $\xi>0$, {\it decreases} as $\xi$ rises from zero, which has the effect of {\it decreasing} $k_c^2$. The denominator in (\ref{kc-xi-DNS-reci}) that depends {\it quadratically} on $\xi$ also {\it decreases}, which however tends to {\it increase} $k_c$. Therefore, for sufficiently small $\xi$, $k_c^2$ should {\it decrease} with increasing $\xi$, but for sufficiently large $\xi>\xi_c|_\text{ reci}$, a finite crossover threshold, $k_c^2$ should start to rise. Thus $k_c^2$ should have a minimum at $\xi=\xi_c|_\text{ reci}>0$.  By setting $\partial k_c^2/\partial\xi=0$ and using (\ref{kc-xi-DNS-reci}), we get
\begin{eqnarray}
 &&\xi_c|_\text{ reci}=   \frac{a+b^2}{b(b^2-a)}\bigg[D_1(a+b^2)-D_2\frac{b^2-a}{a+b^2}\nonumber \\&& + \{[D_1 (a+b^2) - D_2\frac{b^2-a}{a+b^2}]^2 -D_1D_2 \frac{(b^2-a)^2}{a+b^2}\}^{1/2}\bigg].\nonumber \\
\end{eqnarray}
This holds for both $\xi>0$ for $b^2<a$ and $\xi<0$ for $b^2>a$. For real $\xi_c$, we must have the discriminant
\begin{eqnarray}
    \delta_\text{reci}\equiv \bigg[D_1 (a+b^2) - D_2\frac{b^2-a}{a+b^2}\bigg]^2 -D_1D_2 \frac{(b^2-a)^2}{a+b^2}>0.
\end{eqnarray}
We then find
\begin{equation}
    k_c^2|_\text{reci} = \frac{\sqrt{\delta_\text{reci}}}{2\bigg[D_1D_2 -\frac{\xi_c^2|_\text{reci} b^2}{a+b^2}\bigg]},\label{kc-reci}
\end{equation}
which gives the minimum of $k_c^2$ for $\xi>0$ when $b^2<a$, or for $\xi<0$ when $b^2>a$. Now consider a system of linear size $L$, such that $Lk_c|_\text{reci}\ll 1$, or equivalently $L/\lambda_c|_\text{reci}\ll 1$, where the {\it maximum} wavelength $\lambda_c|_\text{reci}\equiv 2\pi/k_c|_\text{reci}$.  In this case, the system should appear to be practically uniform. Thus, for this finite size system, one could observe a re-entrant behavior as $\xi$ is tuned from 0 to large values through $\xi_\text{reci}$. 
A schematic plot of $k_c^2$ as a function of $\xi$ in the fully reciprocal case is shown in Fig.~\ref{kc-xi-plot-reci}. 

We therefore find that the dependence of $k_c^2$ on $\xi$ changes drastically across the curve $b^2=a$. Since the pattern region in the $a-b$ plane is largely confined to $b^2>a$, one can observe the above nonmonotonic behavior typically with $\xi<0$, i.e., for attractive chemotaxis. For $\xi>0$, i.e., for repulsive chemotaxis, any observation of such nonmonotonic behavior should require careful tuning of $a,b$.

\begin{figure}[htb]
\includegraphics[width=4cm]{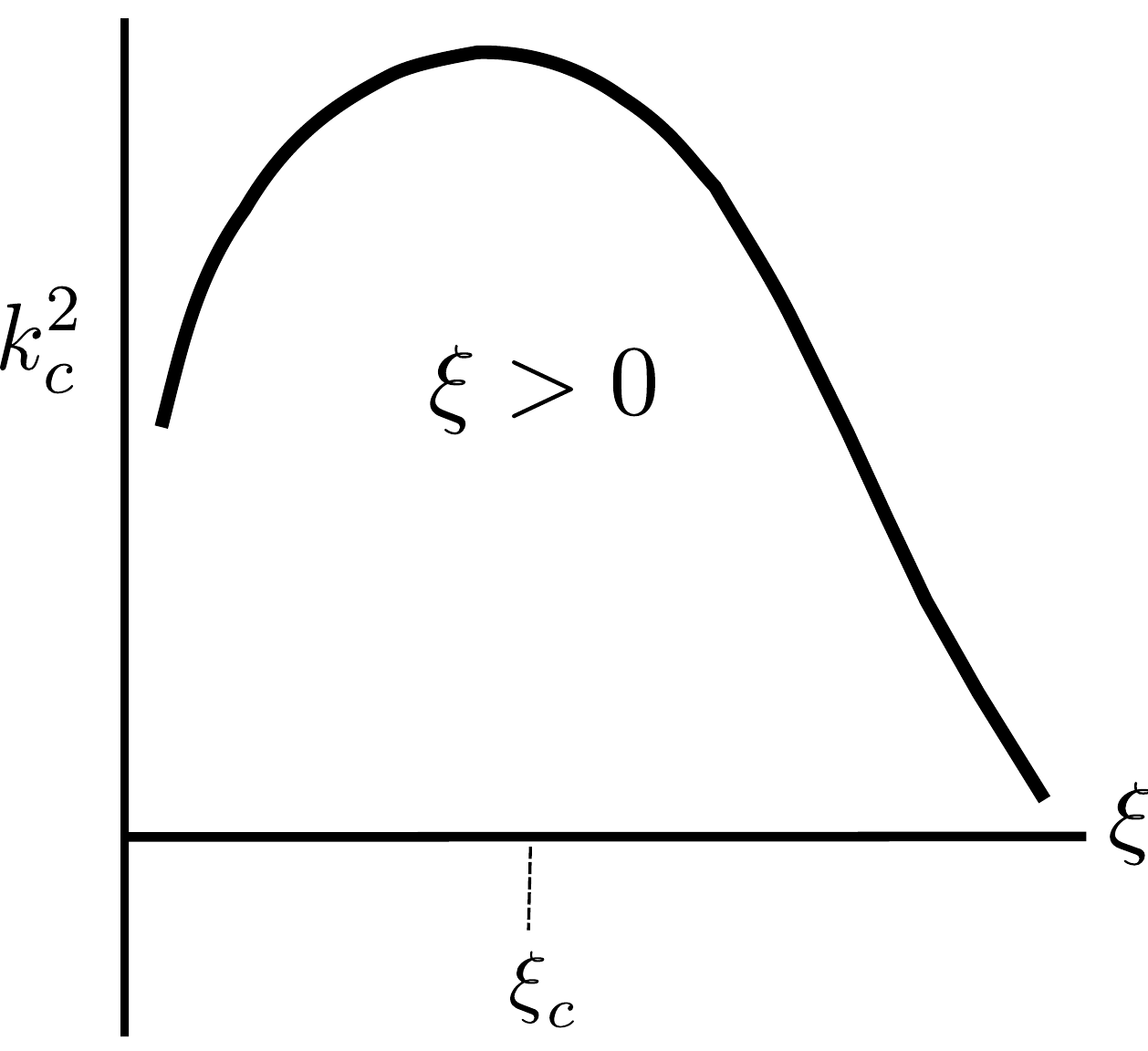}\hfill \includegraphics[width=4cm]{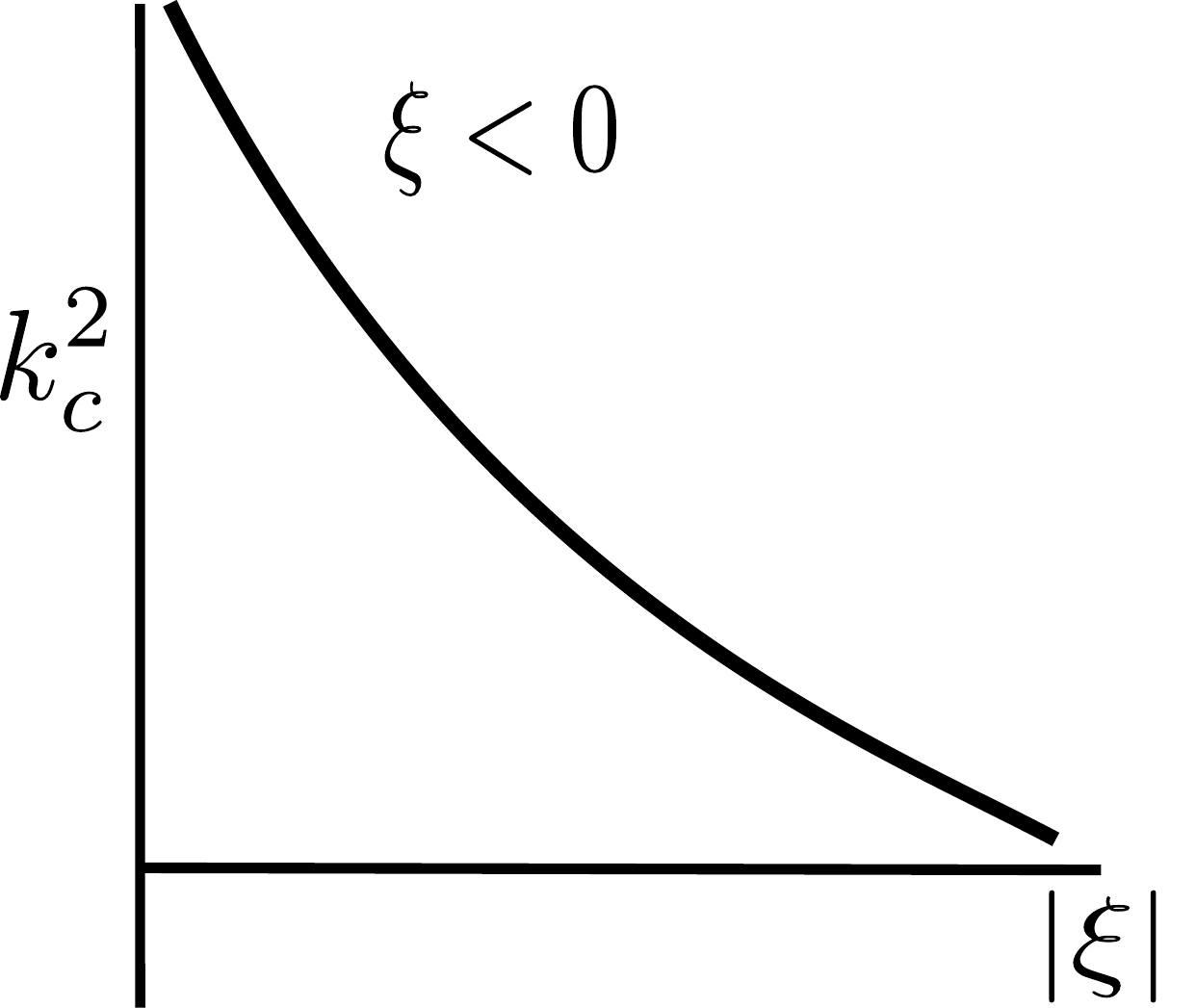}
\caption{Schematic variation of $k_c^2$ with $\xi$ in the extreme nonreciprocal chemotaxis case: (i) $\xi>0$ with $k_c^2$ depending nonmonotonically on $\xi$ and having a maximum at $\xi=\xi_\text{nonreci}$, eventually approaching zero for very large $\xi$, (ii) $\xi<0$ with $k_c^2$
decreasing monotonically with $\xi$, eventually vanishing for large $|\xi|$.}\label{kc-xi-plot-nonreci}
\end{figure}

We now consider the corresponding extreme nonreciprocal case and study the $\xi$-dependence of (\ref{kc-xi-DNS-nonreci}). For $\xi>0$, the numerator of (\ref{kc-xi-DNS-nonreci}) {\it grows} linearly with $\xi$, which {\it enhances} $k_c^2$. However, the denominator of (\ref{kc-xi-DNS-nonreci})  {\it grows} quadratically with $\xi$, which {\it reduces} $k_c^2$. This suggests a maximum of $k_c^2$ at some finite $\xi=\xi_c|_\text{ nonreci}$. A schematic plot of $k_c^2$ as a function of $\xi$ in the fully reciprocal case is shown in Fig.~\ref{kc-xi-plot-nonreci}. 

Setting $\partial k_c^2/\partial\xi=0$, we obtain
\begin{eqnarray}
    &&\xi_c|_\text{nonreci}=\frac{a+b^2}{b(a+3b^2}\bigg[D_1(a+b^2)-D_2\frac{b^2-a}{a+b^2}\nonumber \\&&+\{[D_1(a+b^2)-D_2\frac{b^2-a}{a+b^2}]^2+ (a+3b^2)\frac{D_1D_2b^2}{a+b^2}\}^{1/2}\bigg].\nonumber \\
\end{eqnarray}
This holds for both $b^2>a$ and $b^2<a$. We define
\begin{equation}
    \delta|_\text{nonreci}=\left[D_1(a+b^2)-D_2\frac{b^2-a}{a+b^2}\right]^2+ (a+3b^2)\frac{D_1D_2b^2}{a+b^2}.
\end{equation}
We then find
\begin{equation}
    k_c^2|_\text{nonreci}=\frac{\sqrt{\delta_\text{nonreci}}}{2[D_1D_2+\frac{\xi_\text{nonreci}^2b^2}{a+b^2}]},\label{kc-nonreci}
\end{equation}
which gives the maximum of $k_c^2$ for $\xi>0$. We have argued above that in the fully reciprocal chemotaxis case, for a system of linear size $L\ll \lambda_c|_\text{reci}$, the system can make a transition from a patterned state for low $\xi$ to a patterned state for high $\xi$ via a uniform state at intermediate $\xi\approx \xi_c|_\text{reci}$. Interestingly, the opposite situation can arise in the extreme nonreciprocal case with $\xi>0$. Consider a system of linear finite size $L >\lambda_c|_\text{nonreci}$, where $\lambda_c|_\text{nonreci}\equiv 2\pi/k_c|_\text{nonreci}$, but $L\ll 2\pi k_c^{-1}(\xi=0)$ or $L\ll 2\pi k_c^{-1}\xi\rightarrow \infty$. In this case, for very small of large $\xi$, the system should appear uniform, but for intermediate values of $\xi>0$, in the neighborhood of $\xi_c|_\text{nonreci}$ a pattern should form. We thus note that although nonmonotonic dependence of $k_c^2$ on $\xi$ is possible for fully reciprocal or extreme nonreciprocal cases, the nature of the nonmonotonic dependence is very different in the two cases. 

In a simplified situation, one could consider one species chemotactically interacting with the other one (attractive or repulsive), whereas the second species would not undergo any chemotaxis. This requires setting either $\xi_1$ or $\xi_2$ to zero. In this limit, clearly $\tilde D=D_1D_2>0$ and therefore the reciprocal chemotaxis-induced instability cannot arise. Furthermore, $k_c^2$ in (\ref{qc}) shows that in this simplified case, any nonmonotonic behavior disappears; $k_c^2$ either rises or decreases monotonically with $\xi_1$ or $\xi_2$, whichever is nonvanishing.

The pattern should also depend on the diffusivities $D_1, D_2$, due to the dependence of $k_c^2$ on $D_1, D_2$; see Eq.~\eqref{qc}. The following cases are possible:

(i) $\xi_1=\xi_2=\xi>0$ (fully reciprocal chemotaxis). Since $\tilde D>0$ is the condition for the existence of a pattern with finite selected wavevector $k_c$, we must have
\begin{equation}
    D_1D_2=\frac{\xi^2b^2}{a+b^2}
\end{equation}
as the instability threshold. Thus for
\begin{equation}
    D_1\leq \xi \bigg[\frac{b^2}{D(a+b^2)}\bigg]^{1/2} \label{inst-1}
\end{equation}
all the wavevector modes get unstable. This is the instability found in the reciprocal chemotaxis case, as discussed above.

(ii) $\xi_1=\xi_2=\xi>0$ (fully reciprocal chemotaxis). Since $\tilde D>0$ for a pattern as in the previous case, we once again obtain (\ref{inst-1}) as the condition for which all the wavevector modes become unstable. We thus see that for sufficiently low diffusivities, the fully reciprocal chemotaxis can get unstable. Once again, this is the instability found in the reciprocal chemotaxis case, as discussed above.

(iii) $\xi_1=-\xi_2=\xi$ (extreme nonreciprocal chemotaxis), where $\xi$ can be positive or negative, which form two distinct cases. In both these cases, $\tilde D>0$ {\it always}. Thus, the instability found in the reciprocal case {\it cannot} arise now. We now consider the numerator of (\ref{qc}) with $\xi_1=-\xi_2=\xi$. If $\xi<0$ and for a sufficiently low $D_1$ and a fixed $D$, the numerator in (\ref{qc}) can be negative for specific choices of $a,b$, ruling out any solution for $k_c^2$. Thus, no pattern is possible, and only a uniform state can be realized. On the other hand, if $\xi>0$, for sufficiently small $D_1, D_2$, the numerator in (\ref{qc}) can be dominated by $\xi$, making it negative. As a result, $k_c^2$ is always nonzero, and a pattern should form. 

Lastly, from (\ref{qc}), $k_c^2$ depends linearly on $D_1,D_2$ in the numerator of (\ref{qc}), but depends bilinearly on $D_1,D_2$ in its denominator. Thus,
it is clear that for very high $D_1, D_2$, $k_c^2$ becomes small, eventually vanishing, and hence a uniform state ensues. For intermediate $D_1, D_2$, a finite $k_c^2$ may exist that depends upon the various model parameters, for which a patterned state should follow. Our DNS studies conform to these expectations.  

Before we close this Section, we revisit the linear stability results for the nonchemotactic case, i.e., $\xi_1=0=\xi_2$; see also Ref.~\cite{ab_jkb}. In this limit, the selected wavevector is obtained by setting $\xi_1=0=\xi_2$ in (\ref{qc}) above, giving
\begin{equation}
    k_c^2=-\frac{\Gamma_1^0}{2\sqrt{D_1D_2}},\label{qc0}
\end{equation}
where
\begin{eqnarray}
    \Gamma_1^0=D_1(a+b^2)-D_2\frac{b^2-a}{a+b^2}.\label{Gam1-0}
\end{eqnarray}
The phase boundary between the patterned state and uniform state in the $a-b$ plane is now given by
\begin{equation}
    D_1(a+b^2)-D_2\frac{b^2-a}{a+b^2}= -2\sqrt { D_1D_2}\sqrt{a+b^2}.\label{pattern-boundary-0}
\end{equation}
The phase boundary (\ref{pattern-boundary-0}) intersects the $b$-axis at $b=(-1+\sqrt 2)\sqrt{D}$.
Finally, the intersection between the Hopf line (\ref{hopf-curve}) and the pattern boundary~(\ref{pattern-boundary-0}) can now be explicitly calculated, giving the point of intersection as 
\begin{eqnarray}
    a_c&=& \frac{2D_1D_2}{(D_1-D_2)^2}- \frac{8D_1^2D_2^2}{(D_1-D_2)^4},\label{ac-0}\\
    b_c^2&=& \frac{2D_1D_2}{(D_1-D_2)^2}+ \frac{8D_1^2D_2^2}{(D_1-D_2)^4},\label{bc-0}
\end{eqnarray}
   see also Ref.~\cite{ab_jkb}.

\section{Numerical studies}\label{dns}


In this Section, our aim is to numerically solve Eqs.~(\ref{mod1}) and (\ref{mod2}) to verify and validate the analytical results obtained by linear stability analysis. We use a square geometry with $L=128$ in our DNS studies.


Within the linearized treatments as above, modes $\exp(i{\bf k_c}\cdot {\bf x})$ and $\exp(i\omega_0t)$ are independent. However, nonlinear terms should mix them, potentially giving rise to various scenarios with pure oscillatory, pure stationary patterns, or a mixture of them. In the nonlinear theory, these are expected to be controlled by the appropriate fixed points in the parameter space, as speculated in Ref.~\cite{ab_jkb}. Since our principal focus is on the properties of stationary patterns, our region of interest is the basin of attraction of the putative fixed point that controls the stationary patterns. However, an analytic enumeration of this putative fixed point and its basin of attraction is algebraically extremely complicated and is also not particularly insightful. Instead, we systematically explore various regions of the phase space and look for steady states with stationary patterns. 

\begin{figure}[htb]
\includegraphics[width=\columnwidth]{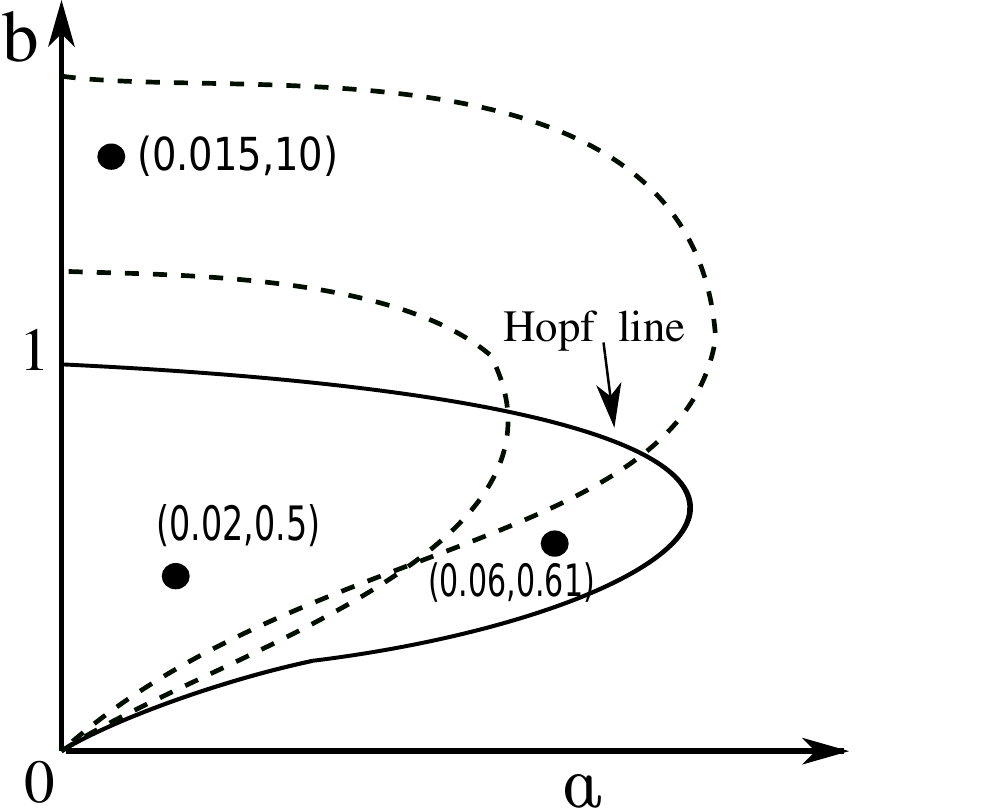}
\caption{Phase diagram in the $(a,b)$ plane for various $D_1$ and $D$. Different points at which the DNS studies were done are marked. 
}\label{phase-diag-1}
\end{figure}

We first explore the patterns and the oscillatory instabilities {\it without} chemotaxis, i.e., $\xi_1=0=\xi_2$. The phase diagram in this case is shown in Fig.~\ref{phase-diag-1}.
In this limit, our DNS studies were done at three different points in the $(a,b)$ plane: 

(i) $a= 0.015, b = 10$, $D_1=0.01$ and $D=1000$. At this point ${\cal Tr}(k^2)<0$. This point falls {\it outside} the region of oscillatory instability in the phase diagram in Fig.~\ref{phase-diag-1}. Thus, oscillatory instability is expected to be absent. However, this point is within the region of linear instability with a finite $k_c$, giving rise to the possibility of steady patterns. 

(ii) $a=0.06, b=0.61$, $D_1=1$ with $D=4,\, 1000$. This point falls {\it within} the region of oscillatory instability but outside the pattern forming region in the phase diagram in Fig.~\ref{phase-diag-1}. Thus, oscillatory instability is expected to be visible. However, this point lies outside the region of linear instability with a finite $k_c$, meaning a steady pattern is {\it not} expected to be observed. 

(iii) $a=0.02$, $b=0.50$, $D_1=0.2$ with $D=200$. This point falls {\it within} both the oscillatory instability and the linear instability with a finite $k_c$ regions in the phase diagram in Fig.~\ref{phase-diag-1}. Thus, the oscillatory instability and steady pattern are expected to be observed.

We now discuss the patterns in the presence of chemotaxis. For our DNS studies, we choose two sets of points in the $(a,b)$ plane: 

(i) $a= 0.015, b = 10$, $D_1=0.01$ with $D=1000$. Since this point falls {\it outside} the region of oscillatory instability in the phase diagrams in Fig.~\ref{phase-diag}, the oscillatory instability may be absent even in the presence of chemotaxis. However, this point lies {\it within} the respective pattern line, giving rise to the possibility that a steady pattern may be observed. Our DNS studies confirm the lack of any oscillatory instabilities, but show steady patterns, which makes it suitable to study the effects of chemotaxis on the patterns. Most of the results in this article are obtained at this point in the phase space.

(ii) $a=0.02, b=0.5$. Since this point falls {\it within} the oscillatory instability region in the phase diagram in Fig.~\ref{phase-diag-1}, both the oscillatory instability and a pattern may be visible even in the presence of chemotaxis. Our DNS studies confirm this and also show steady patterns, which makes it suitable to study the interplay between the two instabilities. 


Below, we present our results from our DNS studies with random initial conditions. We have further explored the role of initial conditions in the ensuing pattern formations by varying qualitatively and macroscopically different initial conditions, which are presented in Appendix~\ref{ini_cond}. 

The model equations (\ref{mod1}) and (\ref{mod2}) are isotropic. Therefore, na\"ively it is expected that the steady states will conform to the symmetries of the governing equations (\ref{mod1}) and (\ref{mod2}). This suggests that any ensuing pattern should contain spot-like structures, which violates translational invariance (as expected naturally for a patterned state, and retain the isotropy in the sense that a patterned state with spots has {\it no} preferred direction. Surprisingly, however, we find that the ensuing patterns may be {\it stripes}, which explicitly violate rotational invariance.

\subsection{Nonchemotactic limit: $\xi_1=0=\xi_2$}

In the nonchemotactic limit ($\xi_1=0=\xi_2$), the linear stability analysis in the previous Section predicts a spatially uniform Hopf bifurcation with a frequency $\omega_0$ as given in \eqref{hopf-freq} and a time-independent saddle-node Turing instability with a preferred wavevector given in (\ref{qc0}) above. DNS studies of the nonlinear PDEs (\ref{mod1}) and (\ref{mod2}) reveal sustained oscillations and/or steady patterns, depending on $a,b$.

\begin{figure}[!ht]
    \centering
\includegraphics[width=\columnwidth]{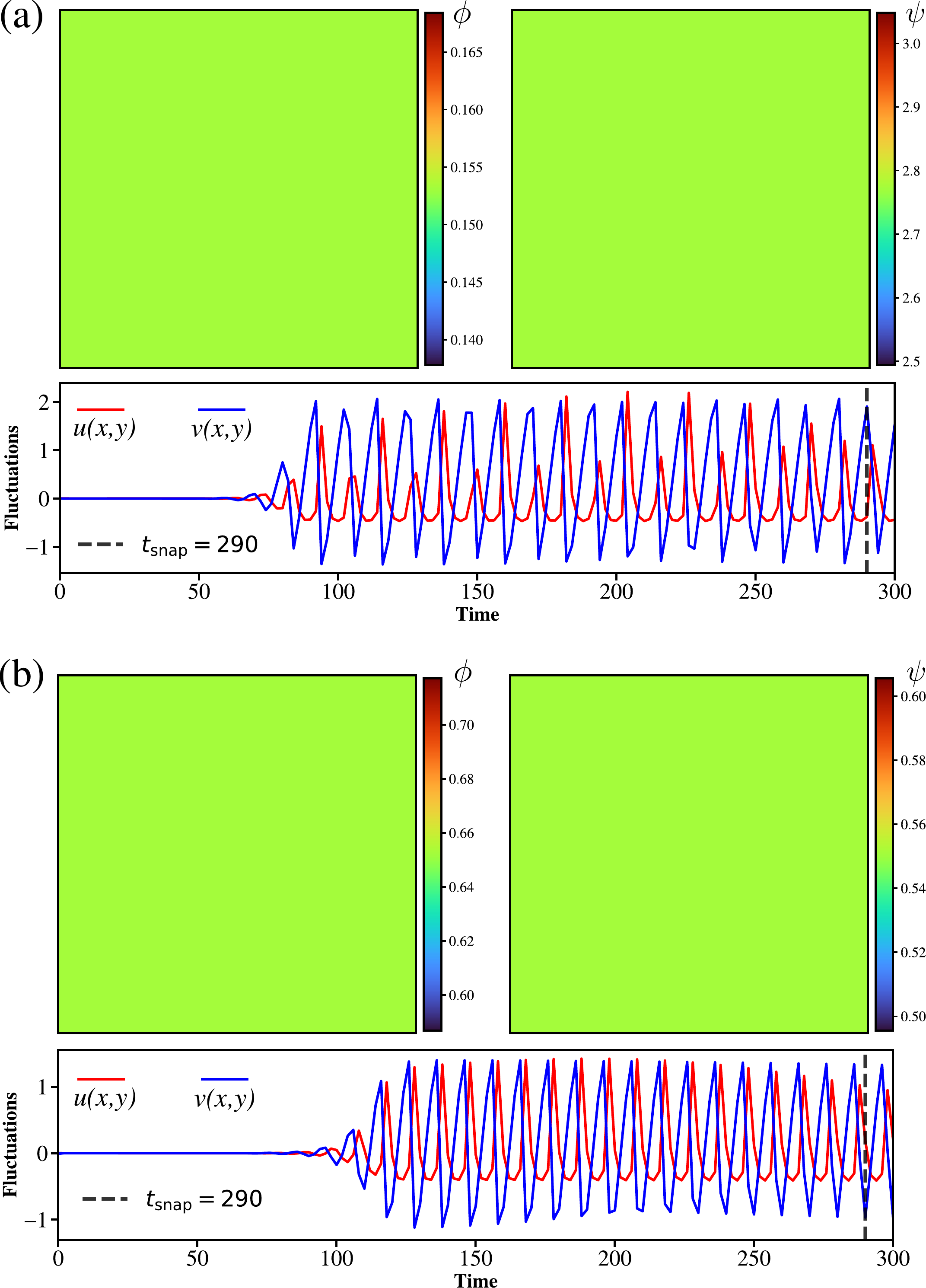}
    \caption{
    (Color online)
    Temporal oscillations in the absence of chemotactic coupling ($\xi_1=\xi_2=0$) without spatial pattern formation.
    Panels (a) and (b) show snapshots of the density fields $\phi$ and $\psi$ at $t_{\mathrm{snap}}=290$, together with the time evolution of local density fluctuations $u(x,t)$ and $v(x,t)$ at a representative spatial location.
    In both cases, the density fields remain spatially homogeneous, while the system exhibits sustained oscillations in time.
    (a)~For $a=0.06$, $b=0.61$, $D_1=1.0$, and $D=4.0$, oscillations emerge after a transient and reach a stable periodic regime.
    (b)~For $a=0.04$, $b=0.61$, $D_1=1.0$, and $D=1000.0$, similar temporal oscillations persist despite the large diffusion contrast.
    These results demonstrate that, in the non-chemotactic limit, the system can sustain temporal oscillations without developing spatial structure.}\label{fig:fig_solo_osci}
\end{figure}

 \begin{figure}[!ht]
    \centering
\includegraphics[width=\columnwidth]{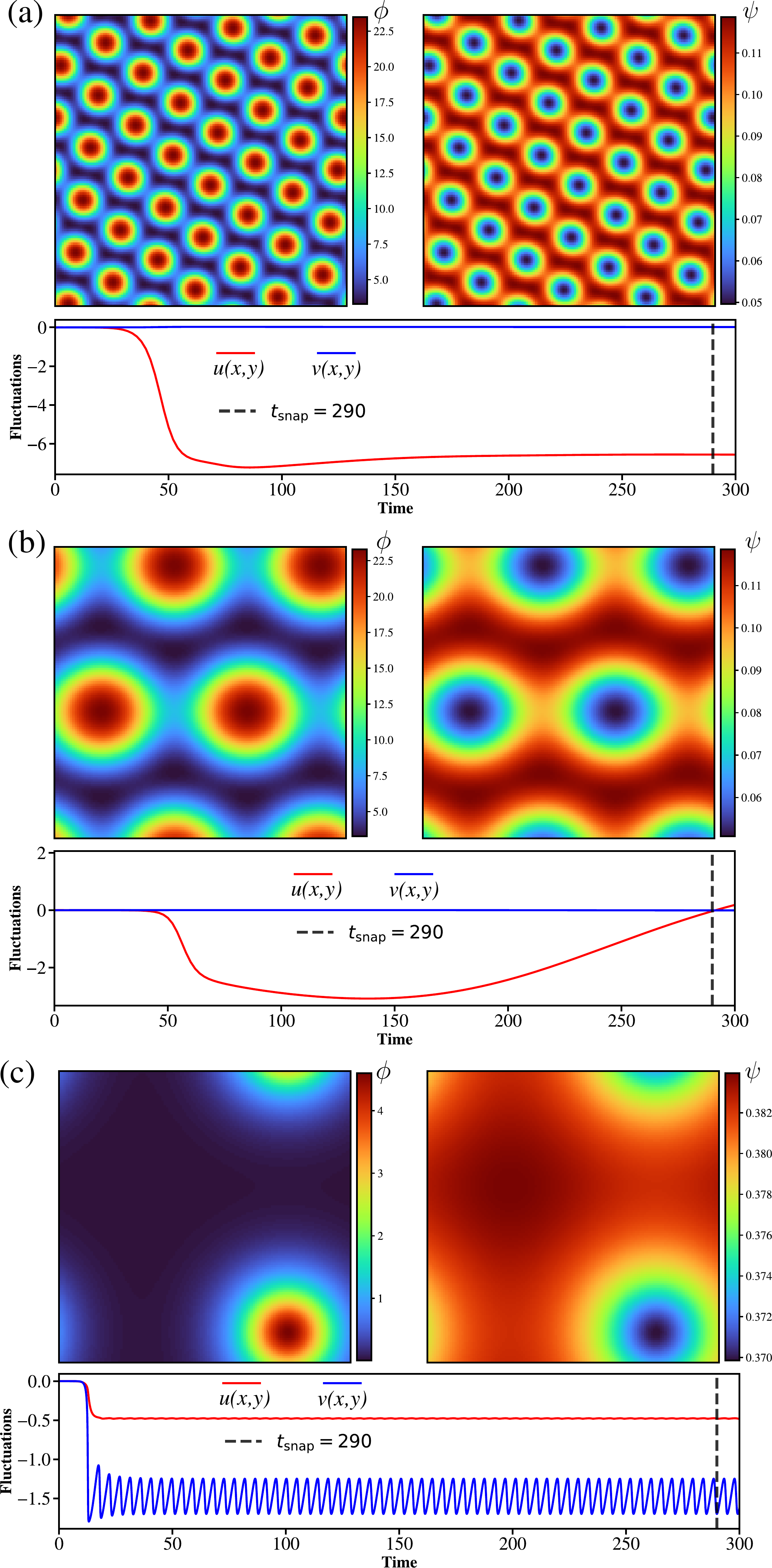}
    \caption{
    (Color online)
    Density fields and temporal dynamics in the absence of chemotactic coupling ($\xi_1=\xi_2=0$).
    Panels (a)--(c) show snapshots of the densities $\phi$ and $\psi$ at $t_{\mathrm{snap}}=290$, together with the time evolution of local fluctuations $u(x,t)$ and $v(x,t)$ at a representative spatial point.
    (a)~For $a=0.015$, $b=10.0$, $D_1=0.01$, and $D=10.0$, the system exhibits stable spot-like patterns with rapid relaxation to a steady state.
    (b)~At intermediate diffusion ($D_1=0.10$, $D=100.0$), ordered structures persist while the dynamics shows slow relaxation toward equilibrium.
    (c)~For different reaction parameters ($a=0.02$, $b=0.50$, $D_1=0.20$, $D=200.0$), the system displays a coexistence of spatially localized structures and sustained temporal oscillations, indicating a regime of oscillatory patterned states.
  }\label{fig:fig_pattern_osci}
\end{figure}
Figure~\ref{fig:fig_solo_osci} shows steady oscillations but spatially homogeneous states (i.e., no patterns). 
Figure~\ref{fig:fig_pattern_osci} shows how varying $D$ for fixed $D_1, a,b$ can give drastically different long time limit behavior of the system, ranging from steady patterns to sustained oscillations but a spatially homogeneous state. 

\begin{figure}[htb]
    \centering
    \includegraphics[width=0.9\columnwidth]{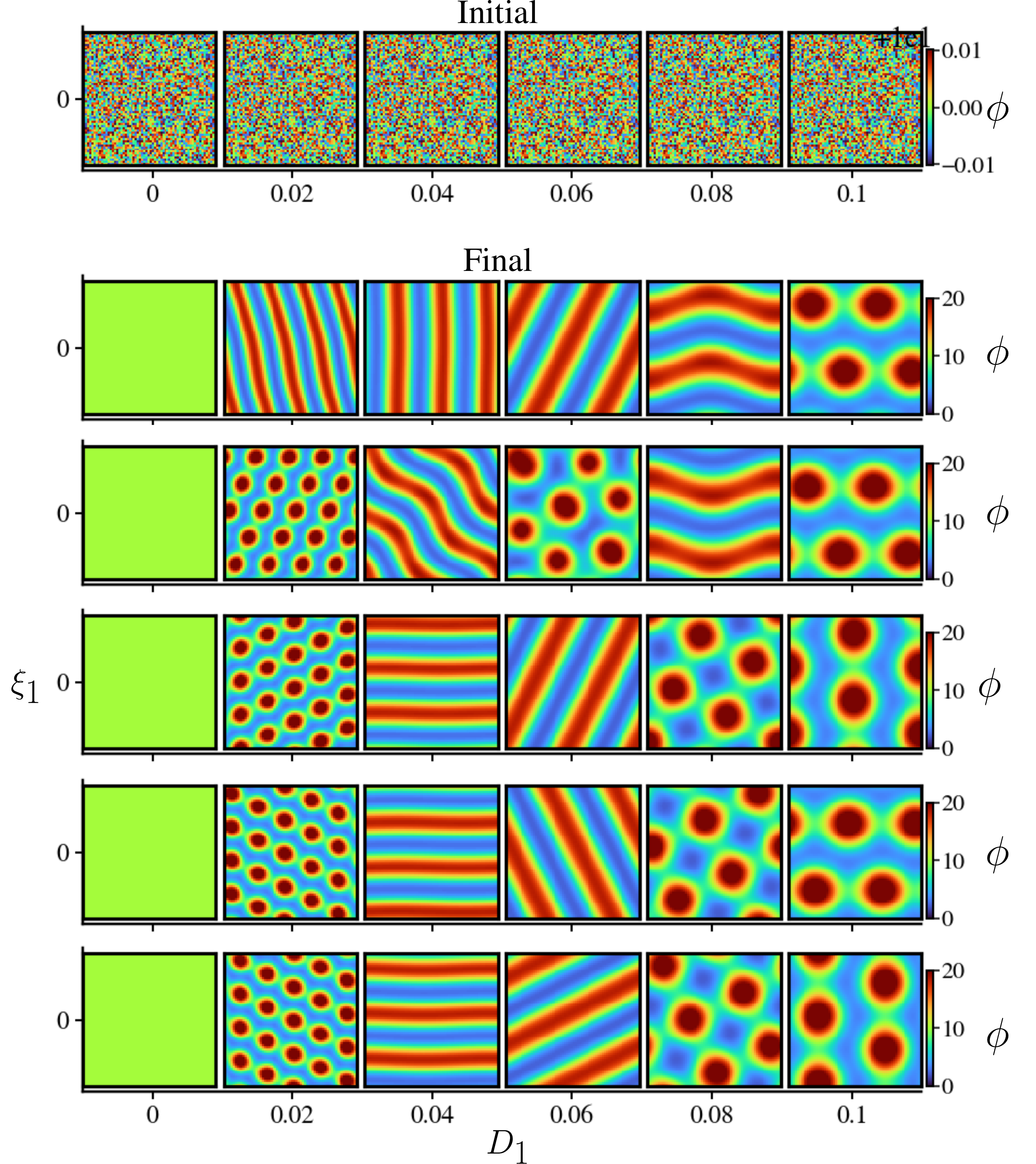}
    \caption{
    (Color online)
    Pattern selection in the nonchemotactic limit ($\xi_1=0$) starting from random initial conditions.
    The top row shows representative initial configurations, while subsequent rows display the corresponding steady states for five independent random seeds.
    Columns correspond to increasing diffusion coefficient $D_1$ with $D=1000D_1$.
    Depending on the initial condition, the system selects distinct morphologies, including stripe and hexagonal spot patterns.
    As $D_1$ is varied, the preferred morphology switches between spots and stripes, indicating multistability and diffusion-controlled mode selection in the absence of chemotaxis.
    Parameters: $a=0.015$ and $b=10.0$.
    }
    \label{fig:fig_nonchemo-spots-stripes}
\end{figure}

We also investigated the nature of the patterns formed in the nonchemotactic limit by DNS of the governing equations with $\xi_1=0=\xi_2$. Intriguingly, by varying the random initial conditions (i.e., by varying the seeds for the random initial conditions), we find that the resulting pattern can be spots or stripes; see Fig.~\ref{fig:fig_nonchemo-spots-stripes}. This means that in the nonchemotactic limit, the system has no preference for stripes or spots. Precise initial conditions dictate the nature of the resulting pattern. This is thus a nonequilibrium equivalent of ``spontaneous symmetry breaking'' in equilibrium systems, in which an equilibrium system spontaneously chooses one among several free energetically equivalent states~\cite{chaikin}. Furthermore, for a given random initial condition, variation of $D_1$ for a fixed $D\equiv D_2/D_1$ can lead to transitions between patterns with spots and stripes. Our numerical results suggest that the $D_1$-dependence of the ensuing patterns can even be {\it nonmonotonic}, i.e., re-entrant transitions between patterns of different types (spots versus stripes) can take place by varying $D_1$. We also find that for $D_1=0, D_2=0$, there is no pattern or oscillation, confirming the simple fact that the pattern is due to the diffusivities, and the point $(a=0.015, b=10.0)$ is outside the region with oscillatory instability in the $(a,b)$ plane.

\subsection{Patterns with chemotaxis}

In this Section, we present the results from our DNS studies on patterns in the presence of chemotaxis. More specifically, our aim is to explore how the nature of patterns depends upon the chemotaxis parameters $\xi_1,\xi_2$.  We numerically solve Eqs.~(\ref{mod1}) and (\ref{mod2}) by using pseudo-spectral methods; see Appendix~\ref{sec:numerical_methods} for technical details. We obtain different steady-state solutions on square geometries for various sets of $\xi_1,\xi_2$ and then explore the signatures of re-entrant transitions predicted by the linear stability analysis. To reduce the number of model parameters in our DNS studies, we set $\xi_1=\xi=\pm\xi_2$, or $\xi_1=\xi\neq 0,\,\xi_2=0$ or $\xi_2=\xi\neq 0, \xi_1=0$ and allow $\xi$ to be positive or negative. The snapshots of the configurations of the various steady state solutions are presented below. An alert reader will notice that the linear stability analysis makes predictions on how the selected or preferred wavevector $k_c$ depends on the chemotaxis parameters $\xi_1,\xi_2$, i.e., how the periodicity of the pattern is affected by $\xi_1,\xi_2$. However, a steady pattern (at the minimal level) is characterized by $k_c$ and the variation in the amplitude of the pattern. Unsurprisingly, a linear stability analysis makes no predictions on the amplitude. However, the solution of the full nonlinear problem should reveal how the pattern amplitudes vary with the model parameters. To quantify the patterns, in particular how they depend on chemotaxis parameters $\xi_1,\xi_2$ and also on  diffusivities, we define $\Delta_\phi$ and $\Delta_\psi$ as

\begin{eqnarray}
   && \Delta_\phi\equiv \sqrt{\sum_{x,y}[\phi(x,y)-\overline \phi]^2\frac{1}{L^2}},\label{Del-phi-def}\\
   && \Delta_\psi\equiv \sqrt{\sum_{x,y}[\phi(x,y)-\overline \psi]^2\frac{1}{L^2}},\label{Del-psi-def}
\end{eqnarray}
to be measured in the steady states,
where $\overline\phi$ and $\overline\psi$ are the spatial mean values of $\phi(x,y)$ and $\psi(x,y)$ respectively in the steady states, and hence time independent. Quantities $\Delta_\phi$ and $\Delta_\psi$ are, in fact, measures of pattern amplitudes (see later): In uniform states, $\Delta_\phi=0=\Delta_\psi$. However, in patterned 
states, $\Delta_\phi>0,\,\Delta_\psi>0$. The variation (including nonmonotonic behavior) of $k_c^2$ with $\xi_1,\xi_2$ in the linear stability analysis may get reflected in the DNS studies results in the variation of the periodicity and amplitudes or both of the pattern with $\xi_1,\xi_2$. In the absence of any comprehensive nonlinear theory of patterns, we should also allow for the fascinating possibility that the two amplitudes for $\phi$ and $\psi$ may behave {\it qualitatively differently} as $\xi_1,\xi_2$ are varied. Indeed, in one of the cases that we studied by using our DNS, this intriguing possibility turns out to be true. 

We investigate the pattern amplitudes and calculate $\Delta_\phi$ and $\Delta_\psi$ in the steady states  using DNS studies of (\ref{mod1}) and (\ref{mod2}) below for a few specific cases. 

\subsubsection{Limiting case: $\xi_1>0,\,\xi_2=0$}

\begin{widetext}

In this limit, Eq.~\eqref{mod1} has a chemotaxis contribution, but Eq.~\eqref{mod2} has none. Nonetheless, nonlinear effects ensure that {\it both} $\phi({\bf x},t)$ and $\psi({\bf x},t)$ are affected by chemotaxis. 
We solve \eqref{mod1} and \eqref{mod2} numerically to calculate $\phi({\bf x},t)$ and $\psi({\bf x},t)$ in the steady states for a range of $\xi_1>0$ with two fixed values of $D_1=0.1,\,0.01$, $\xi_2=0$ and $D=1000$ and $(a,b)=(0.015,10.0)$; see Fig.~\ref{fig:fig_limiting_xi20_xi1pos}. We observe that as $\xi_1>0$ increases, the pattern appears {\it more intense}, i.e., the amplitude becomes larger, but the periodicity, given by $k_c$ that controls the number of spots in the steady states, does not appear to depend significantly on $\xi_1$. Interestingly, both $\phi({\bf x},t)$ and $\psi({\bf x},t)$ show very similar patterns, even though only the $\phi({\bf x},t)$ given by \eqref{mod1} is directly affected by a nonzero $\xi_1$. That $k_c$ is largely unaffected but pattern amplitudes depend sensitively on $\xi_1$ is a behavior that is in contrast to the corresponding linear theory results, which predict that for $\xi_1>0, \xi_2=0$, $k_c^2$ should rise continuously with $\xi_1$.

\begin{figure}[!ht]
    \centering
\includegraphics[width=1.05\columnwidth]{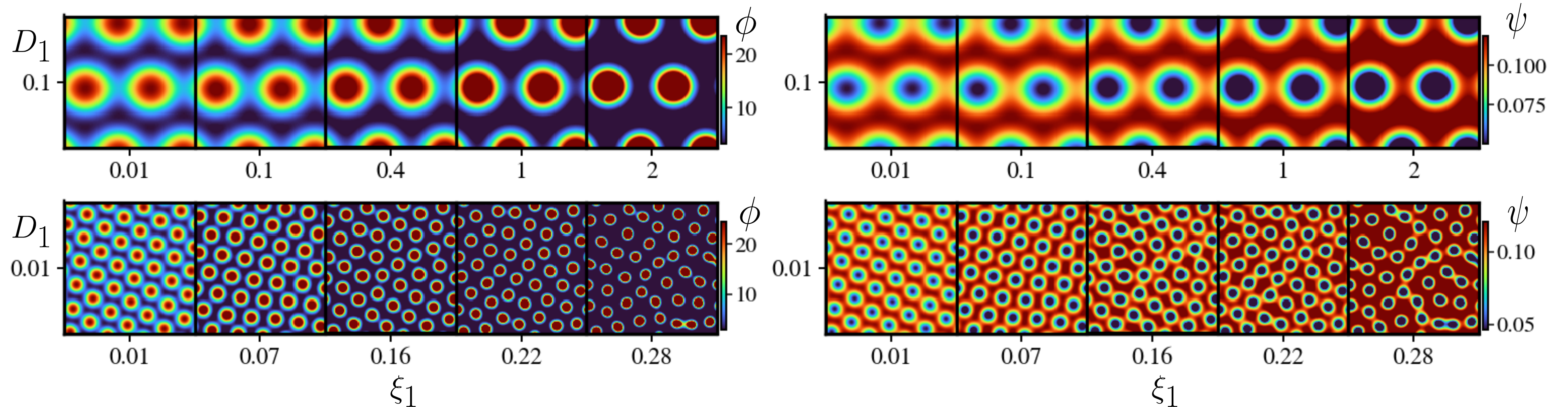}
    \caption{(Color online) Variation of the steady patterns of $\phi$ and $\psi$ for increasing $\xi_1>0,\,\xi_2=0$ for a fixed $D=1000, a=0.015, b = 10.0$. (top) $D_1=0.1$, (bottom) $D_1=0.01$. As $\xi_1$ increases, the ensuing patterns become more intense (i.e., the amplitudes increase), while maintaining the same periodic nature. This is in contrast to the predictions from the linear stability analysis, in which the selected wavevector $k_c^2$ is expected to increase with $\xi_1$. See text.    }
    \label{fig:fig_limiting_xi20_xi1pos}
\end{figure}

The variation in the pattern amplitudes in Fig.~\ref{fig:fig_limiting_xi20_xi1pos} is further quantified by calculating $\Delta_\phi,\,\Delta_\psi$ for various $\xi_1>0$ with $\xi_2=0$ in Fig.~\ref{Delta-xi1-pos-xi2-0} for two values of $D_1=0.1, 0.01$ for a fixed $D=1000$. See Fig.~\ref{Delta-xi1-pos-xi2-0}, which shows that both $\Delta_\phi,\Delta_\psi$ increase with $\xi_1>0$. Thus the dependence of {\it both} $\Delta_\phi$ and $\Delta_\psi$ on $\xi_1>0$ in our DNS studies is similar to that of $k_c^2$ on $\xi_1>0$ in the linear stability analysis, strongly suggesting a phenomenological correspondence  between the dependence of $k_c^2$ in the linear stability analysis and pattern amplitudes in the DNS studies of the full nonlinear PDEs on $\xi_1$.

\begin{figure}
    \centering
    \includegraphics[width=0.43\linewidth]{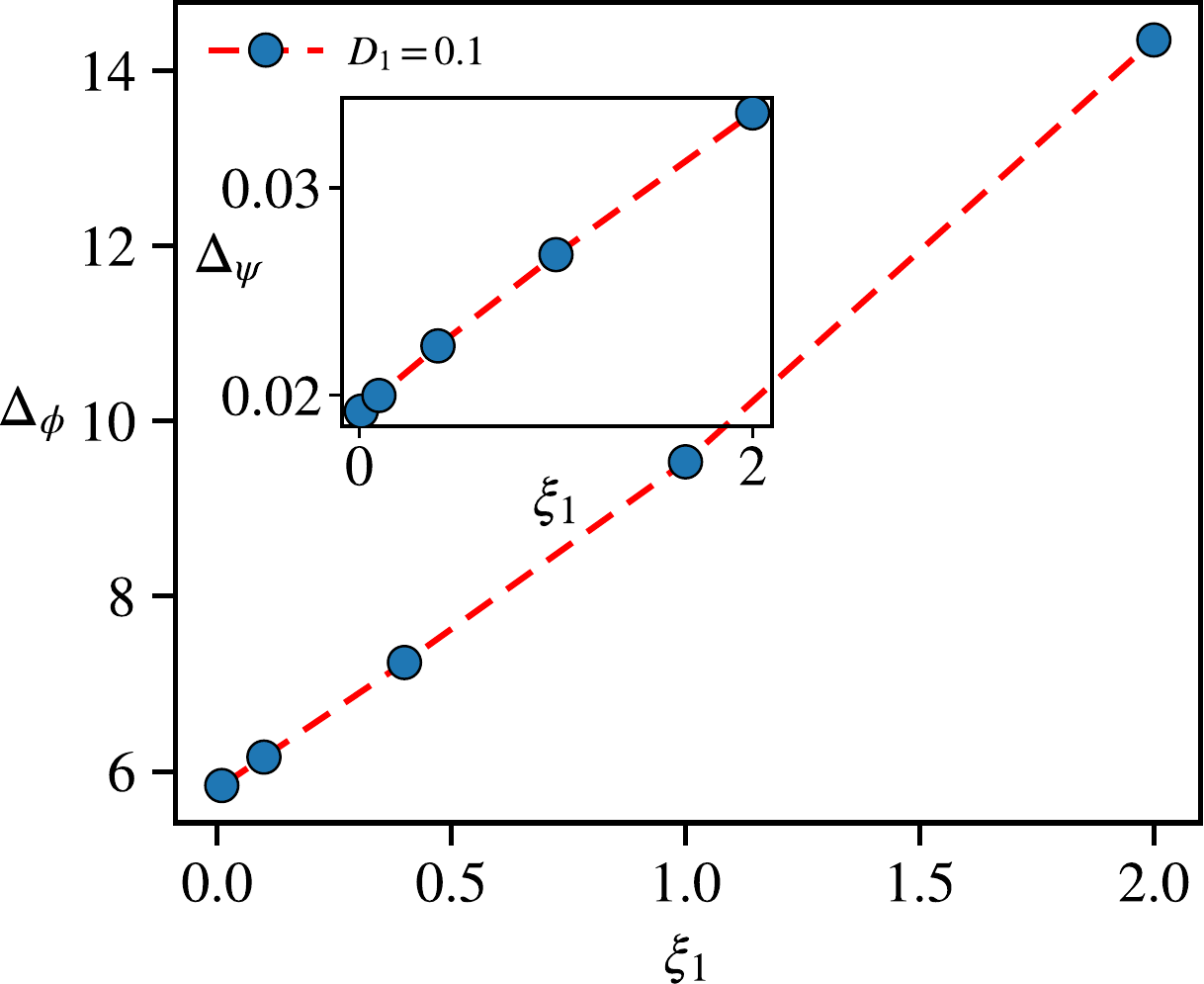}\hfill \includegraphics[width=0.43\linewidth]{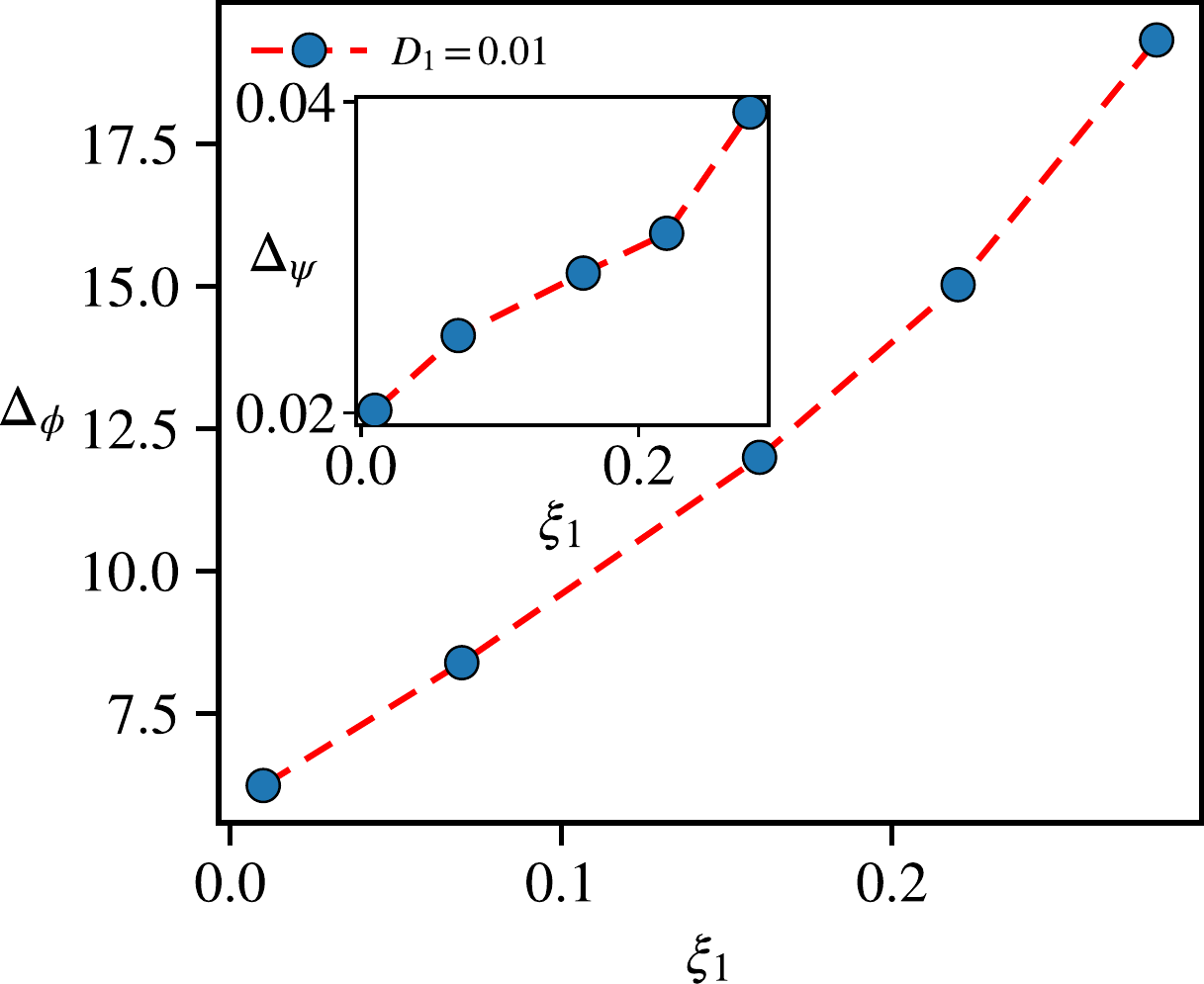}
    \caption{Plot of $\Delta_\phi$ (inset $\Delta_\psi$) versus $\xi_1>0$ with $\xi_2=0$ and $D_1=0.1,\,0.01$, $D=1000$. Both $\Delta_\phi$ and $\Delta_\psi$ increase monotonically with $\xi_1$, giving a sharper pattern for larger $\xi_1$; see text. }
    \label{Delta-xi1-pos-xi2-0}
\end{figure}

\subsubsection{Limiting case: $\xi_1<0,\,\xi_2=0$} 

We now consider the opposite case with $\xi_1=\xi<0$ and $\xi_2=0$. The linear theory predicts that $k_c^2$ monotonically decreases as $|\xi_1|$ rises. Our DNS studies however show that the amplitudes of the patterns for both $\phi$ and $\psi$ monotonically decrease, as $|\xi_1$ increases, ultimately giving a uniform state, i.e., a state with vanishing pattern amplitudes. However, the periodicity of the pattern does not show any significant dependence on $\xi_1$; see Fig.~\ref{fig:fig_limiting_xi20_xi1neg}. This behavior is in contrast to the corresponding linear theory results, which predict that for $\xi_1<0$, the selected wavevector $k_c^2$ should continuously decrease as $\xi_1$ becomes more negative.

\begin{figure}[!ht]
    \centering
    \includegraphics[width=1.05\columnwidth]{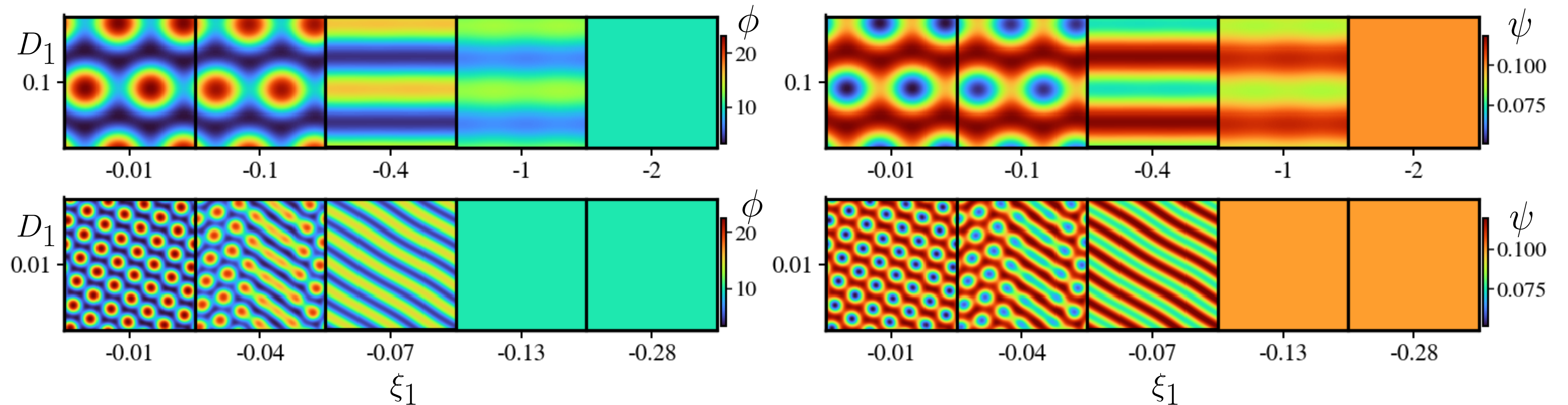}
    \caption{(Color online) Variation of the steady patterns of $\phi$ and $\psi$ for increasing $\xi_1<0,\,\xi_2=0$ for a fixed $D=1000, a=0.015, b = 10.0$. (top) $D_1=0.1$, (bottom) $D_1=0.01$. As $|\xi_1|$ increases (i.e., as $\xi_1$ becoming more negative), the ensuing patterns become {\it less} intense, i.e., the amplitudes progressively decrease, while maintaining the same periodic nature. This is in contrast to the predictions from the linear stability analysis, in which the selected wavevector $k_c^2$ is expected to decrease with increasing $|\xi_1|$. See text.
        }
    \label{fig:fig_limiting_xi20_xi1neg}
\end{figure} 

The variation in the pattern amplitudes in Fig.~\ref{fig:fig_limiting_xi20_xi1neg} is further quantified by calculating $\Delta_\phi,\,\Delta_\psi$ for various $\xi_1<0$ with $\xi_2=0$ in Fig.~\ref{Delta-xi1-neg-xi2-0} for two values of $D_1=0.1, 0.01$ for a fixed $D=1000$. See Fig.~\ref{Delta-xi1-neg-xi2-0}, which shows that both $\Delta_\phi,\Delta_\psi$ decrease with $\xi_1<0$ becomes more negative. Thus the dependence of {\it both} $\Delta_\phi$ and $\Delta_\psi$ on $\xi_1<0$ in our DNS studies is similar to that of $k_c^2$ on $\xi_1<0$ in the linear stability analysis. This strongly suggests a phenomenological correspondence between the dependence of $k_c^2$ in the linear stability analysis and pattern amplitudes in the DNS studies of the full nonlinear PDEs on $\xi_1$.

 \begin{figure}
    \centering
    \includegraphics[width=0.43\linewidth]{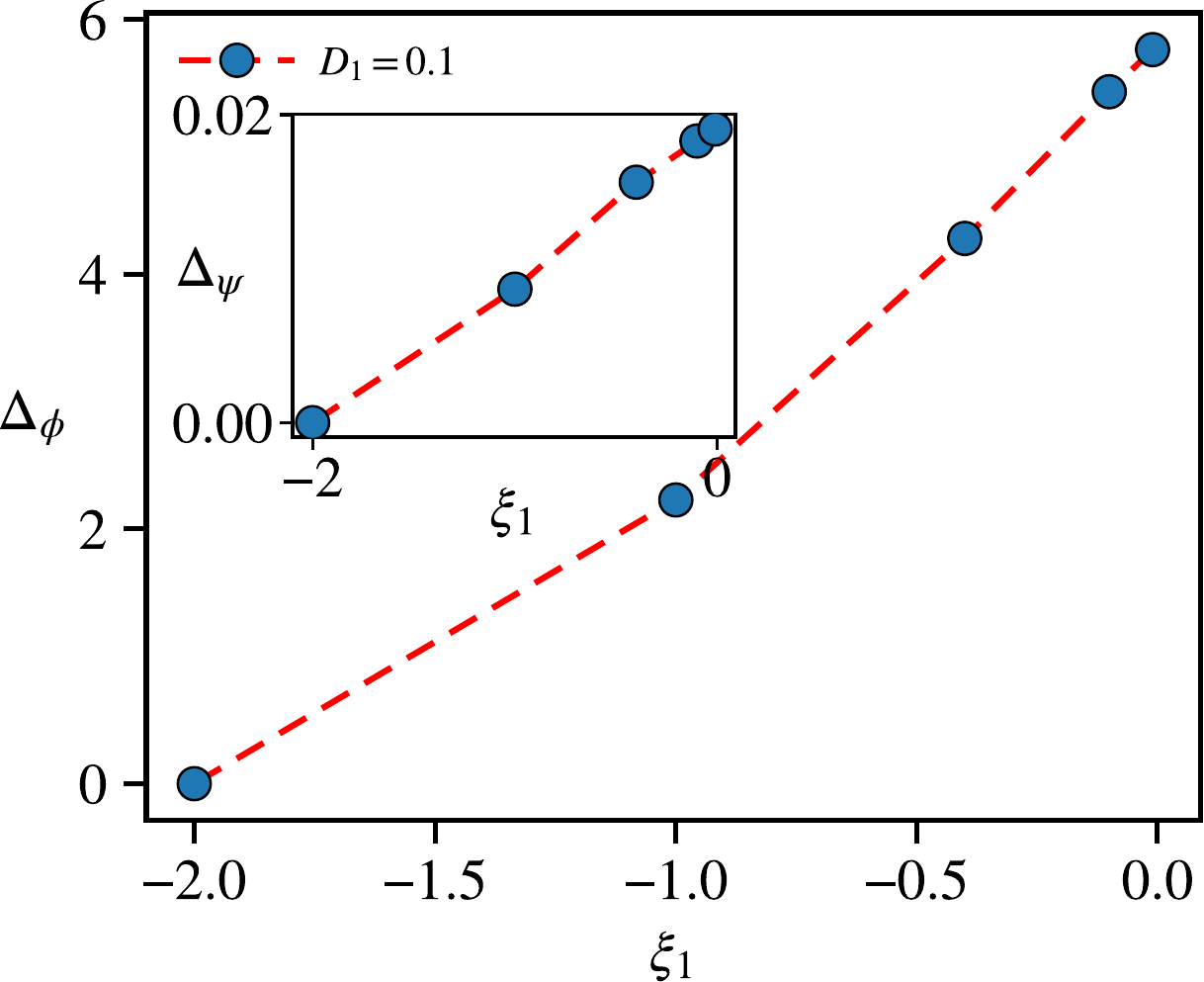} \hfill \includegraphics[width=0.43\linewidth]{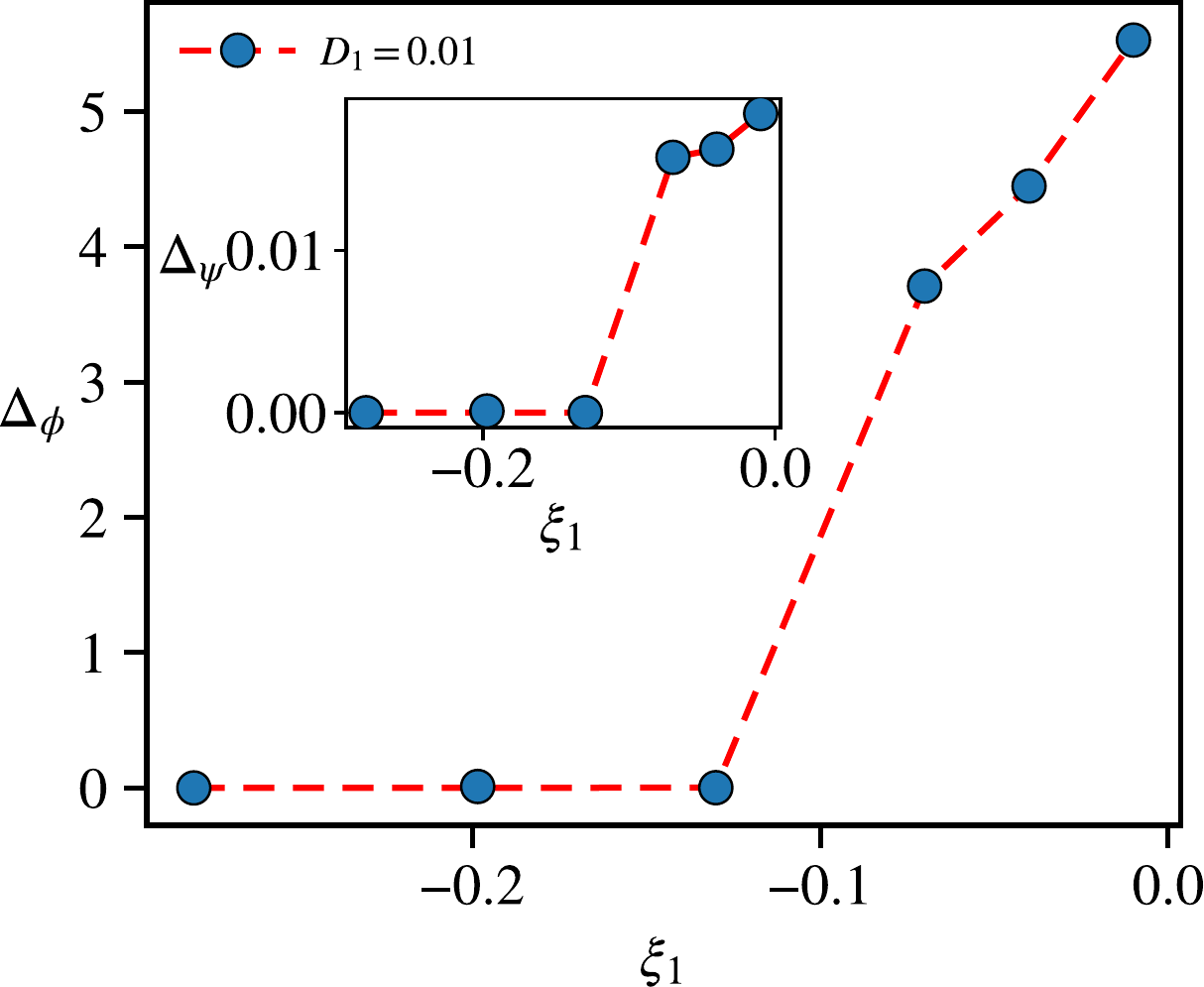} 
    \caption{Plot of $\Delta_\phi$ (inset $\Delta_\psi$) versus $\xi_1<0$ with $\xi_2=0$ and $D_1=0.1,\,0.01$, $D=1000$. Both $\Delta_\phi$ and $\Delta_\psi$ increase monotonically with $|\xi_1|$, giving an increasingly fade pattern as $\xi_1$ becomes more negative; see text.}
    \label{Delta-xi1-neg-xi2-0}
\end{figure}

The results of Fig.~\ref{fig:fig_limiting_xi20_xi1pos} and Fig.~\ref{fig:fig_limiting_xi20_xi1neg} together, or equivalently, Fig.~\ref{Delta-xi1-pos-xi2-0} and Fig.~\ref{Delta-xi1-neg-xi2-0} together suggest  the following phenomenological forms for the amplitudes  
\begin{eqnarray}
    \phi({\bf x},t)=A_\phi({\bf x},t)[\exp(i{\bf \tilde k}_c^{(1)}\cdot {\bf r})+cc],\;A_\phi\sim f_\phi^{(1)}(\xi_1),\\
    \psi({\bf x},t)=A_\psi({\bf x},t)[\exp(i{\bf \tilde k}_c^{(1)}\cdot {\bf r})+cc],\;A_\psi\sim f_\psi^{(1)}(\xi_1),
\end{eqnarray}
where $f_\phi^{(1)}(\xi_1), f_\psi^{(1)}(\xi_1)$ are functions of $\xi$ (ignoring any $\bf x$ and $t$ dependence) with the properties that $f_\phi^{(1)}(\xi_1), f_\psi^{(1)}(\xi_1)$ vanish for $\xi_1$ large negative, and $f_\phi^{(1)}(\xi_1), f_\psi^{(1)}(\xi)$ become large as $\xi_1$ becomes large positive; cc refers to complex conjugate. The preferred wavevector $\tilde k_c^{(1)}$ should be different from its counterpart $k_c$ in the linear theory. However, $\tilde k_c^{(1)}$ is not expected to have a strong dependence on $\xi$, but should depend on the other model parameters.

 \subsubsection{Limiting case: $\xi_1=0,\,\xi_2>0$}

 We now study the case with $\xi_1=0$, $\xi_2>0$. Thus, the $\phi$-dynamics given by \eqref{mod1} has no direct chemotaxis contribution in it, but the $\psi$-dynamics given by \eqref{mod2} has, which makes it complementary to the case $\xi_1>0,\,\xi_2=0$. Nonetheless, nonlinear effects ensure that {\it both} $\phi$- and $\psi-$ dynamics are affected by $\xi_2$. We solve \eqref{mod1} and \eqref{mod2} for a range of $\xi_2>0$ for fixed $\xi_1=0$, two values of $D_1=0.1, 0.01$ and fixed values of the other parameters.  We observe that as $\xi_2=\xi>0$ increases, the ensuing patterns become less and less intense, i.e., their amplitudes decrease monotonically, eventually vanishing, giving uniform states; see Fig.~\ref{fig:fig_limiting_xi10_xi2pos}.  However, the periodicity of the pattern does not show any strong dependence on $\xi_2$. This behavior is in contrast to the corresponding linear theory results, which predict that for increasing $\xi_2>0$, $k_c^2$ should continuously decrease.

\begin{figure}[!ht]
    \centering
    \includegraphics[width=1.05\columnwidth]{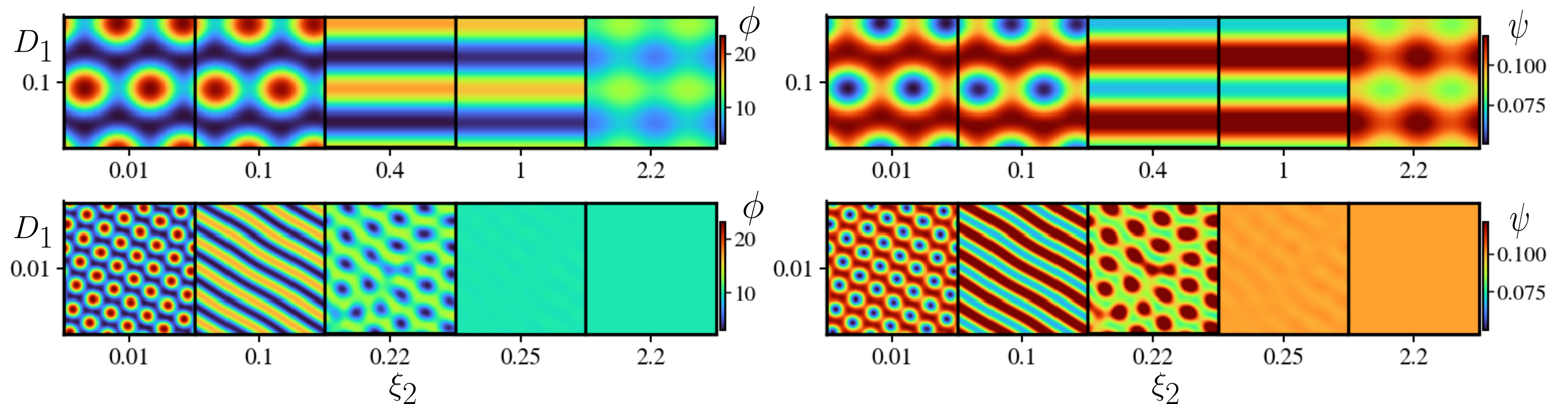}
    \caption{(Color online) Variation of the steady patterns of $\phi$ and $\psi$ for increasing $\xi_1=0,\,\xi_2>0$ for a fixed $D=1000, a=0.015, b = 10.0$. (top) $D_1=0.1$, (bottom) $D_1=0.01$. As $\xi_2$ increases, the ensuing patterns become {\it less} intense (i.e., the amplitudes monotonically decrease), while maintaining the same periodic nature. This is in contrast to the predictions from the linear stability analysis, in which the selected wavevector $k_c^2$ is expected to decrease with $\xi_1$. See text. }
    \label{fig:fig_limiting_xi10_xi2pos}
\end{figure}

The variation in the pattern amplitudes in Fig.~\ref{fig:fig_limiting_xi10_xi2pos} is further quantified by calculating $\Delta_\phi,\,\Delta_\psi$ for various $\xi_2$ with $\xi_1=0$ in Fig.~\ref{Delta-xi1-0-xi2-pos} for two values of $D_1=0.1, 0.01$ for a fixed $D=1000$. See Fig.~\ref{Delta-xi1-0-xi2-pos}, which shows that both $\Delta_\phi,\Delta_\psi$ decrease with $\xi_2>0$ increases. Thus the dependence of {\it both} $\Delta_\phi$ and $\Delta_\psi$ on $\xi_2>0$ in our DNS studies is similar to that of $k_c^2$ on $\xi_2>0$ in the linear stability analysis.  This again strongly suggests a phenomenological correspondence  between the dependence of $k_c^2$ in the linear stability analysis and pattern amplitudes in the DNS studies of the full nonlinear PDEs on $\xi_2$.

\begin{figure}
    \centering
    \includegraphics[width=0.43\linewidth]{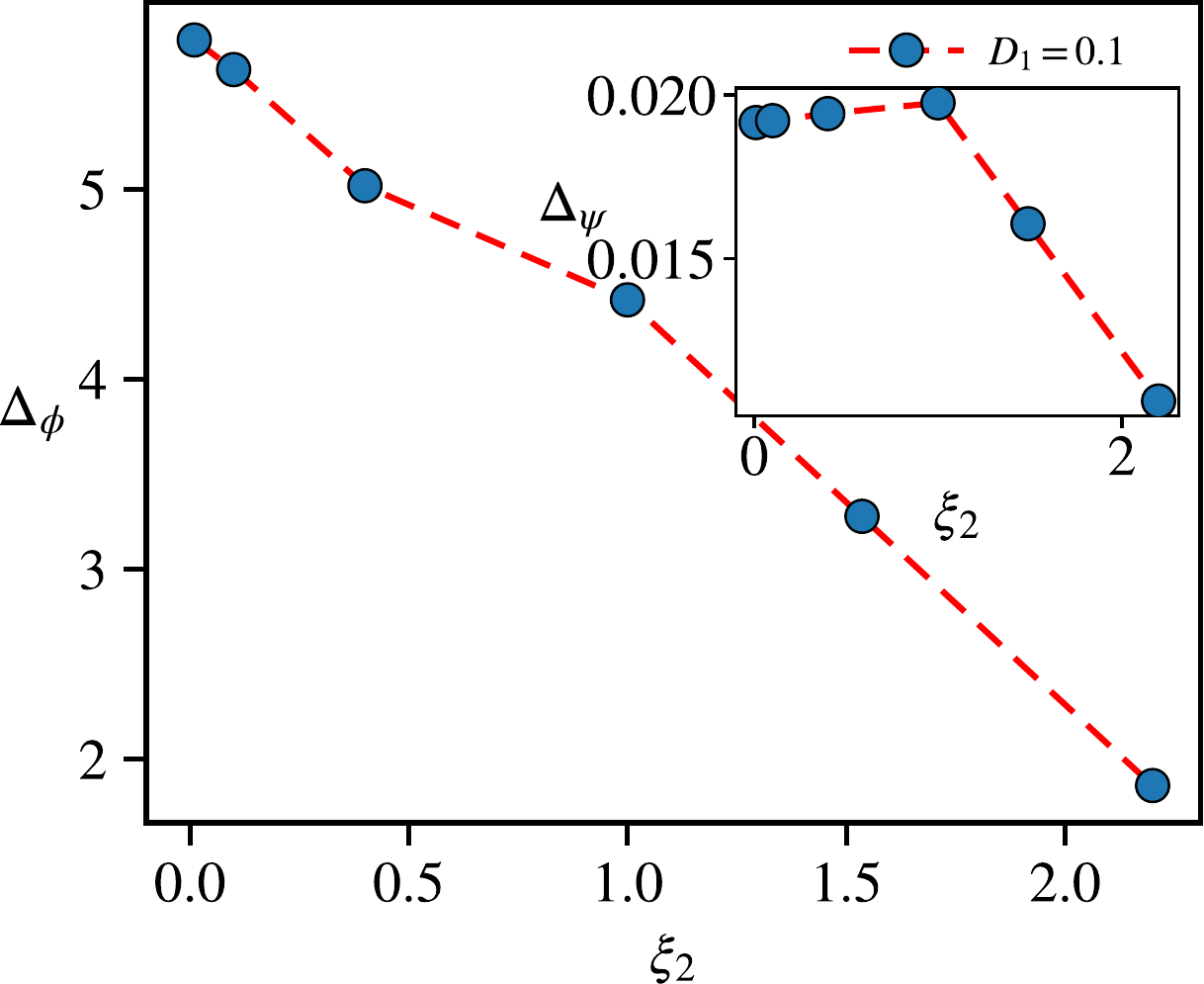} \hfill \includegraphics[width=0.43\linewidth]{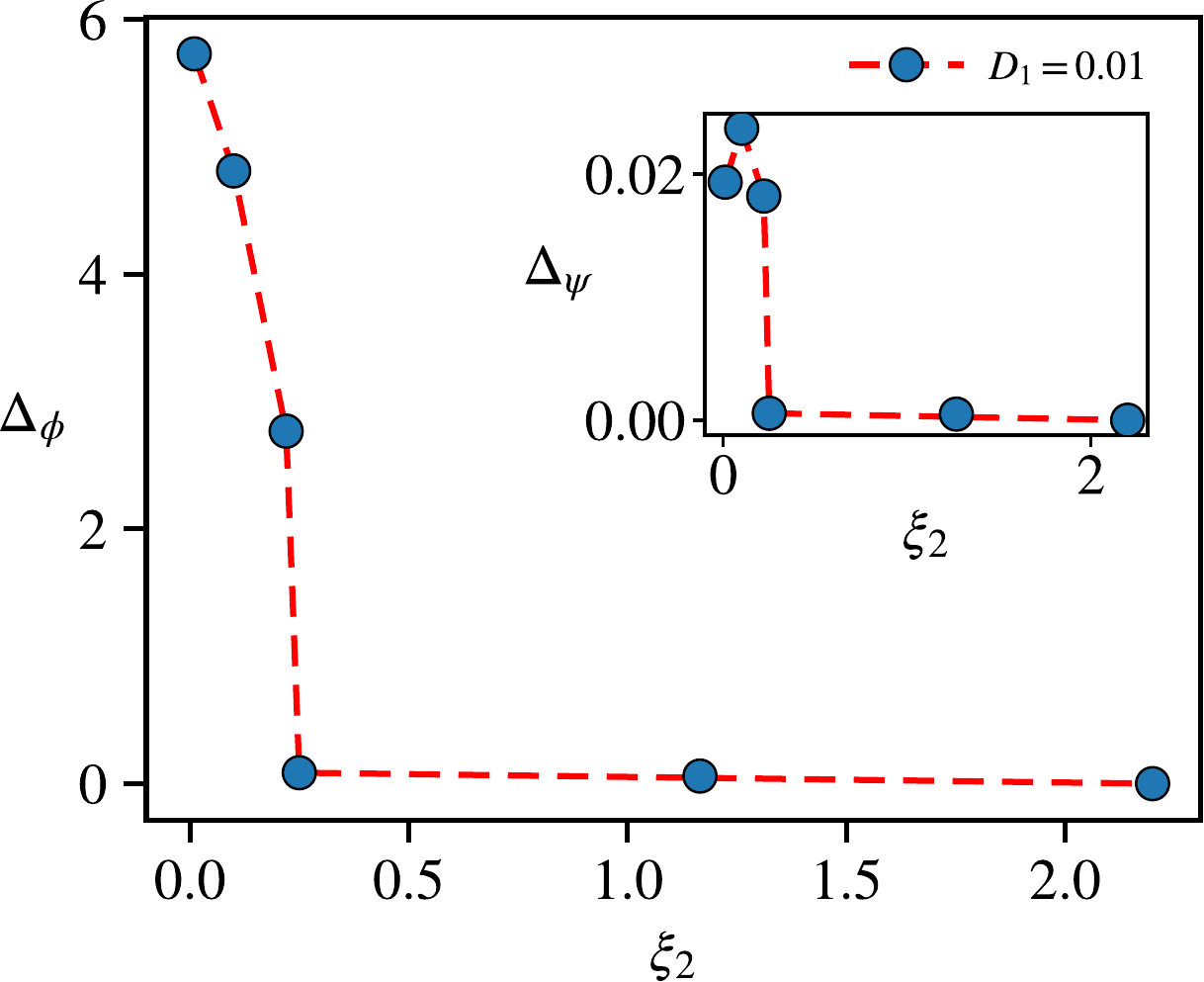}
    \caption{Plot of $\Delta_\phi$ (inset $\Delta_\psi$) versus $\xi_2>0$ with $\xi_1=0$ and $D_1=0.1,\,0.01$, $D=1000$. Both $\Delta_\phi$ and $\Delta_\psi$ decrease monotonically as $\xi_2$ increases, giving an increasingly fade pattern as $\xi_2$ becomes larger; see text.}
    \label{Delta-xi1-0-xi2-pos}
\end{figure}

 \subsubsection{Limiting case: $\xi_1=0,\,\xi_2<0$}

 Next, we study the case with $\xi_1=0$, $\xi_2=\xi<0$. Then again $\phi$-dynamics given by \eqref{mod1} has no direct chemotaxis contribution in it, but the $\psi$-dynamics given by \eqref{mod2} has, which makes it complementary to the case $\xi_1<0,\,\xi_2=0$. Nonetheless, nonlinear effects ensure that {\it both} $\phi$- and $\psi-$ dynamics are affected by $\xi_2$. As with $\xi_2>0$, we solve \eqref{mod1} and \eqref{mod2} for a range of $\xi_2<0$ for fixed $\xi_1=0$, two values of $D_1=0.1, 0.01$ and fixed values of the other parameters.  We find that as $\xi_2=\xi<0$ increases, the ensuing patterns become more and more intense, i.e., their amplitudes increase monotonically; see Fig.~\ref{fig:fig_limiting_xi10_xi2neg}.  However, the periodicity of the pattern appears to be largely unaffected by $\xi_2$. This behavior is in contrast to the corresponding linear theory results, which predict that, for increasing $\xi_2=\xi<0$, $k_c^2$ should continuously increase. 

\begin{figure}[!ht]
    \centering
    \includegraphics[width=1.05\columnwidth]{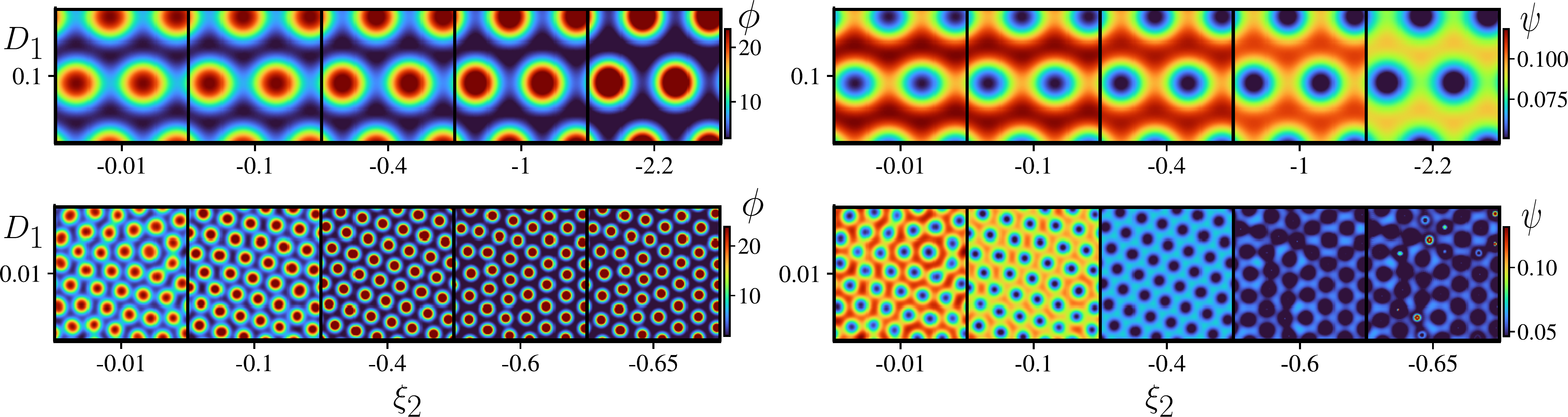}
    \caption{(Color online) Variation of the steady patterns of $\phi$ and $\psi$ for increasing $\xi_1=0,\,\xi_2<0$ for a fixed $D=1000, a=0.015, b = 10.0$. (top) $D_1=0.1$, (bottom) $D_1=0.01$. As $|\xi_2|$ increases (i.e., $\xi_2$ becomes more negative), the ensuing patterns become {\it more} intense (i.e., the amplitudes monotonically increase), while maintaining the same periodic nature. This is in contrast to the predictions from the linear stability analysis, in which the selected wavevector $k_c^2$ is expected to decrease with $\xi_1$. See text.
        }
    \label{fig:fig_limiting_xi10_xi2neg}
\end{figure}

The variation in the pattern amplitudes in Fig.~\ref{fig:fig_limiting_xi10_xi2neg} are further quantified by calculating $\Delta_\phi,\,\Delta_\psi$ for various $\xi_2$ with $\xi_1=0$ in Fig.~\ref{Delta-xi1-0-xi2-neg} for two values of $D_1=0.1, 0.01$ for a fixed $D=1000$. See Fig.~\ref{Delta-xi1-0-xi2-neg}, which shows that while $\Delta_\phi$ {\it increases}, $\Delta_\psi$ {\it decreases} as $\xi_2<0$ becomes more negative. Thus the dependence of $\Delta_\phi$ on $\xi_1<0$ in our DNS studies is similar to that of $k_c^2$ on $\xi_1$ in the linear stability analysis. But the dependence of $\Delta_\psi$ is just the opposite. This establishes an intriguing complex correspondence between the linear stability result on $k_c^2$ and DNS studies on the pattern amplitudes.

 \begin{figure}
    \centering
    \includegraphics[width=0.43\linewidth]{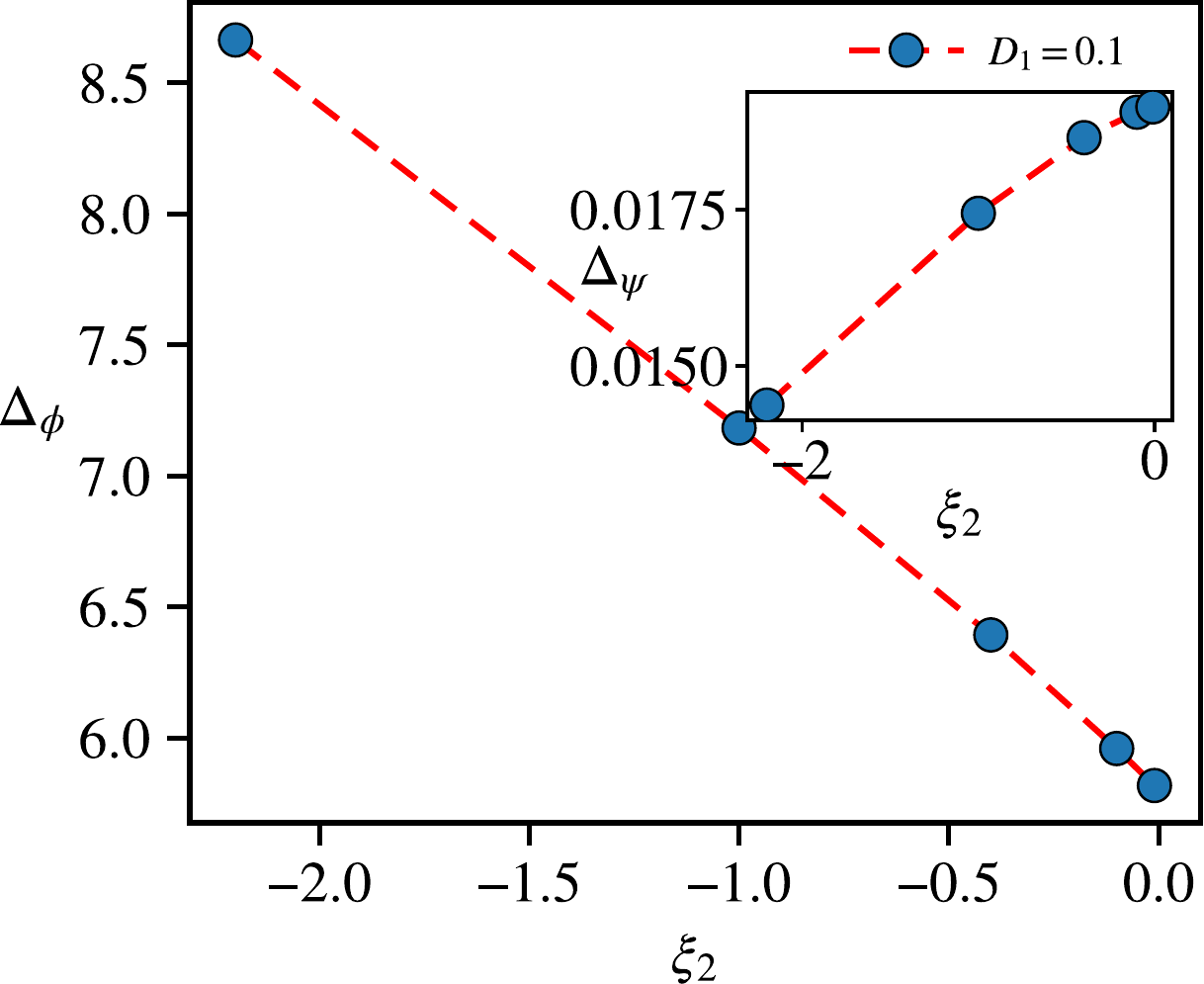} \hfill \includegraphics[width=0.43\linewidth]{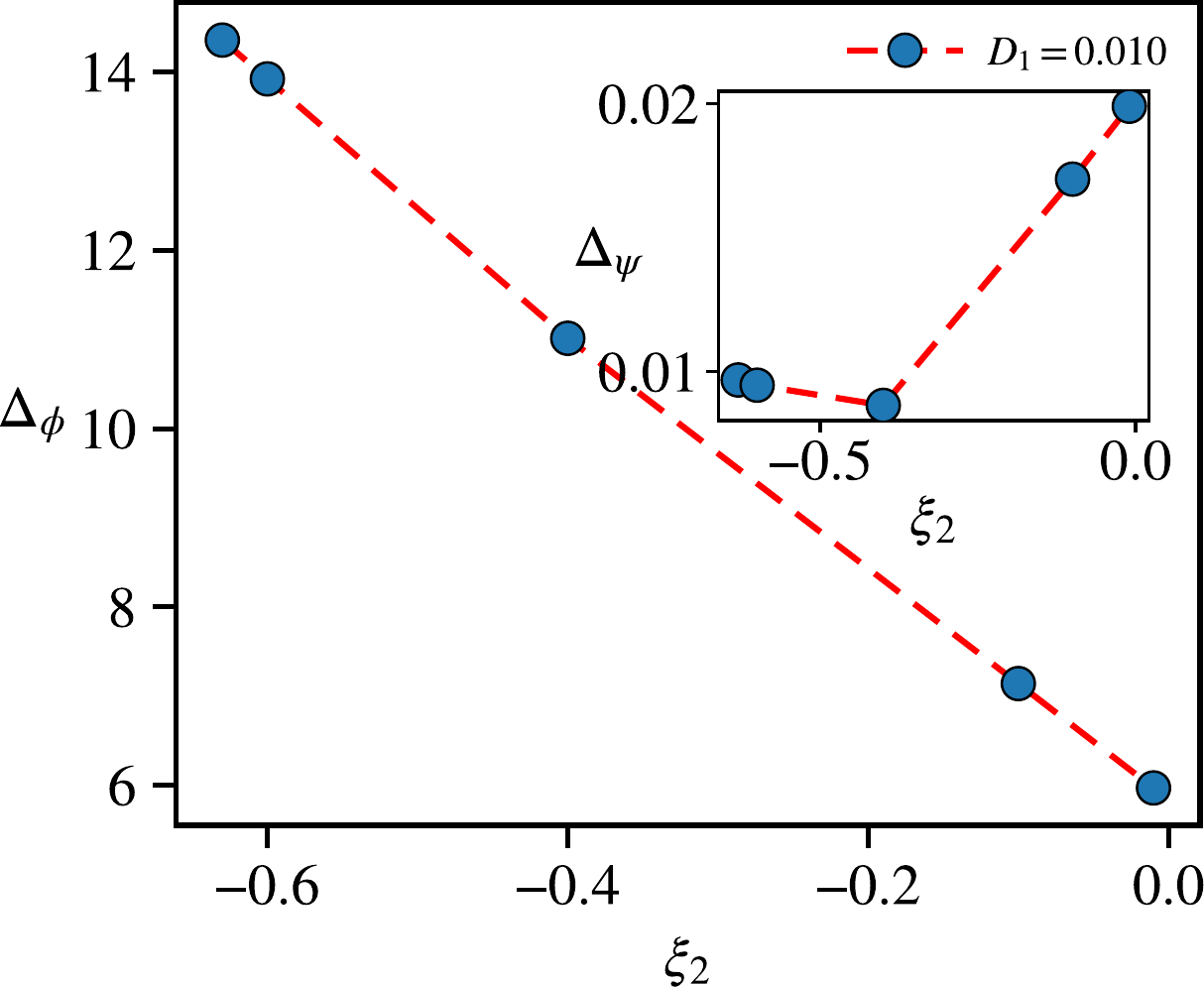}
    \caption{Plot of $\Delta_\phi$ (inset $\Delta_\psi$) versus $\xi_1<0$ with $\xi_2=0$ and $D_1=0.1,\,0.01$, $D=1000$. Notice that $\Delta_\phi$ increases but $\Delta_\psi$ decreases monotonically with $|\xi_2|$, giving an increasingly sharper pattern for $\phi$ but an increasingly fade pattern for $\psi$, as $\xi_2$ becomes more negative; see text.}
    \label{Delta-xi1-0-xi2-neg}
\end{figure}

As in the previous case, the results of Fig.~\ref{fig:fig_limiting_xi10_xi2pos} and Fig.~\ref{fig:fig_limiting_xi10_xi2neg} together, or equivalently, Fig.~\ref{Delta-xi1-0-xi2-pos} and Fig.~\ref{Delta-xi1-0-xi2-neg} together suggest  the following phenomenological forms for the amplitudes  
\begin{eqnarray}
    \phi({\bf x},t)=A_\phi({\bf x},t)[\exp(i{\bf \tilde k}_c^{(2)}\cdot {\bf r})+cc],\;A_\phi\sim f_\phi^{(2)}(\xi),\\
    \psi({\bf x},t)=A_\psi({\bf x},t)[\exp(i{\bf \tilde k}_c^{(2)}\cdot {\bf r})+cc],\;A_\psi\sim f_\psi^{(2)}(\xi),
\end{eqnarray}
where $f_\phi^{(2)}(\xi), f_\psi^{(2)}(\xi_2)$ are functions of $\xi_2$ (ignoring any $\bf x$ and $t$ dependence) with the properties that $f_\phi^{(2)}(\xi), f_\psi^{(2)}(\xi)$ vanish for $\xi_2$ being large positive, and $f_\phi^{(2)}(\xi_2)$ becomes large and $f_\psi^{(2)}(\xi_2)$ becomes small as $\xi_2$ becomes large negative. The preferred wavevector $\tilde k_c^{(2)}$ should be different from its counterpart $k_c$ in linear theory. However, $\tilde k_c^{(2)}$ is not expected to have a strong dependence on $\xi$, but should depend on the other model parameters.

In the following, we study the steady patterns when {\it both} $\xi_1$ and $\xi_2$ are nonzero.

\subsection{Patterns in the extreme nonreciprocal chemotaxis: $\xi_1=\xi=-\xi_2$}

We now study the extreme nonreciprocal chemotaxis case with $\xi_1=\xi=-\xi_2$, which corresponds to the chasing phenomena discussed above. This has two distinct subcases for the two different signs of $\xi$, which we discuss below.

\subsubsection{ $\xi>0$}

With $\xi_1=\xi=-\xi_2>0$, both $\phi$- and $\psi$-dynamics are affected by the mutual chemotactic interactions.  We solve \eqref{mod1} and \eqref{mod2} for a range of $\xi>0$ for two values of $D_1=0.1, 0.01$ and fixed values of the other parameters. In this case, linear stability analysis predicts the selected wavevector $k_c^2$ to depend on $\xi$ nonmonotonically - $k_c^2$ first rises with $\xi$, then decreases, ultimately vanishing in the limit of large $\xi$, giving a spatially uniform state. DNS studies of \eqref{mod1} and \eqref{mod2}, however, show that with increasing $\xi$, the amplitude of the pattern first increases, i.e., the pattern first gets more intense, and then on further increasing $\xi$, the amplitude decreases, the pattern fades away, eventually disappearing for very large $\xi$. The periodicity of the pattern does not appear to have a strong dependence on $\xi$. See Fig.~\ref{fig:fig_NR_Xipos}. Although not shown in Fig.~\ref{fig:fig_NR_Xipos}, it clearly suggests that if $D_1,D$ are chosen such that pattern disappears for $\xi\rightarrow 0$, as $\xi$ is increased a pattern should form, which ultimately disappears if $\xi$ is further increased beyond a threshold value. This is the re-entrant behavior from uniform state to uniform state via a pattern as $\xi$ is continuously increased, as mentioned earlier in this article.



\begin{figure}[!ht]
    \centering
    \includegraphics[width=1.05\columnwidth]{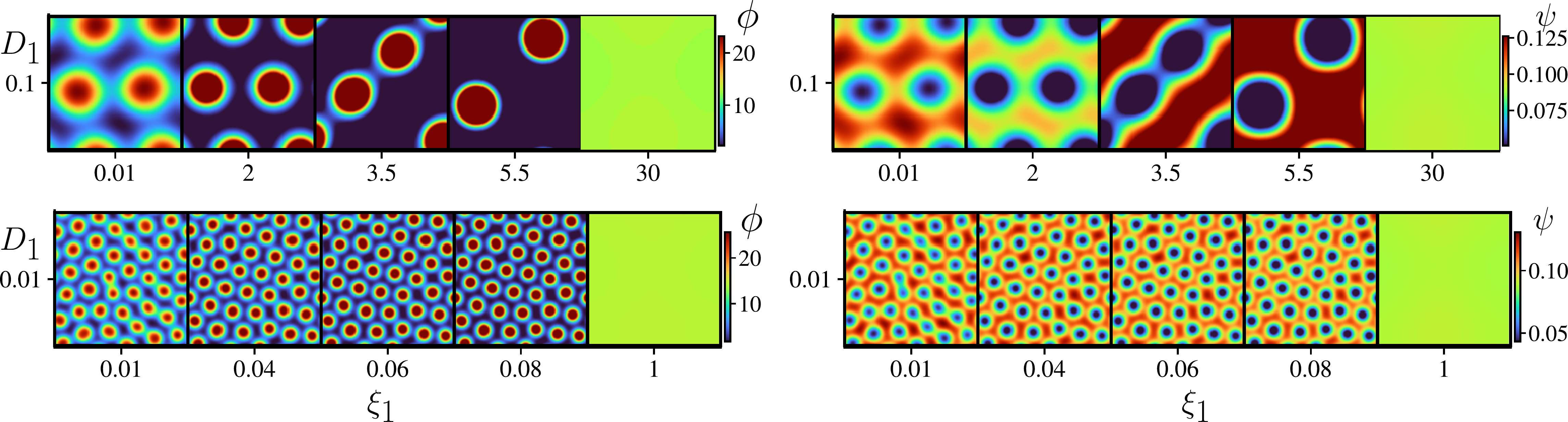}
    \caption{(Color online)  Steady state snapshots of the density fields $\phi$ (left) and $\psi$ (right) for different values of the chemotactic couplings $\xi_1=\xi=-\xi_2>0$ and diffusion coefficients in the extreme nonreciprocal chemotaxis case.
    Columns correspond to increasing coupling strength $\xi=0.01$, $0.17$, and $0.49$, while rows give two sets of diffusion coefficient $D_1$:
    $D_1=0.01$ (top) and $0.1$ (bottom) for a fixed ratio $D=1000$. For both the values of $D_1$, we find that for smaller values of $\xi$, the pattern consists of fuzzy spots, which evolve into spots with well-defined boundaries as $\xi$ increases. Eventually, for very high $\xi$, spatially uniform states are obtained. Thus, the amplitudes of the patterns depend nonmonotonically on $\xi$.
    Parameters: $a=0.015$, $b=10$, $\xi_1>0$ and $\xi_2<0$.} 
    \label{fig:fig_NR_Xipos}
\end{figure}

The variation in the pattern amplitudes in Fig.~\ref{fig:fig_NR_Xipos} are further quantified by calculating $\Delta_\phi,\,\Delta_\psi$ for various $\xi_1=\xi=-\xi_2>0$  in Fig.~\ref{Delta-xi1-pos-xi2-neg} for two values of $D_1=0.1, 0.01$ for a fixed $D=1000$. See Fig.~\ref{Delta-xi1-pos-xi2-neg}, which shows that both $\Delta_\phi$ and$\Delta_\psi$ display a novel nonmonotonic behavior: both of them first increase and then decrease as $\xi>0$ becomes larger. This nonmonotonic dependence of $\Delta_\phi,\Delta_\psi$ on $\xi>0$ in the DNS studies is qualitatively same as that observed for the nonmonotonic dependence of $k_c^2$ on $\xi>0$ in the extreme nonreciprocal case of the linear stability analysis. This establishes a direct correspondence between the linear stability analysis result on $k_c^2$ and our DNS studies results on the pattern amplitudes found in the extreme nonreciprocal chemotaxis.

\begin{figure}
    \centering
    \includegraphics[width=0.43\linewidth]{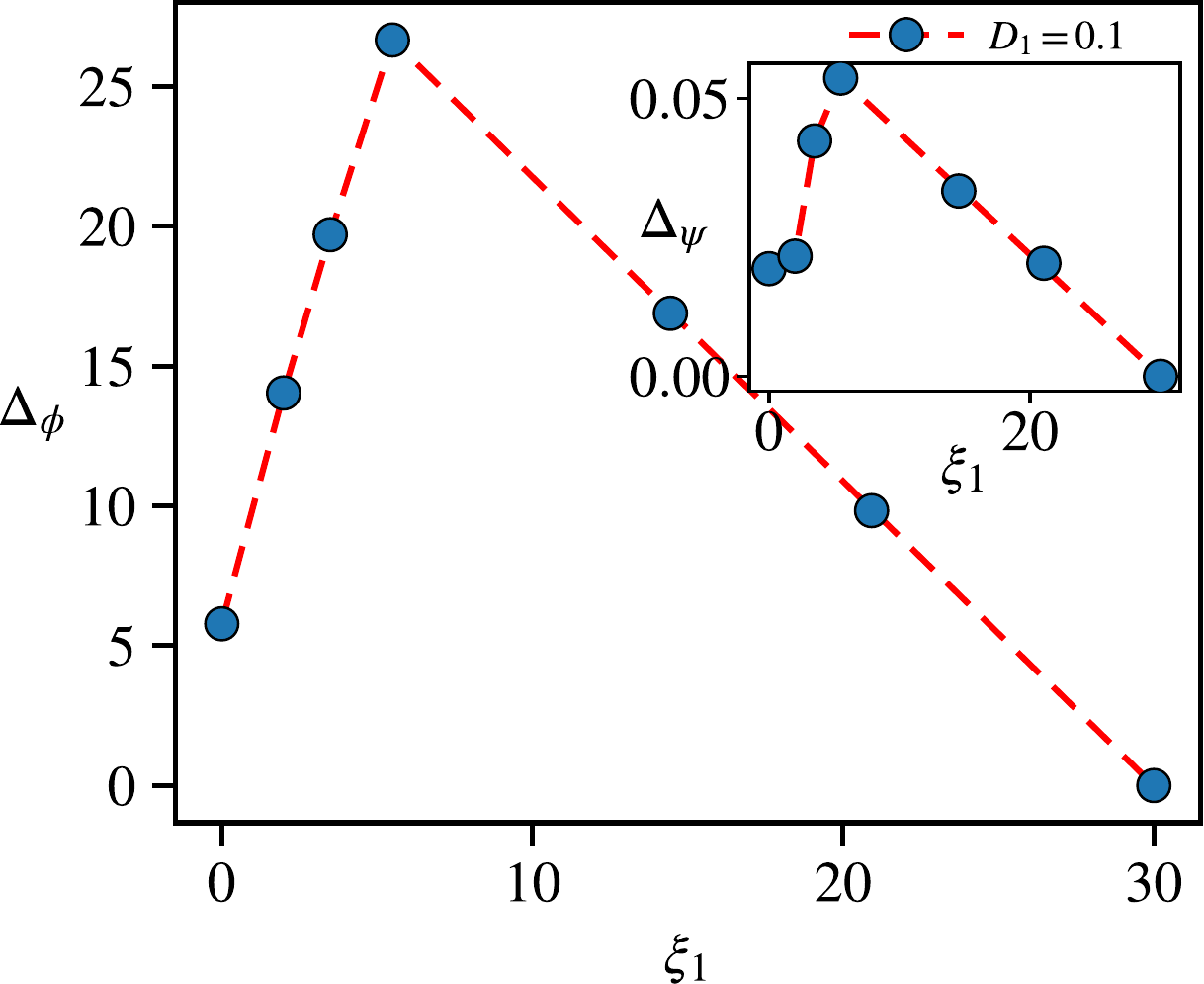} \hfill \includegraphics[width=0.43\linewidth]{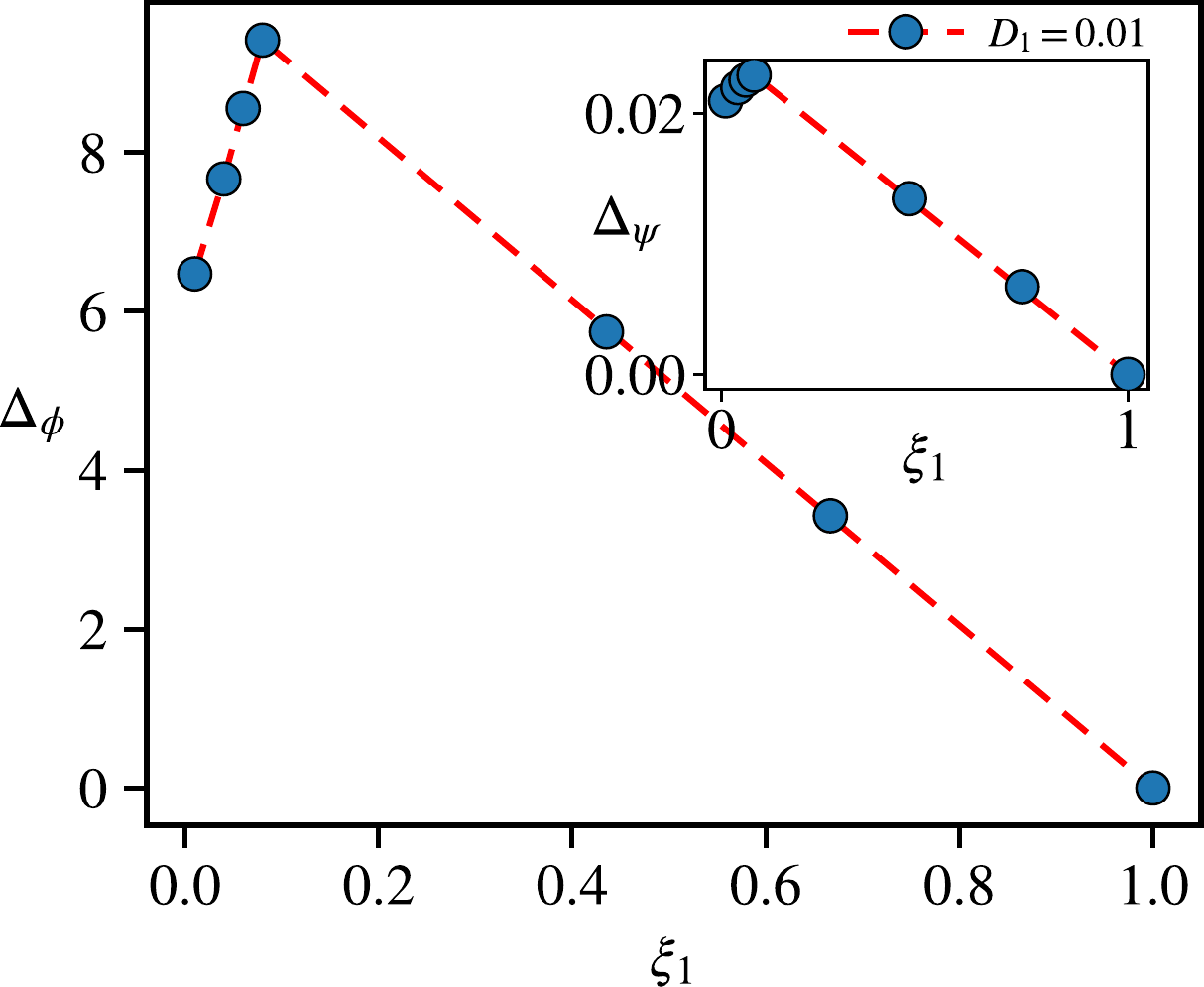}
    \caption{Plot of $\Delta_\phi$ (inset $\Delta_\psi$) versus $\xi_1=\xi=-\xi_2>0$ with  $D_1=0.1,\,0.01$, $D=1000$. Notice that both $\Delta_\phi$ and$\Delta_\psi$ display a novel nonmonotonic behavior: both of them first increase and then decrease as $\xi>0$ increases; see text.}
    \label{Delta-xi1-pos-xi2-neg}
\end{figure}

\subsubsection{ $\xi<0$} 

We now consider the complementary case, i.e., $\xi_1=\xi=-\xi_2<0$. Similar to the previous case, we numerically solve \eqref{mod1} and \eqref{mod2} and use pseudo-spectral methods for a range of $\xi<0$ for two values of $D_1=0.1, 0.01$ and fixed values of the other parameters. In this case, linear stability analysis predicts the selected wavevector $k_c^2$ to depend on $\xi$ monotonically - it decreases as $|\xi|$ increases,  ultimately vanishing in the limit of large $|\xi|$, giving a spatially uniform state. See Fig.~\ref{fig:fig_NR_Xineg}.

\begin{figure}[!ht]
    \centering
    \includegraphics[width=1.05\columnwidth]{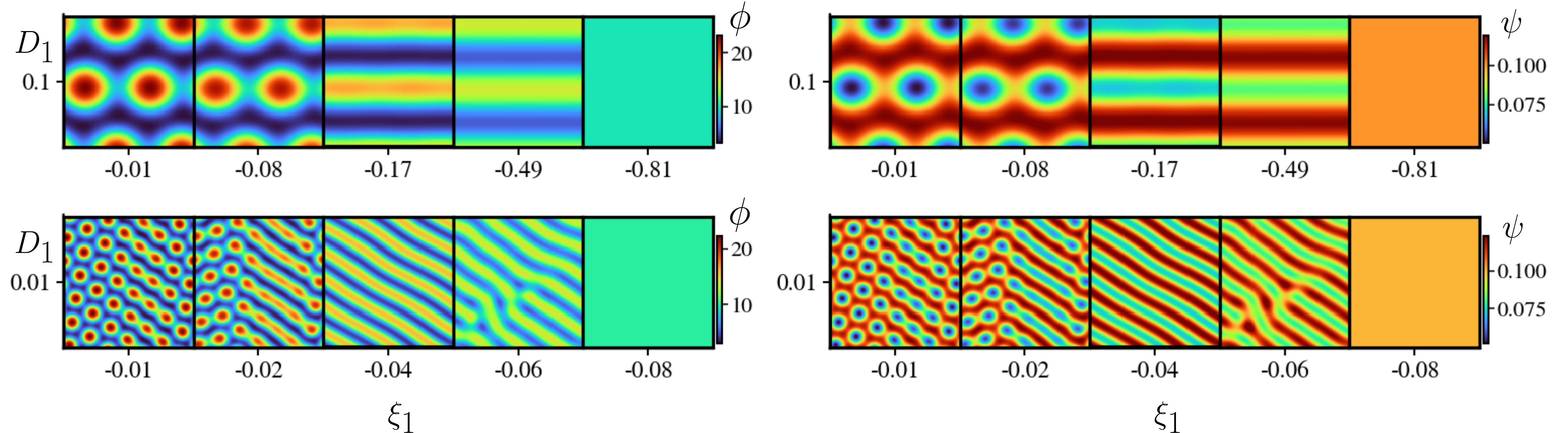}
    \caption{(Color online) Steady state snapshots of the fields $\phi$ and $\psi$ for different chemotactic couplings $\xi_1=\xi=-\xi_2<0$ in the extreme nonreciprocal chemotaxis case.
    Columns correspond to increasing coupling strength $\xi=-0.01$, $-0.17$, and $-0.49$, while rows give two values of the diffusion coefficient $D_1=0.01$ (top) and $D_1=0.1$ (bottom) for a fixed ratio $D$.
    As $|\xi|$ increases, the pattern morphologies change from spots to bands, ultimately giving rise to spatially uniform states. Thus, stronger chemotactic interactions lead to spatial homogenization.
    Parameters: $a=0.015$, $b=10$, $\xi_1<0$ and $\xi_2>0$.}
    \label{fig:fig_NR_Xineg}
\end{figure}

The variation in the pattern amplitudes in Fig.~\ref{fig:fig_NR_Xineg} is further quantified by calculating $\Delta_\phi,\,\Delta_\psi$ for various $\xi_1=\xi=-\xi_2>0$  in Fig.~\ref{Delta-xi1-neg-xi2-pos} for two values of $D_1=0.1, 0.01$ for a fixed $D=1000$. See Fig.~\ref{Delta-xi1-neg-xi2-pos}, which shows that both $\Delta_\phi$ and$\Delta_\psi$ decrease monotonically as $\xi<0$ becomes more negative. This monotonic dependence of $\Delta_\phi,\Delta_\psi$ on $\xi<0$ in the DNS studies is qualitatively same as that observed for the monotonic dependence of $k_c^2$ on $\xi<0$ in the extreme nonreciprocal case of the linear stability analysis. This establishes a direct correspondence between the linear stability analysis result on $k_c^2$ and DNS studies results on the pattern amplitudes found in the extreme nonreciprocal chemotaxis.

\begin{figure}
    \centering
    \includegraphics[width=0.43\linewidth]{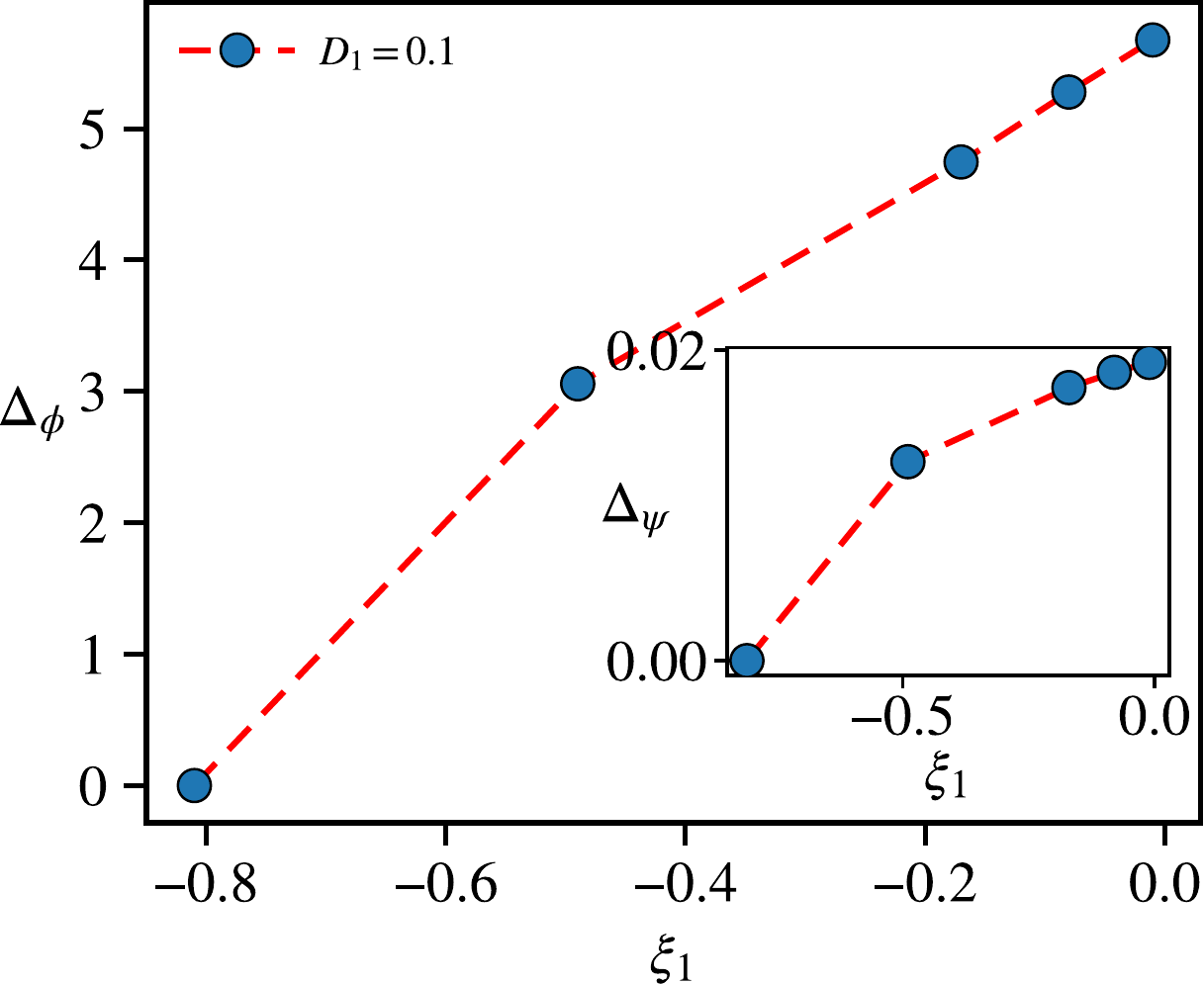} \hfill \includegraphics[width=0.43\linewidth]{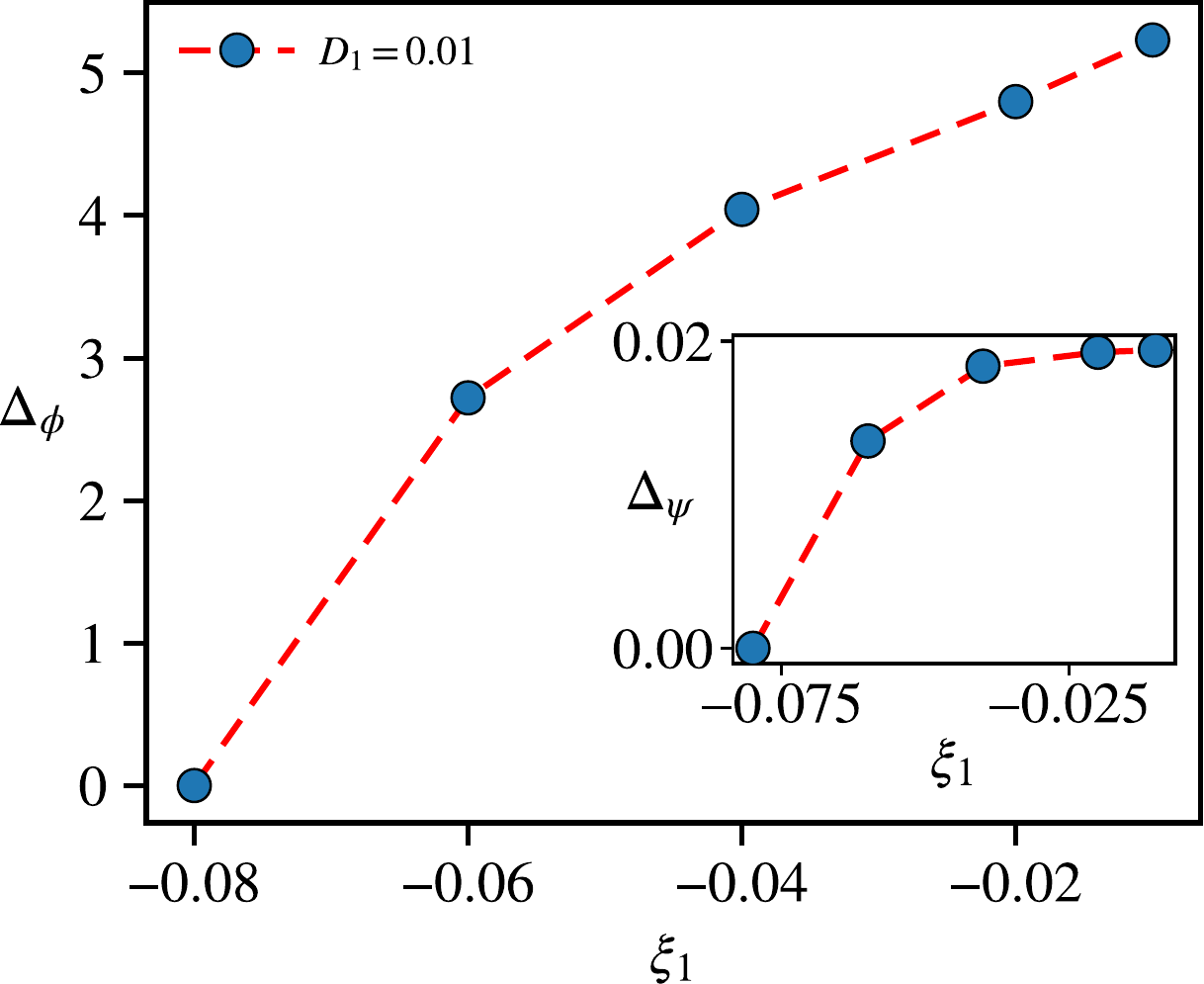} 
    \caption{Plot of $\Delta_\phi$ (inset $\Delta_\psi$) versus $\xi_1=\xi=-\xi_2>0$ with  $D_1=0.1,\,0.01$, $D=1000$. Notice that both $\Delta_\phi$ and$\Delta_\psi$ decrease monotonically as $\xi<0$ becomes larger negative. See text. }
    \label{Delta-xi1-neg-xi2-pos}
\end{figure}

Thus, we find that the cases $\xi>0$ and $\xi<0$ are mutually very distinct from each other. Also, the predictions from the linear stability analysis differ strongly from the DNS studies of \eqref{mod1} and \eqref{mod2}: The linear stability analysis predicts that $k_c^2$ strongly depends on $\xi$, while the DNS studies, which include nonlinear effects,  find that the pattern amplitudes depend on $\xi$, qualitatively in the {\it same} way as $k_c^2$ is predicted to depend on $\xi$ in the linear stability analysis. On the other hand, our DNS studies do not show any significant dependence of the pattern periodicity on $\xi$. 
Our results establish a direct correspondence between the linear stability analysis result on $k_c^2$ and DNS studies results on the pattern amplitudes found in the extreme nonreciprocal chemotaxis.

The results of Fig.~\ref{fig:fig_NR_Xipos} and Fig.~\ref{fig:fig_NR_Xineg} together, or equivalently, Fig.~\ref{Delta-xi1-pos-xi2-neg} and Fig.~\ref{Delta-xi1-neg-xi2-pos} together suggest  the following phenomenological forms of the amplitudes  
\begin{eqnarray}
    \phi({\bf x},t)=A^\text{NR}_\phi({\bf x},t)[\exp(i{\bf \tilde k}_c^{(\text{NR})}\cdot {\bf r})+cc],\;A^\text{NR}_\phi\sim f^\text{NR}_\phi(\xi),\\
    \psi({\bf x},t)=A^\text{NR}_\psi({\bf x},t)[\exp(i{\bf \tilde k}_c^{(\text{NR})}\cdot {\bf r})+cc],\;A^\text{NR}_\psi\sim f^\text{NR}_\psi(\xi),
\end{eqnarray}
where $f^\text{NR}_\phi(\xi), f^\text{NR}_\psi(\xi)$ are functions of $\xi$ (ignoring any $\bf x$ and $t$ dependence) with the following properties: $f^\text{NR}_\phi(\xi), f^\text{NR}_\psi(\xi)$ are nonmonotonic functions of $\xi$ for $\xi>0$, i.e., for small $\xi$ these functions increase as $\xi$ increases, if $\xi$ increases further, these functions start to decrease and eventually vanish for very large $\xi$ vanish. On the other hand, for $\xi<0$, $f^\text{NR}_\phi(\xi), f^\text{NR}_\psi(\xi)$ monotonically decrease as $|\xi|$ becomes larger. The preferred wavevector $\tilde k_c^{(\text{NR})}$ should be different from its counterpart $k_c$ in linear theory. However, $\tilde k_c^{(\text{NR})}$ is not expected to have a strong dependence on $\xi$ (positive or negative), but should depend on the other model parameters. Here, superscript NR implies nonreciprocal chemotaxis.

\subsection{Patterns in the fully reciprocal chemotaxis: $\xi_1=\xi=\xi_2$}

We now study the fully reciprocal chemotaxis case with $\xi_1=\xi=\xi_2$, which corresponds to the two species being mutually attracted ($\xi>0$) or repelled ($\xi<0$) as discussed above. We discuss these two cases separately below.

\subsubsection{ $\xi>0$}

In this case, both species attract each other due to mutual chemotaxis. Linear stability analysis predicts that for  $b^2>a$,  $k_c^2$ increases monotonically, ultimately diverging at a finite upper threshold on $\xi>0$ that depends on the other model parameters. DNS studies of  \eqref{mod1} and \eqref{mod2} for $\xi>0$, $D_1=0.01, 0.1$ and $b^2>a$ with fixed values of the other parameters, however, show that pattern amplitudes increase monotonically with $\xi$, eventually diverging at a finite upper threshold on $\xi$. This signals that the system is becoming unstable. See Fig.~\ref{fig:fig_Reci_Xipos}.

\begin{figure}[!ht]
    \centering
    \includegraphics[width=1.05\columnwidth]{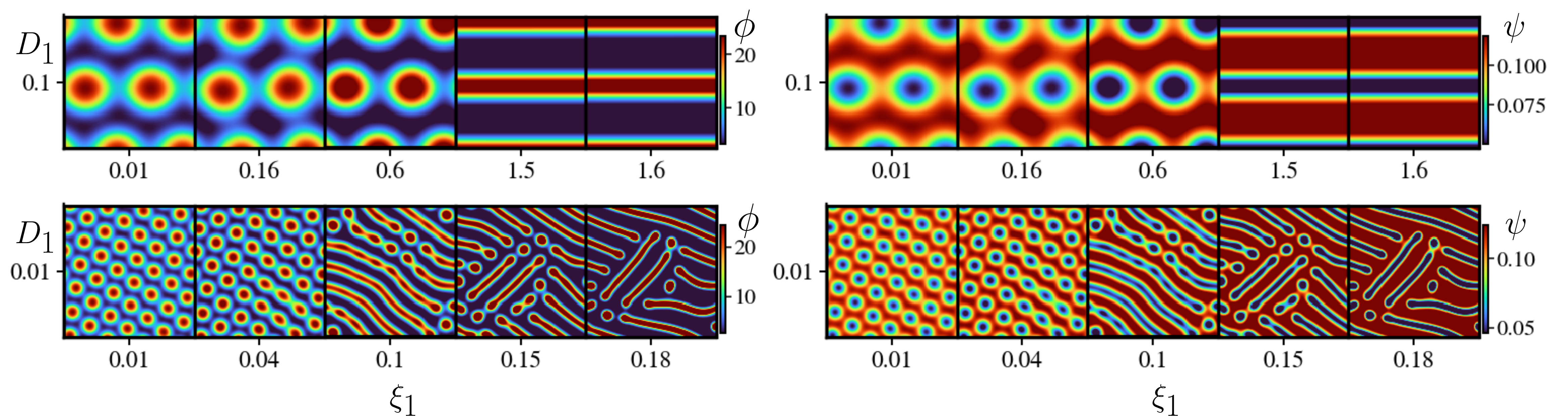}
    \caption{(Color online) { Steady state patterns of  $\phi$ and $\psi$ in the extreme reciprocal chemotactic coupling case $\xi_1=\xi=\xi_2>0$, where both species  attract  each other.}
    Columns correspond to increasing coupling strength $\xi=0.01$, $0.17$, and $0.49$, while rows compare two values of $D_1=0.01$ (top), $D_1= 0.1$ (bottom) for a fixed $D$ and $b^2>a$. As $\xi$ increases, the patterns become more intense, i.e., the amplitudes increase, along with the pattern morphology changes from spots to stripe-like structures for higher $\xi$. Beyond certain upper thresholds on $\xi$ that depend on $D_1$ and also other model parameters, the system becomes unstable (not shown).
    Parameters: $a=0.015$ and $b=10$.}
    \label{fig:fig_Reci_Xipos}
\end{figure}

The variation in the pattern amplitudes in Fig.~\ref{fig:fig_Reci_Xipos} is further quantified by calculating $\Delta_\phi,\,\Delta_\psi$ for various $\xi_1=\xi=\xi_2>0$  in Fig.~\ref{Delta-xi1-pos-xi2-pos} for two values of $D_1=0.1, 0.01$ for a fixed $D=1000$. See Fig.~\ref{Delta-xi1-pos-xi2-pos}, which shows that both $\Delta_\phi$ and$\Delta_\psi$  increase monotonically as $\xi>0$ increases. This monotonic dependence of $\Delta_\phi,\Delta_\psi$ on $\xi>0$ in the DNS studies is qualitatively same as that observed for the monotonic dependence of $k_c^2$ on $\xi>0$ in the extreme reciprocal case of the linear stability analysis. Therefore, our results establish a direct correspondence between the linear stability analysis result on $k_c^2$ and DNS studies results on the pattern amplitudes found in the fully reciprocal chemotaxis.

\begin{figure}
    \centering
    \includegraphics[width=0.43\linewidth]{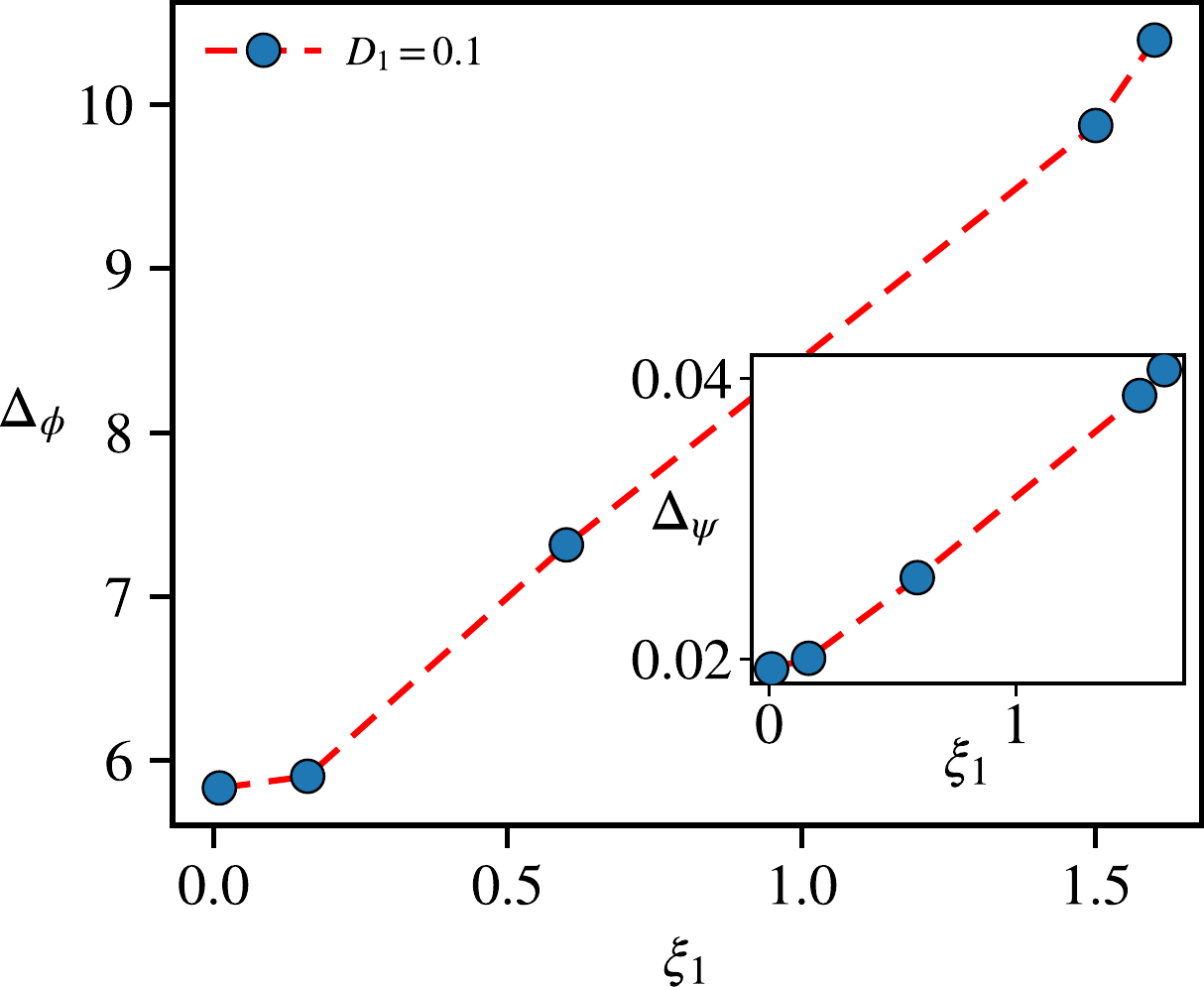} \hfill \includegraphics[width=0.43\linewidth]{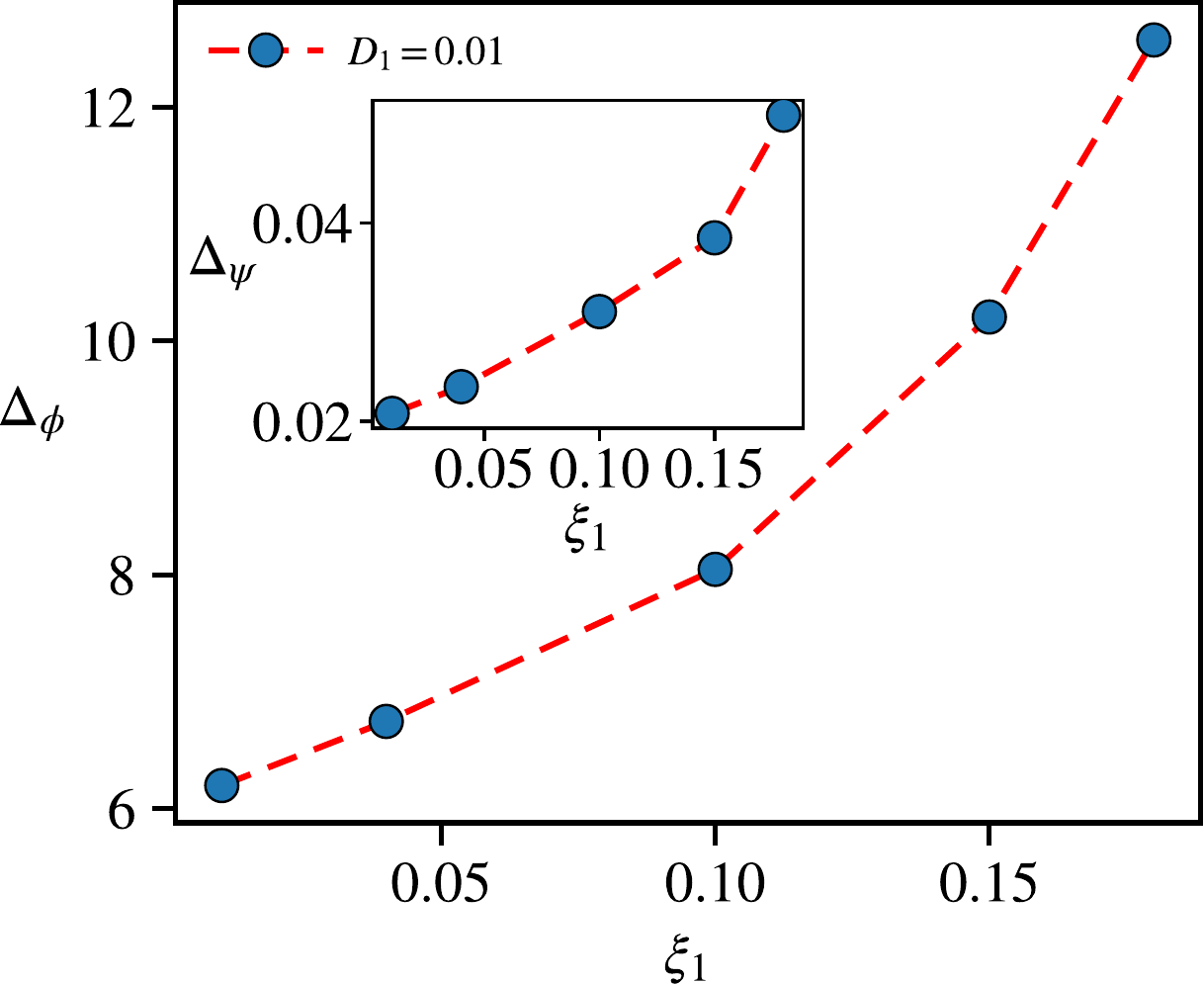}
    \caption{Plot of $\Delta_\phi$ (inset $\Delta_\psi$) versus $\xi_1=\xi=-\xi_2>0$ with  $D_1=0.1,\,0.01$, $D=1000$ in the fully reciprocal chemotaxis case. Notice that both $\Delta_\phi$ and$\Delta_\psi$ increase monotonically as $\xi>0$ increases. See text.}
    \label{Delta-xi1-pos-xi2-pos}
\end{figure}

\subsubsection{ $\xi<0$}

We now study the opposite case where $\xi_1=\xi=\xi_2<0$. In this case, both species mutually repel each other by chemotaxis.  Linear stability analysis predicts that for $b^2>a$, $k_c^2$ in the present case should first decrease as $|\xi|$ increases, then as $|\xi|$ crosses a critical value, $k_c^2$ increases with increasing $|\xi|$, eventually diverging at a finite upper threshold (which depends on the other model parameters) on $\xi$, signaling an instability induced by strong chemotaxis. In contrast, DNS studies of the model equations \eqref{mod1} and \eqref{mod2} for $\xi<0$, $D_1=0.01, 0.1$ and $b^2>a$ with fixed values of the other parameters, however, show that the pattern amplitudes first decrease as $|\xi|$ increases; then as $|\xi|$ crosses a critical value, the amplitudes start increasing, eventually diverging at a finite upper threshold on $|\xi|$ (not shown), signaling a strong chemotaxis-induced instability. The periodicity of the pattern does not appear to have a strong dependence on $|\xi|$. Although not shown in Fig.~\ref{fig:fig_Reci_Xinega}, it is possible to have a situation where there is a pattern for $|\xi|\rightarrow 0$, with the pattern steadily fading away, effectively vanishing at some intermediate $|\xi$ as $|\xi|$ increases, ultimately reappearing and becoming unstable, as $|\xi|$ increases further, finally crossing beyond a threshold. This is a re-entrant behavior quite distinct from its counterpart in the extreme nonreciprocal case with $\xi>0$.

\begin{figure}[!ht]
    \centering
    \includegraphics[width=1.05\columnwidth]{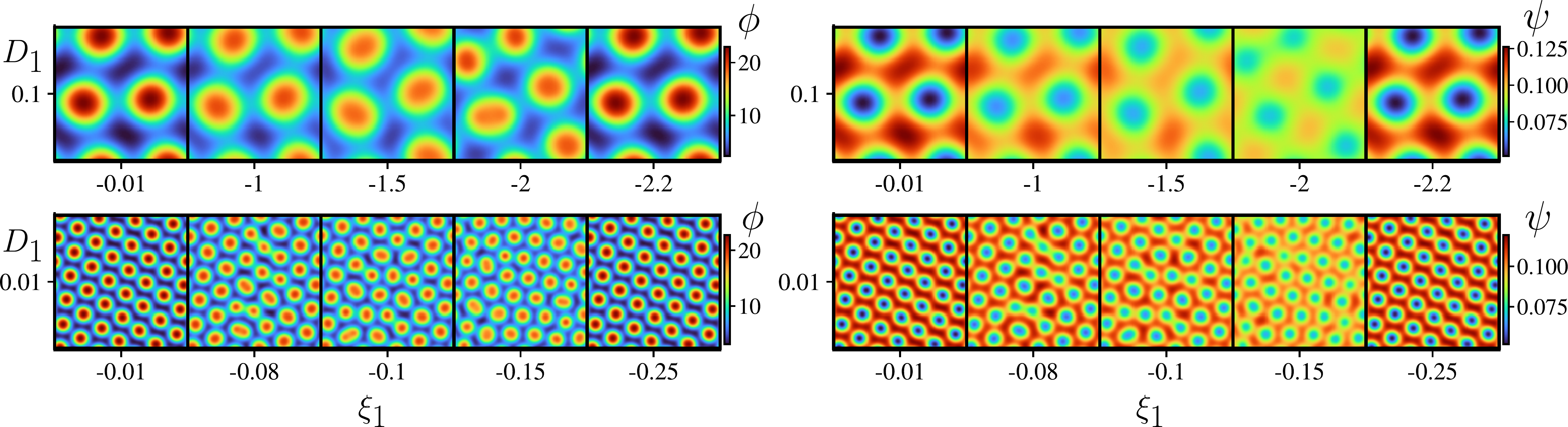}
    \caption{(Color online) { Steady state snapshots of the densities $\phi$ and $\psi$ for the fully reciprocal chemotaxis case with $\xi_1=\xi_2=\xi<0$, where both species repel each other equally.}
    Columns correspond to increasing magnitude of the coupling strength $\xi=-0.01$, $-0.17$, and $-0.49$, while rows compare two values of $D_1=0.01$ (top), $D_1= 0.1$ (bottom) for a fixed $D$ and $b^2>a$.
    Weak repulsive coupling maintains periodic spot-like structures, but as $|\xi|$ increases, chemotactic repulsion smooths out local density variations, leading to progressively more uniform patterns.
    Parameters: $a=0.015$ and $b=10$.}
    \label{fig:fig_Reci_Xinega}
\end{figure}

The variation in the pattern amplitudes in Fig.~\ref{fig:fig_Reci_Xinega} is further quantified by calculating $\Delta_\phi,\,\Delta_\psi$ for various $\xi_1=\xi=\xi_2<0$  in Fig.~\ref{Delta-xi1-neg-xi2-neg} for two values of $D_1=0.1, 0.01$ for a fixed $D=1000$. See Fig.~\ref{Delta-xi1-neg-xi2-neg}, which shows that both $\Delta_\phi$ and$\Delta_\psi$ display an intriguing nonmonotonic behavior with $\xi<0$: As $\xi<0$ becomes more negative,  both $\Delta_\phi$ and$\Delta_\psi$ first decrease and then increase. This nonmonotonic dependence of $\Delta_\phi,\Delta_\psi$ on $\xi<0$ in the DNS studies is qualitatively same as that observed for the nonmonotonic dependence of $k_c^2$ on $\xi<0$ in the extreme reciprocal case of the linear stability analysis. Once again, our results establish a direct correspondence between the linear stability analysis result on $k_c^2$ and DNS studies results on the pattern amplitudes found in the fully reciprocal chemotaxis.

Overall, our results highlight the significant difference between $\xi_1=\xi_2>0$ and $\xi_1=\xi_2<0$ in the fully reciprocal chemotaxis cases. 

\begin{figure}
    \centering
    \includegraphics[width=0.43\linewidth]{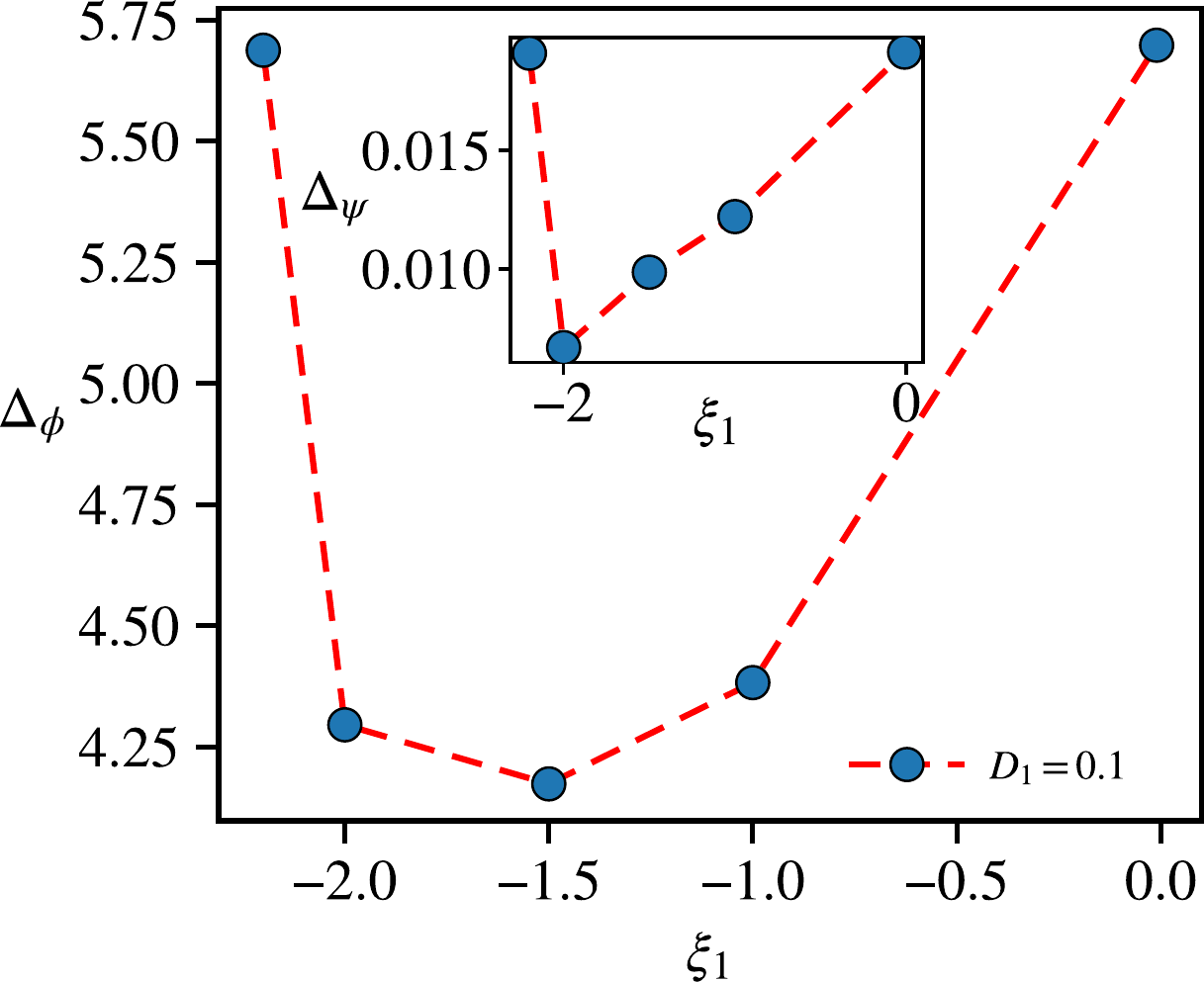} \hfill \includegraphics[width=0.43\linewidth]{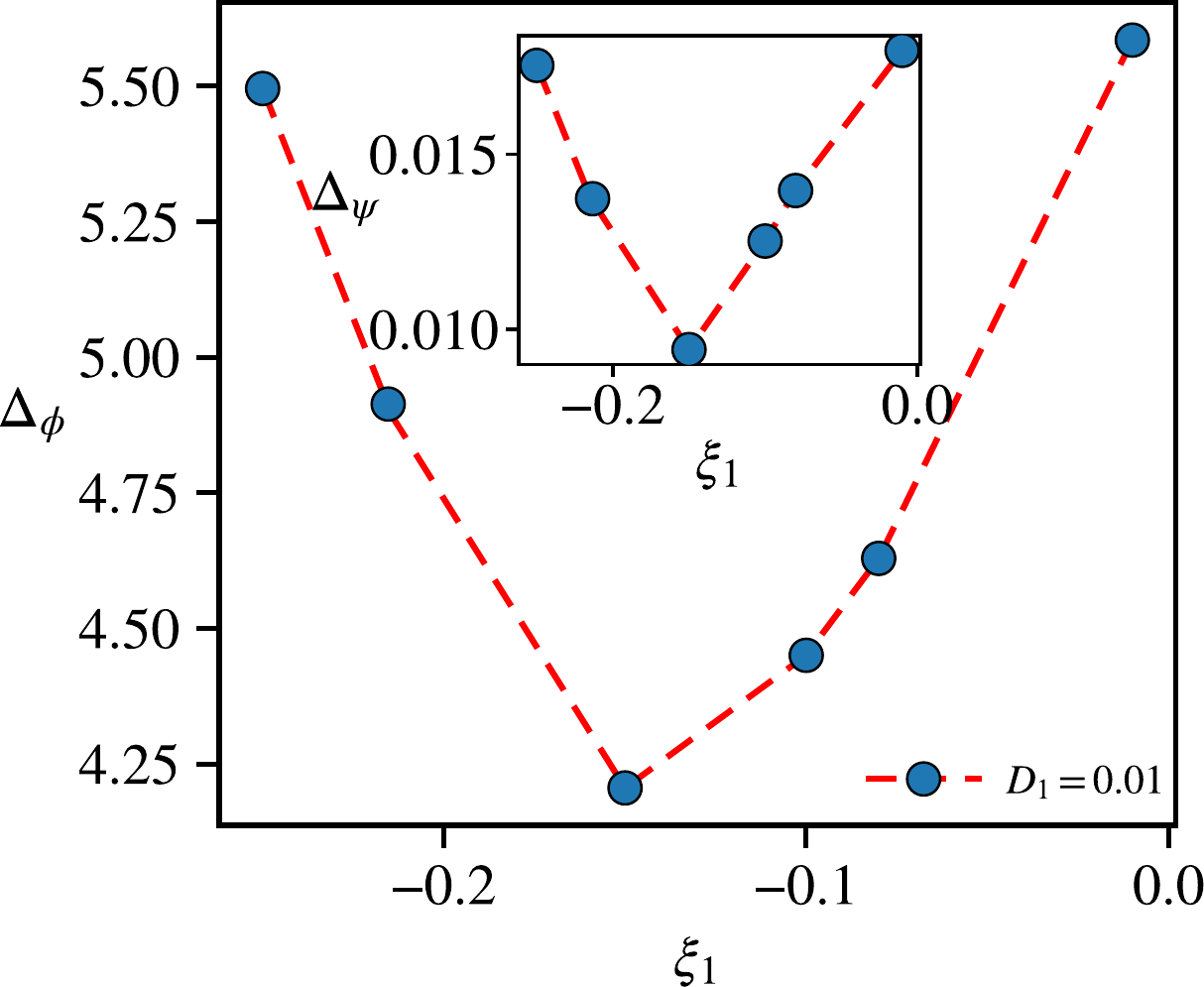}
    \caption{Plot of $\Delta_\phi$ (inset $\Delta_\psi$) versus $\xi_1=\xi=\xi_2<0$ with  $D_1=0.1,\,0.01$, $D=1000$ in the fully reciprocal chemotaxis case. Notice that both $\Delta_\phi$ and$\Delta_\psi$ display an intriguing nonmonotonic behavior with $\xi<0$: As $\xi<0$ becomes more negative,  both $\Delta_\phi$ and$\Delta_\psi$ first decrease and then increase. See text.}
    \label{Delta-xi1-neg-xi2-neg}
\end{figure}
\end{widetext}

Similar to the extreme nonreciprocal cases studied above, we write down the following phenomenological forms of the amplitudes in the fully reciprocal cases considered here. Indeed, 
the results of Fig.~\ref{fig:fig_Reci_Xipos} and Fig.~\ref{fig:fig_Reci_Xinega} together suggest  the phenomenological forms of the amplitudes  
\begin{eqnarray}
    \phi({\bf x},t)=A^\text{R}_\phi({\bf x},t)[\exp(i{\bf \tilde k}_c^{(\text{R})}\cdot {\bf r})+cc],\;A^\text{R}_\phi\sim f^\text{R}_\phi(\xi),\\
    \psi({\bf x},t)=A^\text{R}_\psi({\bf x},t)[\exp(i{\bf \tilde k}_c^{(\text{R})}\cdot {\bf r})+cc],\;A^\text{R}_\psi\sim f^\text{R}_\psi(\xi),
\end{eqnarray}
where $f^\text{R}_\phi(\xi), f^\text{R}_\psi(\xi)$ are functions of $\xi$ (ignoring any $\bf x$ and $t$ dependence) with the following properties: $f^\text{R}_\phi(\xi), f^\text{R}_\psi(\xi)$ are monotonic functions of $\xi$ for $\xi>0$, i.e.,  these functions increase as $\xi$ increases,  and eventually vanish for very large $\xi$ vanish. On the other hand, for $\xi<0$, $f^\text{R}_\phi(\xi), f^\text{R}_\psi(\xi)$ are nonmonotonic functions of $\xi<0$, 
monotonically decreasing as $|\xi|$ becomes larger. The preferred wavevector $\tilde k_c^{(\text{R})}$ should be different from its counterpart $k_c$ in linear theory. However, $\tilde k_c^{(\text{R})}$ is not expected to have a strong dependence on $\xi$ (positive or negative), but should depend on the other model parameters. Here, superscript R implies fully reciprocal chemotaxis.

\section{Transition between spots and stripes}

We now study transitions between different pattern morphologies, e.g., between spots and stripes, and possible tuning of these transitions by varying the relevant parameters. We study these by DNS of the model equations \eqref{mod1} and \eqref{mod2}, and identify $\xi$ and $D$ as the key control parameters. We first discuss the consequence of varying $\xi_1=\xi=\pm\xi_2$ for fixed $D_1$ and $D$.

\begin{figure*}[!ht]
    \centering
\includegraphics[width=2\columnwidth]{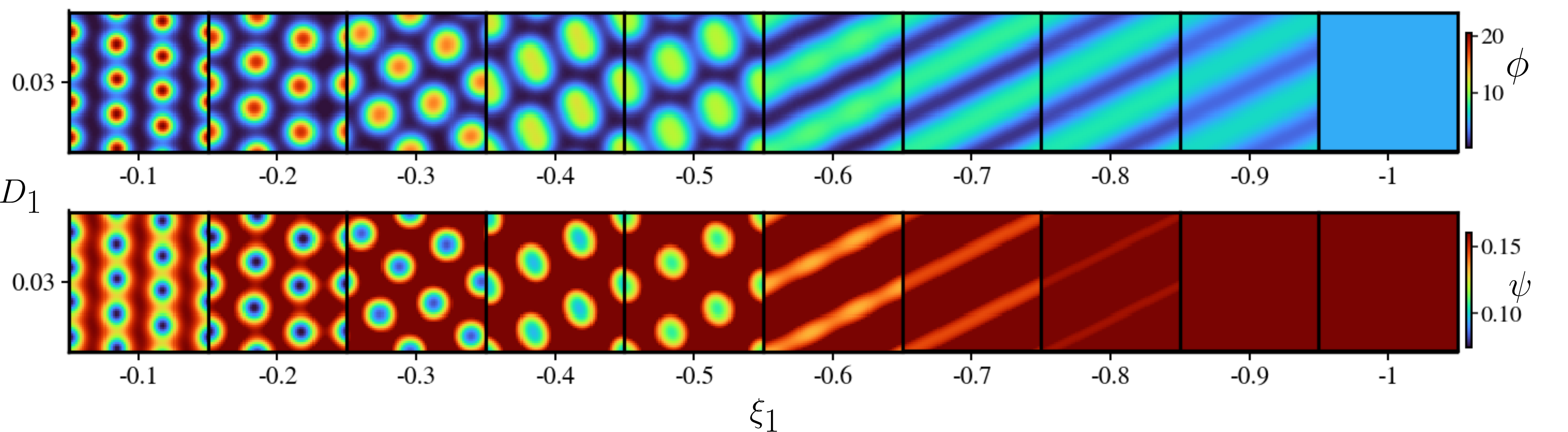}
    \caption{(Color online) Snapshot of the steady state densities illustrating the transition from spot-like to stripe-like morphologies as the coupling $\xi_1=\xi=-\xi_2<0$ is varied in the extreme nonreciprocal case.
   The top and bottom rows show the densities of species $\phi$ and $\psi$, respectively, at fixed diffusion coefficient $D_1=0.03$.
  As $|\xi_1|$ increases, the system undergoes structural transitions from spot arrays to elongated stripe patterns and eventually to nearly homogeneous states.
The two species exhibit distinct responses due to nonreciprocity, with $\psi$ showing stronger smoothing and earlier suppression of spatial modulations.
Parameters: $a=0.015$, $b=5$, and $D_2=1000D_1$. }
\label{fig:fig_NR_spots_stripe_transition}
\end{figure*}

We generally find that in the extreme nonreciprocal case with $\xi<0$ for small $|\xi|$, spots appear as steady patterns. As $|\xi|$ increases, keeping the other model parameters unchanged, the spots become stripes. For even higher $|\xi|$, spatially uniform states ensue. This can already be seen in Fig.~\ref{fig:fig_NR_Xineg} in the previous Section. See also Fig.~\ref{fig:fig_NR_spots_stripe_transition} below for more detailed results from the DNS studies on transitions between various pattern morphologies as $\xi_1=\xi=-\xi_2<0$ is varied in the extreme nonreciprocal case, displaying the robustness of the phenomenon. Similarly, in the fully reciprocal case with $\xi>0$, spots appear for small $\xi$, whereas stripes are observed as $\xi$ is increased, before chemotaxis induced instabilities set in above a threshold on $\xi$. 

We also find that for a fixed $\xi$, varying $D_1$ can lead to structural transitions in the pattern. However, we observe a crucial difference between the transitions controlled by $\xi$ and those controlled by $D_1$. In the former case, the transitions are always from spots to stripes as $|\xi|$ increases. On the other hand, by increasing $D_1$, a variety of transitions are possible, e.g., spots to stripes and vise versa. The robustness of this behavior is confirmed by using various seeds for the random initial conditions.  This holds for both extreme nonreciprocal and reciprocal chemotaxis cases. See Fig.~\ref{fig:fig_NRxiNeg0-01},
Fig.~\ref{fig:fig_RxiNeg0-01} in Appendix~\ref{diff-eff}.

\begin{figure}[htb]
    \includegraphics[width=\columnwidth]{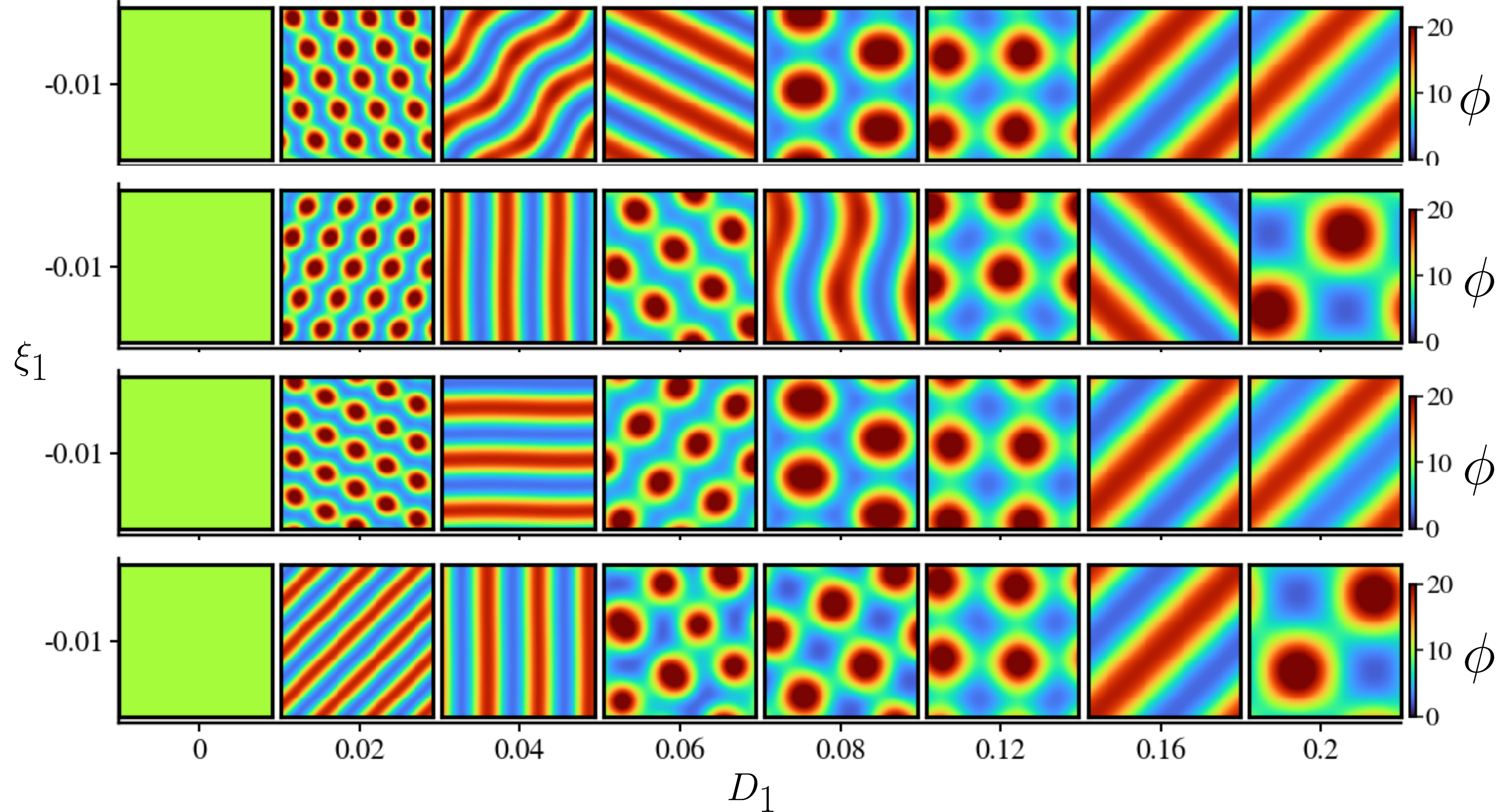}
    \caption{Steady state snapshots of the density field $\phi$ for $\xi_1=\xi=-\xi_2=-0.01$ for the extreme nonreciprocal chemotaxis case, shown as a function of the diffusion coefficient $D_1$ (horizontal axis), with $D_2=1000D_1$.
    Each row corresponds to a different random initial condition (10 independent seeds), illustrating variability in pattern selection.
    For small $D_1$, the system remains nearly homogeneous, while increasing $D_1$ leads to the emergence of ordered structures including stripes and spot arrays.
    Parameters: $a=0.015$ and $b=10$. }
    \label{fig:fig_NRxiNeg0-01}
\end{figure}

\begin{figure}[t]
    \includegraphics[width=\columnwidth]{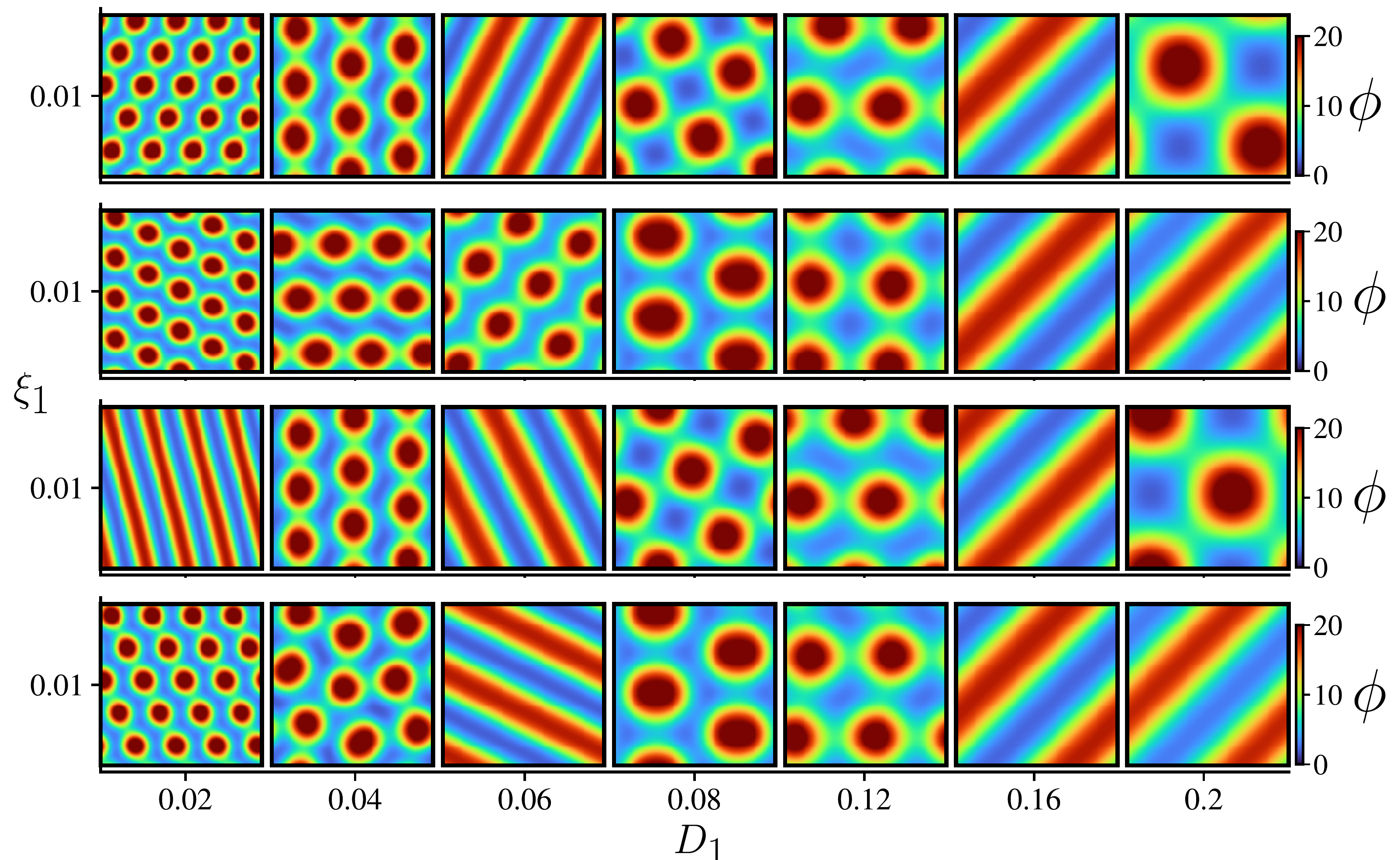}
    \caption{Steady state snapshots of the density field $\phi$ for $\xi_1=\xi=-\xi_2=0.01$ for the extreme nonreciprocal chemotaxis case, shown as a function of the diffusion coefficient $D_1$ (horizontal axis), with $D_2=1000D_1$.
    Each row corresponds to a different random initial condition (5 independent seeds), illustrating variability in pattern selection.
    For small $D_1$, the system remains nearly homogeneous, while increasing $D_1$ leads to the emergence of ordered structures including stripes and spot arrays.
    Parameters: $a=0.015$ and $b=10$.}
    \label{fig:fig_NRxiPos0-01}
\end{figure}

\begin{figure}[htb]
    \centering
    \includegraphics[width=\columnwidth]{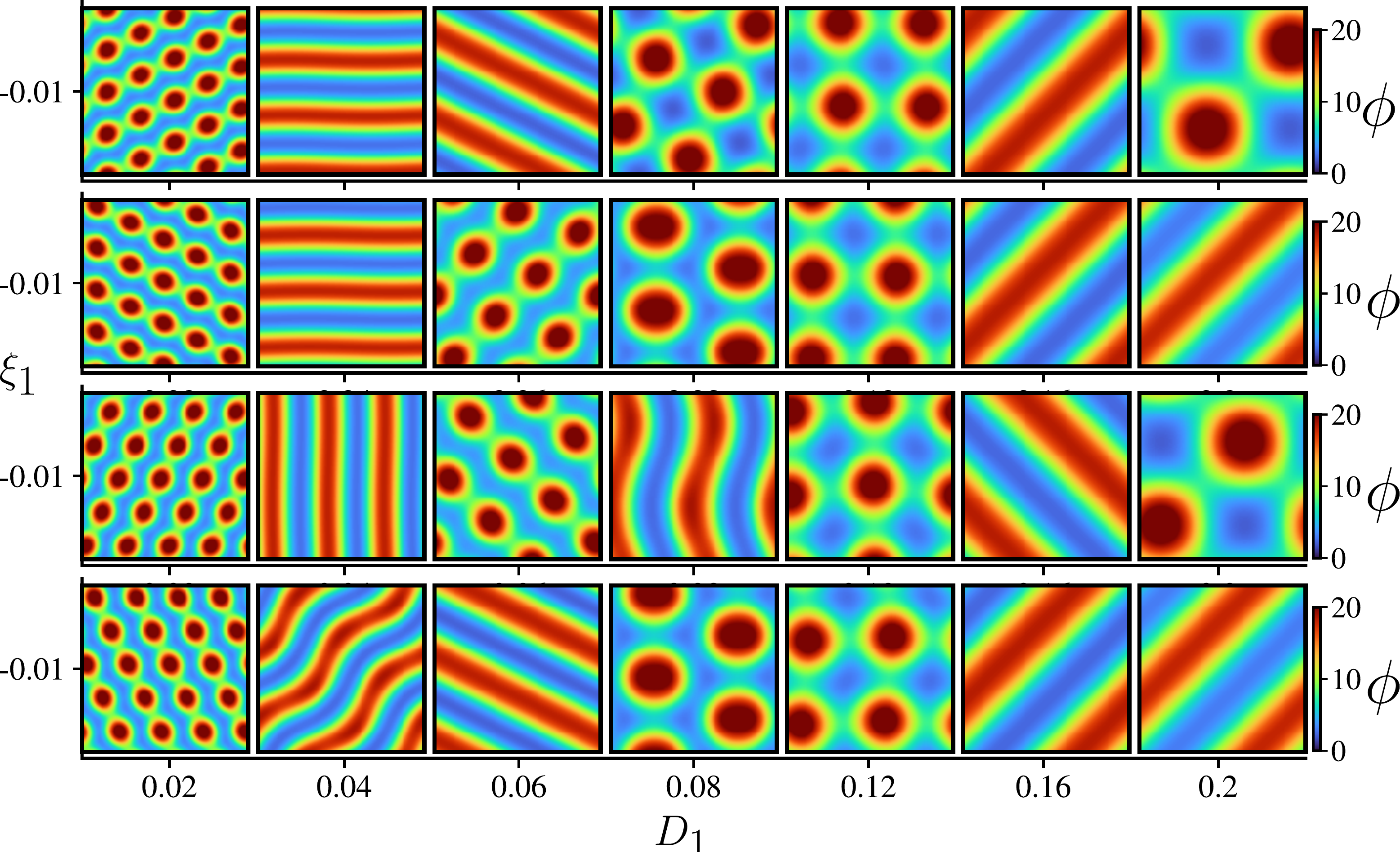}
    \caption{Steady state snapshots of the density field $\phi$ for negative coupling $\xi_1=\xi=-0.01$, for the extreme reciprocal chemotaxis case, shown as a function of the diffusion coefficient $D_1$ (horizontal axis), with $D_2=1000D_1$.
    Each row corresponds to a different random initial condition (5 independent seeds), illustrating variability in pattern selection.
    As $D_1$ increases, the system exhibits re-entrant transitions between spot-like and stripe-like morphologies.
    Parameters: $a=0.015$ and $b=10$.}
    \label{fig:fig_RxiNeg0-01}
\end{figure}

\begin{figure}[htb]
    \centering
    \includegraphics[width=\columnwidth]{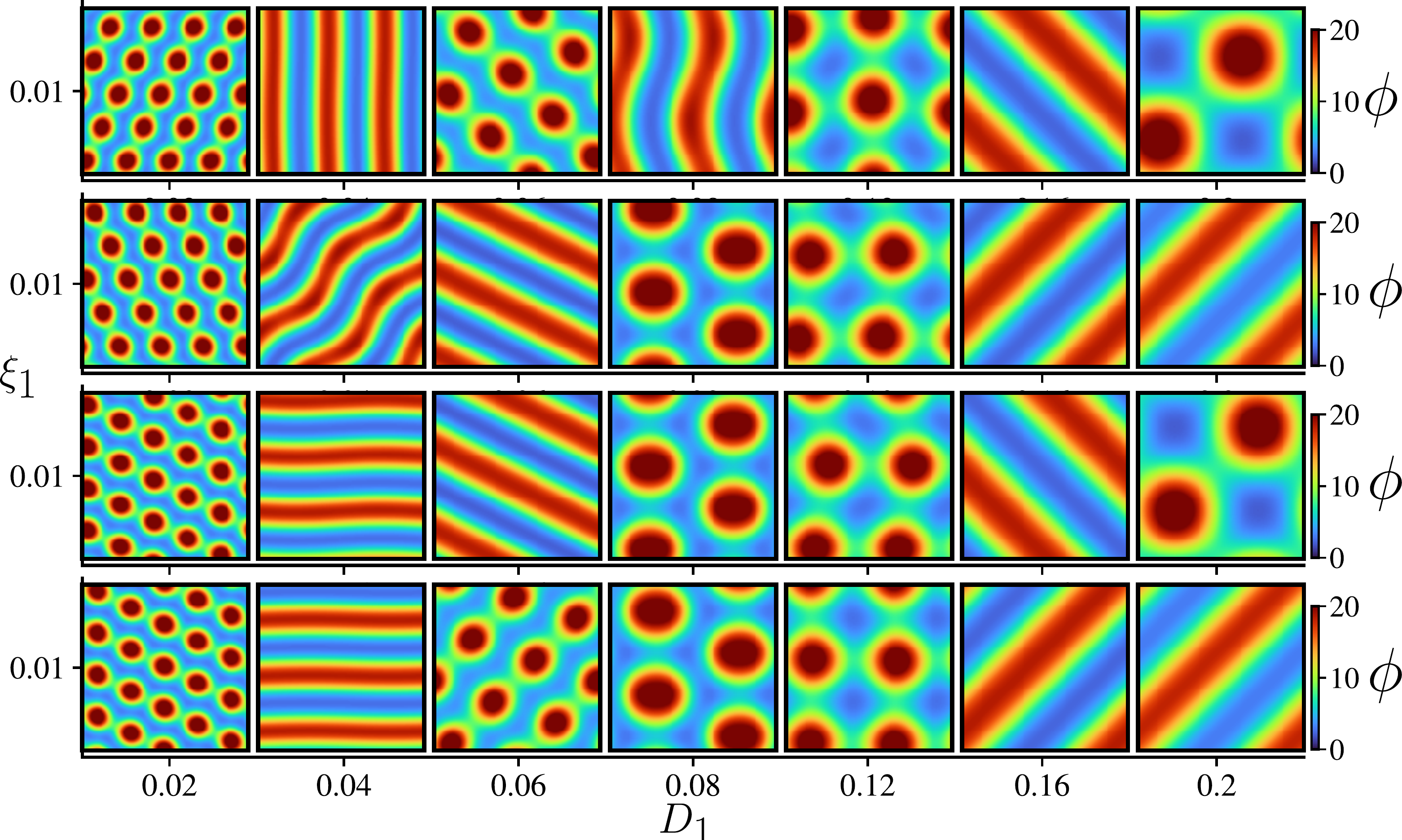}
    \caption{Steady state snapshots of the density field $\phi$ for negative coupling $\xi_1=\xi=0.01$, for the extreme reciprocal chemotaxis case, shown as a function of the diffusion coefficient $D_1$ (horizontal axis), with $D_2=1000D_1$.
    Each row corresponds to a different random initial condition (5 independent seeds), illustrating variability in pattern selection.
    As $D_1$ increases, the system exhibits re-entrant transitions between spot-like and stripe-like morphologies.
    Parameters: $a=0.015$ and $b=10$.}
    \label{fig:fig_RxiPos0-01}
\end{figure}

 \section{Linear amplitude equations}\label{ampli}

 We have found that in spite of the isotropy of the model, the emerging pattern can be stripes, i.e., strongly anisotropic. We now set up the linear amplitude equation for stripe patterns and use it to show that such striped states are {\it linearly stable}. To start with, we have found that $k=k_c$ is a preferred wavevector that becomes the fundamental mode. This means $u,v\sim \exp (\pm i{\bf k}_c\cdot {\bf r}$ are solutions of the linearized equations (\ref{modlin1}) and (\ref{modlin2}). This may be verified as follows. The solvability condition for the coupled linear equations (\ref{modlin1}) and (\ref{modlin2}) is

 \begin{eqnarray}
     \begin{vmatrix}
         -\frac{b^2-a}{a+b^2}+D_1 k_c^2 & -(a+b^2)+\xi_1bk_c^2\\
         \frac{2b^2}{a+b^2}+\frac{\xi_2bk_c^2}{a+b^2} & a+b^2 +D_2 k_c^2
     \end{vmatrix}
     =0,
 \end{eqnarray}
which gives
\begin{eqnarray}
   && k_c^4\bigg[D_1 D_2 - \frac{\xi_1\xi_2b^2}{a+b^2}\bigg]+k_c^2 \bigg [D_1 (a+b^2)-D_2 \frac{b^2-a}{a+b^2}\nonumber \\&&-\frac{2\xi_1b^3}{a+b^2}+\xi_2b\bigg]-(b^2-a)+2b^2=0,
\end{eqnarray}
by using the value of $k_c^2$ in (\ref{qc}), Q.E.D. To explore the linear stability of these solutions, we allow ``small'' deviations from the fundamental mode solutions and set up their linearized equations. To proceed, we assume solutions of the form
\begin{eqnarray}
    u({\bf r},t)&=& A_u ({\bf r},t)[\exp (i{\bf k}_c\cdot {\bf r})] + \text{cc},\label{u-linear-amplitude}\\
    v({\bf r},t)&=& A_v ({\bf r},t)[\exp (i{\bf k}_c\cdot {\bf r})] +\text{cc},\label{v-linear-amplitude}
\end{eqnarray}
where $A_u({\bf r},t), A_v({\bf r},t)$ are slowly varying functions of position $\bf r$. This essentially means that $A_u({\bf \Delta q},t), A_v({\bf \Delta q},t)$, which are functions of wavevector $\Delta \bf q$ and spatial Fourier transforms of $A_u({\bf r},t), A_v({\bf r},t)$ are such functions of wavevector $\bf \Delta q$, where $|\Delta {\bf q}|\ll k_c$, i.e.,  typical variations in $A_u({\bf r},t), A_v({\bf r},t)$  take place over scales $\sim 2\pi/|\Delta {\bf q}|\gg 2\pi/{k_c}$, that scale of variation of the fundamental modes.  Thus, there is a clear length scale separation between these two scales. 

Consider a point $(a=a_c+\epsilon_1,b=b_c+\epsilon_2)$ in the $a-b$ plane, with $a_c,b_c$ being on the boundary of the patterned state. We consider $\epsilon_1,\epsilon_2$ to be such that $(a,b)$ lies in the pattern region, but close to the pattern boundary (i.e., $\epsilon_1,\epsilon_2$ are assumed to be small). Focusing on striped patterns, which we assume to be perpendicular to the $x$-axis, i.e., the stripes are extended along the $y$-axis, we write
\begin{equation}
    {\bf k}_c= k_c \hat x.
\end{equation}
Then 
\begin{equation}
    A_{u,v}({\bf q},t)=A_{u,v}\exp[i\Delta {\bf q}\cdot r][\exp (iq_c x) +\exp(-iqc x)],\label{a-uv}
\end{equation}
where $\Delta {\bf q}\equiv {\bf q}- k_c \hat x$ is ``small'', i.e., $|\Delta q|/k_c \ll 1$. Let
\begin{equation}
    {\bf q}=(k_c + q_x)\hat x + q_y\hat y,
\end{equation}
for stripes extended along the $x$-axis. Therefore,
\begin{eqnarray}
    q= |{\bf q}| = [(k_c+ q_x)^2+ q_y^2]^{1/2}=k_c\bigg[1+\frac{2q_x}{k_c}+\frac{q_y^2}{k_c^2}\bigg]^{1/2}.
\end{eqnarray}
Expanding for small $q_x,q_y$, we then find
\begin{equation}
    q\approx k_c + q_x + \frac{q_y^2}{2k_c},
\end{equation}
giving
\begin{equation}
    |\Delta {\bf q}| = q_x + \frac{q_y^2}{2k_c}. \label{Delta-q}
\end{equation}
Substituting (\ref{a-uv}) in (\ref{modlin1}) and (\ref{modlin2}), we obtain
\begin{eqnarray}
    \frac{\partial A_u}{\partial t}&=& \frac{b^2-a}{a+b^2}A_u + (a+b^2)A_v - D_1 [k_c^2 + (\Delta q)^2]A_u \nonumber \\&&- \xi_1 b [k_c^2+(\Delta q)^2] A_v,\\
    \frac{\partial A_v}{\partial t}&=& -(a+b^2)A_v - \frac{2b^2}{a+b^2}A_u - D_2 [k_c^2 + (\Delta q)^2]A_v \nonumber \\
    && - \frac{\xi_2b}{a+b^2}[k_c^2 + (\Delta q)^2]. 
\end{eqnarray}
We then set $a=a_c+\epsilon_1,\,b=b_c+\epsilon_2$, as mentioned above, and expand for small $\epsilon_1,\epsilon_2$. To the lowest (linear) order in $\epsilon_1,\epsilon_2$, we find

\begin{widetext}
 
\begin{eqnarray}
    \frac{\partial A_u}{\partial t}&=& \bigg[\frac{\epsilon_1+2b_c\epsilon_2}{a_c+b_c^2}-\frac{b^2-a_c}{(a_c+b_c^2)^2}(\epsilon_1 + 2b_c\epsilon_2)\bigg]A_u+ (\epsilon_1 + 2b_c\epsilon_2) A_v -\xi_1\epsilon_2k_c^2 A_v - D_1\bigg (q_x+\frac{q_y^2}{2k_c}\bigg)^2 A_u - \xi_1b \bigg (q_x+\frac{q_y^2}{2k_c}\bigg)^2 A_v, \label{amp-u1}\\
    \frac{\partial A_v}{\partial t}&=& -(\epsilon_1 + 2b_c\epsilon_2)A_v - \frac{4b_c\epsilon_2}{a_c+b_c^2}A_u + \frac{2b_c^2}{(a_c+b_c^2)^2}{\epsilon_1+2b_c\epsilon_2}A_u - D_2 \bigg (q_x+\frac{q_y^2}{2k_c}\bigg)^2A_u - \frac{\xi_2b_c}{a_c +b_c^2}\bigg (q_x+\frac{q_y^2}{2k_c}\bigg)^2 A_u,\label{amp-v1}
\end{eqnarray}

\end{widetext}
 where we have eliminated the terms that survive for  $\epsilon_1=0=\epsilon_2$ (i.e., at the phase boundary) by using the conditions valid on the pattern phase boundary, {\it viz.}

\begin{eqnarray}
    \frac{b_c^2-a_c}{a_c+b_c^2}A_u + (a_c+b_c^2)A_v - D_1 k_c^2 A_u - \xi_1 b_c k_c^2 A_v =0,\\
    -(a_c+b_c^2)A_v - \frac{2b_c^2}{a_c+b_c^2}A_u -D_2 k_c^2 A_v - \frac{\xi_2 b_c}{a_c+b_c^2}k_c^2 A_u=0.
\end{eqnarray}
 Equations~(\ref{amp-u1}) and (\ref{amp-v1}) are the linear amplitude equations for (\ref{modlin1}) and (\ref{modlin2}) to the lowest order in smallness, where we have neglected terms which are ${\cal O}(\epsilon_a (\Delta q)^2),\,a=1,2$, as both $\epsilon_a$ and $(\Delta q)^2$ are assumed to be small. Very close to the phase boundary but in the pattern forming region, $\epsilon_1\approx 0,\epsilon_2\approx 0$. In that limit, (\ref{amp-u1}) and (\ref{amp-v1}) reduce to
 \begin{eqnarray}
   \frac{\partial A_u}{\partial t}&=&   - D_1\bigg (q_x+\frac{q_y^2}{2k_c}\bigg)^2 A_u - \xi_1b \bigg (q_x+\frac{q_y^2}{2k_c}\bigg)^2 A_v, \label{amp-u}\\
   \frac{\partial A_v}{\partial t}&=& - D_2 \bigg (q_x+\frac{q_y^2}{2k_c}\bigg)^2A_u - \frac{\xi_2b_c}{a_c +b_c^2}\bigg (q_x+\frac{q_y^2}{2k_c}\bigg)^2 A_u.\label{amp-v}
 \end{eqnarray}
We now use (\ref{amp-u}) and (\ref{amp-v}) to examine the linear stability of $A_u, A_v$. Given that $A_u=0=A_v$ is the only fixed point of (\ref{amp-u}) and (\ref{amp-v}), if this point is linearly stable (unstable), it should imply that a striped pattern is also linearly stable (unstable). Assuming a time dependence of the form $A_u,A_v\sim \exp (\Lambda_\text{amt} t)$, we find
\begin{eqnarray}
    \Lambda_\text{amp}&=&\frac{1}{2}\bigg(q_x + \frac{q_y^2}{2k_c}\bigg)^2\bigg[-(D_1+D_2)\nonumber\\&&\pm\bigg\{(D_1-D_2)^2+\frac{4\xi_1\xi_2b_c^2}{a_c+b_c^2}\bigg\}^{1/2}\bigg]\nonumber\\
    &=&\frac{1}{2}\bigg(q_x + \frac{q_y^2}{2k_c}\bigg)^2\bigg[-(D_1+D_2)\pm\bigg\{(D_1+D_2)^2\nonumber \\&-&4D_1D_2 +\frac{4\xi_1\xi_2b_c^2}{a_c+b_c^2}\bigg\}^{1/2}\bigg].\label{band-stability1}
\end{eqnarray}
 We thus see that if 
 \begin{equation}
     \frac{\xi_1\xi_2b_c^2}{a_c+b_c^2}>D_1 D_2 \label{band-stability}
 \end{equation}
 one of the solutions for $\Lambda_\text{amp}$ is positive, implying linear instability. However, when this condition for instability is satisfied, we also have $k_c^2<0$ [{\it cf.} Eq.~(\ref{qc} above], making the pattern itself {\it nonexistent}. For $\xi_1\xi_2<0$, $k_c^2>0$, and both the solutions for $\Lambda_\text{amp}$ are negative, giving linear stability. In other words, at least in the linear theory, if a striped pattern is created with a real $k_c$, it remains linearly stable. Therefore, in the linear stability analysis, whenever a pattern can exist for a particular range of the relevant model parameters, a linearly stable striped pattern can be formed for that particular range of model parameters. With $k_c$ as the preferred or selected wavevector of the stripes, the corresponding wavelength is $\lambda_c=2\pi/k_c$. This gives the number of stripes to be $n_\text{stripe}=L/\lambda_c$, where $L$ is the linear size of the system, assuming a square geometry. Since $k_c$ (and hence $\lambda_c$) is a function of $D_1,D_2,\xi_1,\xi_2$, $n_\text{stripe} $ can be controlled by tuning these model parameters.   While the linear stability of the stripes supports the observed striped patterns in the DNS studies of the full model equations [see Sec .~\ref {dns} above],  there is a crucial difference. Our DNS studies show that by varying the chemotaxis parameters $\xi_1,\xi_2$ and also diffusivity $D_1$, one can achieve interconversions of stripes and spots. Lastly, by using (\ref{band-stability1}), we can conclude that with nonreciprocal chemotaxis,

 (i) $\xi_1\xi_2<0$ and $|\xi_1\xi_2|b_c^2<a_c+b_c^2)(D_1-D_2)^2$, both the roots are real and negative, giving damping of fluctuations {\it without} propagation,

 (ii) $\xi_1\xi_2<0$ and $\xi_1\xi_2|b_c^2>a_c+b_c^2)(D_1-D_2)^2$, both roots are real and negative, giving damping of fluctuations {\it with} propagation.

\begin{widetext}

 \section{Summary of the main results}\label{table-results}

We now give a short summary of the main results on how variation in the chemotaxis parameters $\xi_1,\xi_2$ affects the patterns as obtained from linear stability analysis and DNS studies in a tabular form in Table~\ref{table1} below.


\begin{table}[h!]
\centering
\renewcommand{\arraystretch}{1.35}
\begin{tabular}{|c|c|c|}
\hline

\makecell{\textbf{Parameters}} 
& \makecell{\textbf{Linear stability analysis}} 
& \makecell{\textbf{Direct numerical simulations (DNS)}} \\
\hline

\makecell{$\xi_1>0,\xi_2=0$} 
& \makecell{$k_c^2$ increases monotonically \\ with $\xi_1$} 
& \makecell{$\Delta_\phi$ and $\Delta_\psi$ or  pattern amplitudes increase \\ monotonically with $\xi_1$} \\
\hline

\makecell{$\xi_1<0,\xi_2=0$} 
& \makecell{$k_c^2$ decreases monotonically \\ with $|\xi_1|$} 
& \makecell{$\Delta_\phi$ and $\Delta_\psi$ or pattern amplitudes decrease \\ monotonically with $|\xi_1|$\\ As $|\xi_1|$ increases 
patterns change from \\spots to stripes before finally disappearing} \\
\hline

\makecell{$\xi_1=0,\xi_2>0$} 
& \makecell{$k_c^2$ decreases monotonically \\ with $\xi_2$} 
& \makecell{$\Delta_\phi$ and $\Delta_\psi$ or pattern amplitudes decrease \\ monotonically with $\xi_2$\\ As $\xi_2$ increases 
patterns change from \\spots to stripes before finally disappearing} \\
\hline

\makecell{$\xi_1=0,\xi_2<0$} 
& \makecell{$k_c^2$ increases monotonically \\ with $|\xi_2|$} 
& \makecell{$\Delta_\phi$ decreases but $\Delta_\psi$ increases\\ monotonically with $|\xi_2|$} \\
\hline

\makecell{$\xi_1=\xi=-\xi_2>0$ \\ (extreme nonreciprocal chemotaxis)} 
& \makecell{$k_c^2$ depends nonmonotonically on $\xi$: \\ increases then decreases, \\ eventually vanishing} 
& \makecell{$\Delta_\phi$ and $\Delta_\psi$ or pattern amplitudes \\first increase then decrease with $\xi>0$, \\ eventually vanishing. As $\xi$ increases 
patterns change\\ from spots to stripes before finally vanishing} \\
\hline

\makecell{$\xi_1=\xi=-\xi_2<0$ \\ (extreme nonreciprocal chemotaxis)} 
& \makecell{$k_c^2$ decreases monotonically, \\ eventually vanishing as $|\xi|$ increases} 
& \makecell{$\Delta_\phi$ and $\Delta_\psi$ or pattern amplitudes decrease monotonically, \\ eventually vanishing as $|\xi|$ increases.\\ As $|\xi|$ increases 
patterns change from \\spots to stripes before finally vanishing.} \\
\hline

\makecell{$\xi_1=\xi=\xi_2>0$ with $b^2>a$ \\ (fully reciprocal chemotaxis)} 
& \makecell{$k_c^2$ increases monotonically, \\ diverging for finite $\xi$ \\ (chemotaxis-induced instability)} 
& \makecell{$\Delta_\phi$ and $\Delta_\psi$ or pattern amplitudes \\increase monotonically,  diverging for finite $\xi$ \\ (chemotaxis-induced instability)\\ As $\xi$ increases 
patterns change from \\spots to stripes before finally the instability setting in} \\
\hline

\makecell{$\xi_1=\xi=\xi_2<0$ with $b^2>a$ \\ (fully reciprocal chemotaxis)} 
& \makecell{$k_c^2$ nonmonotonic: decreases then increases, \\ diverging at finite $|\xi|$ \\ (chemotaxis-induced instability)} 
& \makecell{$\Delta_\phi$ and $\Delta_\psi$ or pattern amplitudes \\decrease then increase,  diverging at finite $|\xi|$ \\ (chemotaxis-induced instability)} \\
\hline

\end{tabular}
\caption{Changes in the nature of the patterns as the chemotaxis parameters are varied. Results from the linear stability analysis and direct numerical simulation (DNS) studies are listed.}
\label{table1}
\end{table}


\end{widetext}
 
 \section{Conclusion and outlook}\label{summ}

We have thus explored how mutual chemotaxis affects the linear instabilities and pattern formations in two-component reaction diffusion models.  We have studied the specific example of an extended version of the diffusive Selkov model, where the two species, in addition to nonconserving onsite reactions, also mutually interact via the mechanism of chemotaxis. The latter is a number conserving process, parametrized by $\xi_1,\xi_2$, which gives the strength of the chemotaxis interactions in the dynamics of the two species, and can be individually positive or negative. Thus, chemotaxis can be mutually attractive ($\xi_1>0,\xi_2>0$) or repulsive ($\xi_1<0,\xi_2<0$), which are collectively called reciprocal chemotaxis with $\xi_1\xi_2>0$. It can also be ``attraction-repulsion'' type, for which one has either $\xi_1>0,\xi_2<0$ or $\xi_1<0,\xi_2>0$, which are collectively called nonreciprocal chemotaxis with $\xi_1\xi_2<0$.  

We study the model by linear stability analysis, complemented by direct numerical simulations (DNS) of the dynamical equations, which are partial differential equations. Our linear stability analysis reveals that
the extended Selkov model has an oscillatory but spatially uniform instability (a Hopf bifurcation), which unsurprisingly is identical to the same in the original Selkov model for glycolysis.  It also admits a saddle-node (Turing) instability at a finite wavevector, which is the preferred wavevector $k_c$, but at zero frequency. We show that in the reciprocal case, there is a novel chemotaxis-induced instability for sufficiently strong chemotaxis parameters. Linear stability analysis further reveals the complex dependence of $k_c^2$ on $\xi_1,\xi_2$. For reasons of simplicity in the explicit calculations, we calculate $k_c^2$ in the fully reciprocal case with $\xi_1=\xi_2=\xi$ and extreme nonreciprocal case with $\xi_1=-\xi_2=\xi$. In each of the cases, $\xi$ can be positive or negative. We find that within the linear stability analysis, $k_c^2$ displays a striking nonmonotonic dependence on $\xi$ for fixed diffusivities, the details of which again depend on whether the chemotaxis is reciprocal or nonreciprocal in nature, and also on $a,b$, which parameterize the on site reactions between the two species.  For example, in the fully reciprocal chemotaxis case with $b^2>a$ and $\xi<0$ or with $b^2<a$ and $\xi>0$, $k_c^2$ shows a nonmonotonic behavior with $\xi$, where $k_c^2$ first decreases with $\xi$ and then increases, eventually diverging at a finite threshold on $\xi$, which is the chemotaxis induced instability mentioned earlier. This means that in a sufficiently large system with reciprocal chemotaxis, it is possible to have patterns for very low and also high values of $\xi$ with an intervening uniform phase, giving a novel re-entrant transition. On the other hand, for $b^2>a$ and $\xi>0$ or $b^2<a$ and $\xi<0$, $k_c^2$ increases {\it monotonically} with $\xi$, ultimately diverging at a finite threshold on $\xi$. The dependence of $k_c^2$ on $\xi$ in the extreme nonreciprocal case is just the opposite. With  $\xi>0$, $k_c^2$ first increases with $\xi$, but then begins to decrease after reaching a maximum, eventually disappearing for very large $\xi$. 
In contrast, with $\xi<0$ in the extreme nonreciprocal case, we find $k_c^2$ to decrease monotonically as $\xi$ increases. In this extreme nonreciprocal case, the behavior of $k_c^2$ as a function of $\xi$ is independent of the sign of $b^2-a$. Similar complex dependence of $k_c^2$ on the diffusivities is found by varying them for fixed chemotaxis parameters.

To complement the results from our linear stability analysis, we have solved Eqs.~(\ref{mod1}) and (\ref{mod2}) numerically by using the pseudo-spectral method with random initial conditions in square geometries. The numerical solutions of the fields in the steady states are presented as snapshots for different values of the model parameters. We first numerically study the nonchemotactic limit of Eqs.~(\ref{mod1}) and (\ref{mod2}),i.e., by setting $\xi_1=0=\xi_2$. In this limit, there are regions in the parameter space spanned by $a, b$, where the system either shows sustained oscillations but remains spatially uniform at any given time, or shows steady patterns, which are spots or stripes, or shows {\it both} sustained oscillations and spatial nonuniformity, i.e., a pattern. Next, since our focus is to study the role of chemotaxis on steady patterns, we have chosen a point $(a,b)$  in the parameter space that shows only steady patterns for various values of $\xi_1,\xi_2$, and no oscillations. Our DNS studies further show that patterns can be spots or stripes. 

The results from our DNS studies clearly corroborate the linear stability analysis predictions of a chemotaxis-induced instability in the reciprocal case and the nonmonotonic dependence of $k_c^2$ on $\xi$ in both fully reciprocal and extreme nonreciprocal cases. In particular, for the purposes of quantitative analysis of the DNS results on the patterns, we have defined two quantities $\Delta_\phi,\Delta_\psi$ for $\phi$ and $\psi$, which vanish in the uniform states, but is positive definite in the patterned states. Thus, $\Delta_\phi,\Delta_\psi$ may be used as the proxies for the pattern amplitudes (which also vanish in the uniform states). From our DNS studies, we calculate $\Delta_\phi,\Delta_\psi$ as functions of $\xi_1,\xi_2$, which show complex dependence including nonmonotonic behavior. However, there is a crucial difference. Unlike in the linear stability analysis, the dependence (monotonic or nonmonotonic) of the patterns on the chemotaxis parameters is reflected in  the pattern amplitudes. 
The nature of the dependence of $k_c^2$ on the chemotaxis parameters in the linear stability analysis is found to have a relationship to the dependence of the pattern amplitudes on the chemotaxis parameters in the DNS studies. Indeed, our linear stability analysis results and the DNS studies results together reveal an intriguing phenomenological correspondence between the two. Furthermore, our DNS studies detect complex dependence of the pattern amplitudes (monotonic or nonmonotonic) on the diffusivities for fixed chemotaxis parameters, which has a direct correspondence on the monotonic or nonmonotonic dependence of $k_c^2$ on the diffusivities for fixed chemotaxis parameters. This gives us confidence to assert the veracity of the general complex dependence of the patterns on the chemotaxis parameters and diffusivities. While we are able to phenomenologically establish a novel correspondence between $\Delta_\phi,\Delta_\psi$ in the DNS studies and $k_c^2$ calculated in the linear stability analysis as functions of $\xi_1,\xi_2$, a microscopic theory for this starting from the equations of motion for $\phi,\psi$ is still lacking, and should be an important future direction of research.

Our finding of striped patterns in our DNS studies is surprising, as it violates rotational invariance of the equations of motion (\ref{mod1}) and (\ref{mod2}) at the macroscopic scales. However, the existence of stable striped patterns is rationalized by studying the linear stability of striped patterns, which gives the stability conditions of striped patterns that coincide with the linear stability condition for the existence of a pattern itself. Interestingly, our DNS studies also reveal intriguing transitions between spots and stripes as the chemotaxis parameter $\xi$ is varied. We argue that nonlinear chemotaxis effects, which are not captured by linear stability analysis, should be responsible for this. Our DNS studies also show that the nature of the ensuing pattern can also depend on the type of initial conditions chosen.  We have also investigated the role of geometry by considering rectangular geometries in our DNS studies, which suggest the importance of the number of stripes as a characteristic of the pattern. A full understanding of this requires additional studies.
We have numerically explored the role of  noise on the steady patterns. We have found that for low enough noise, the patterns remain stable, largely indistinguishable from their forms without any noise. However, as the noise strength increases, the patterns are eventually destroyed, giving uniform states. 


Finally, we have performed our DNS studies in the presence of annealed noises, which represent fluctuations due to contact with a fluctuating environment. Our DNS studies reveal that for low enough noises, the steady patterns are virtually identical to those formed in the absence of any noise, whereas stronger noises unsurprisingly destroys patterns. 

While our quantitative results are obtained from one specific model, the qualitative features of our results, {\it viz.}, the nonmonotonic dependence of the pattern on the chemotaxis parameters, should hold in generic nonconserving reaction diffusion models with chemotaxis. Furthermore, our studies can be generalized in straightforward ways to account for a separate chemoattractant or chemorepellant usually present in biological chemotaxis. We expect the qualitative features of our results to remain valid even when such an explicit chemical is included in the model, so long as the production and degradation rates of the chemical dominate over its diffusion. These expected generalities of our results should allow us to experimentally test these results in a variety of systems with artificial chemotaxis; see, e.g., Ref.~\cite{hartmunt}.

We have illustrated a novel chemotaxis induced instability in the reciprocal chemotaxis case within the linear stability analysis. Our DNS studies capture a clear signature of it. However, from either our linear stability analysis or DNS studies, we are unable to comment on the eventual steady states that the systems ultimately reach through this chemotaxis induced instability, as even our DNS studies do not appear to reach a steady state. It is possible that additional linear stabilizing terms, e.g., hyperdiffusion terms are required for systematic studies of these instabilities. The resulting steady states may also be chaotic~\cite{chaos_pattern}. More work is needed to determine the nature of the resulting steady states. We hope that our present studies will provide impetus for future research in this direction.

In the present work, we have investigated instabilities and pattern formation within a continuum PDE framework. An attractive alternative approach is provided by agent-based models, which can capture pattern formation from the dynamics of individual constituents and may offer complementary insights into the underlying mechanisms~\cite{agent}. Exploring such models constitutes a promising direction for future research. Another interesting question to examine is the nature of the transitions between homogeneous and patterned states, and to understand how these transitions are influenced by the chemotactic parameters.

Our DNS studies reveal the existence of a region in the phase space spanned by $a,b$, where the system can display both sustained spatial and temporal modulations, i.e., patterns superposed with oscillators. Detailed studies should be able to uncover the nature of the interplay of the patterns and oscillations, and how that depends on the chemotaxis parameters.

Our results can be verified by considering various two-component chemotaxis systems, in vivo or in vitro, characterized by different values of $\xi_1$ and $ \xi_2$. The pattern can also be affected by the mean densities of the two species, which serve as control parameters and, in effect, are proxies for the parameters $ a$ and $ b$.

The study of patterns in live cell membranes is a biologically relevant question. Recent hydrodynamic theory for active membrane fluctuations reveals that the interplay of active stresses and active permeation flows can lead to pattern formation~\cite{mukherjee2025flat, debayan1}. It will be interesting to study chemotaxis on such fluctuating membranes.

From the perspective of nonequilibrium statistical mechanics, our study highlights the nontrivial consequences of the interplay between nonconserving onsite reactions and conserving chemotactic dynamics in determining the long-time states of the system. The observed dependence of the pattern amplitudes on the chemotactic parameters has motivated us to propose phenomenological expressions for these amplitudes as functions of the chemotactic couplings at fixed diffusivities. Analogous phenomenological forms can also be formulated when the diffusivities, or their ratio, are varied, while keeping the chemotaxis parameters fixed. A systematic derivation of these proposed forms directly from the underlying equations of motion remains an important theoretical challenge.

Recent years have witnessed significant progress in understanding pattern formation in mass-conserving reaction–diffusion systems or nonreciprocal Cahn-Hilliard systems, leading to the development of generic theoretical frameworks for analyzing patterns in the presence of conservation laws; see, e.g., \cite{frey_natphys, ramin3, marchetti, ramin4, erwin_2026_1, erwin_2026_2, erwin_2026_3,thiele}. An interesting direction for future work is to generalize these frameworks to pattern-forming systems that lack conservation laws, such as the one considered here. We hope our study will provide to future experimental studies on relevant biological and artificial in vitro realizations of this system.



\section{Acknowledgment}
M.K. and A.B thank S. Chatterjee for useful comments and suggestions.
A.B. thanks Alexander von Humboldt Stiftung (Germany) for partial financial support through their research group linkage programme (2024) and ANRF (India) for partial financial support through the ARG (MATRICS) programme (file no.: ANRF/ARGM/2025/000461/TS).

\section{Data availability}
The computational data and simulation outputs generated in this work are available from the corresponding author upon reasonable request. The source code used for the analysis will be deposited on GitHub at [\href{https://github.com/mkarmakar094}{github.com/mkarmakar094}] upon publication.

\appendix
\section{Details of the numerical techniques}
\label{sec:numerical_methods}

To numerically integrate the coupled nonlinear partial differential equations governing the chemotactic reaction–diffusion system, we employ a pseudo-spectral scheme with periodic boundary conditions in both spatial directions. The model fields, $\phi(\mathbf{r},t)$ and $\psi(\mathbf{r},t)$, represent the  concentrations of the two species, and evolve according to
\begin{align}
\partial_t \phi &= -\phi + a \psi + \phi^2 \psi + D_1\nabla^2 \phi 
   + \xi_1 \nabla \!\cdot\! (\phi \nabla \psi), \label{eq:phi_eq} \\
\partial_t \psi &= b - a \psi - \phi^2 \psi + D_2 \nabla^2 \psi
   + \xi_2 \nabla \!\cdot\! (\psi \nabla \phi), \label{eq:psi_eq}
\end{align}
where $a$ and $b$ are kinetic parameters of the local reaction term, $D$ denotes the diffusivity ratio of $\psi$ to $\phi$, and $\xi_{1,2}$ quantify the strength and polarity of chemotactic coupling between the two fields.  
To study the emergence of spatiotemporal patterns, we expand both fields around the spatially uniform steady state $(\phi_0,\psi_0)$ satisfying
\begin{align}
 -\phi_0 + a \psi_0 + \phi_0^2 \psi_0 &= 0, &
 b - a \psi_0 - \phi_0^2 \psi_0 &= 0.
\end{align}
For numerical convenience, we introduce small fluctuations $u$ and $v$ defined by $\phi = \phi_0 + u$ and $\psi = \psi_0 + v$.  The mean values of these fluctuations are constrained to vanish at all times, ensuring the total density conservation in Fourier space, i.e., $\hat{u}(\mathbf{k}=0) = \hat{v}(\mathbf{k}=0) = 0$. We now linearize Eqs.~(\ref{eq:phi_eq}) and (\ref{eq:psi_eq}) around the fixed points (\ref{fps}). We find
\begin{eqnarray}
 \frac{\partial u}{\partial t} &=& \frac{b^2-a}{b^2+a} u + av + b^2 v + D_1\nabla^2 u+\xi_1 b\nabla^2 v,\label{modlin1} \\
 \frac{\partial v}{\partial t} &=& -(b^2+a)v - \frac{2ub}{b^2+a} + D_2\nabla^2 v\nonumber \\&&+\xi_2\frac{b}{a+b^2}\nabla^2 v,\label{modlin2}
\end{eqnarray}
where $u=\phi-\phi^*,\,v=\psi - \psi^*$. The obtained nonlinear PDEs (\ref{modlin1}) and (\ref{modlin2}) are used for the pseudo-spectral integration scheme.
Spatial derivatives are computed in Fourier space using the Fast Fourier Transform (FFT) library \texttt{FFTW3}, which allows us to represent the Laplacian and divergence operators as multiplications by $-k^2$ and $i\mathbf{k}\!\cdot$, respectively.  
The nonlinear reaction and chemotactic terms are evaluated in real space, and the fields are transformed back and forth between real and Fourier domains at each time step.  
We adopt a semi-implicit time-stepping scheme: the linear diffusion terms are treated implicitly, while nonlinear reaction and chemotactic terms are advanced explicitly. This provides stability at small time steps without significant numerical damping. The update rule for the Fourier coefficients $\hat{u}(\mathbf{k},t)$ and $\hat{v}(\mathbf{k},t)$ is schematically written as
\begin{align}
\hat{u}(\mathbf{k},t+\Delta t) &= 
   \frac{\hat{u}(\mathbf{k},t) + \Delta t\,\hat{N}_u(\mathbf{k},t)}
        {1 + \Delta t\,k^2}, \\
\hat{v}(\mathbf{k},t+\Delta t) &= 
   \frac{\hat{v}(\mathbf{k},t) + \Delta t\,\hat{N}_v(\mathbf{k},t)}
        {1 + \Delta t\,D k^2},
\end{align}
where $\hat{N}_{u,v}$ denotes the Fourier transforms of the nonlinear reaction and chemotactic contributions computed in real space.
Periodic boundary conditions are imposed in both directions of a square domain of size $L_x = L_y = 2\pi$, discretized uniformly on an $N_x \times N_y$ grid.  
The initial conditions consist of small random perturbations around the uniform state:
\[
u(x,y,0),\,v(x,y,0) \in [-\varepsilon,\varepsilon],
\]
with $\varepsilon \ll 1$.  The mean of these perturbations is subtracted at each initialization step to ensure zero spatial average.  
All numerical routines are implemented in \texttt{C} using contiguous memory allocation for two-dimensional arrays to ensure cache efficiency.  
The Fourier transforms and precomputed FFTW plans handle their inverse, and all nonlinear quantities are computed in real space.  
The simulation time step $\Delta t$ is typically chosen as $10^{-5}$ to ensure numerical stability and convergence. The evolution of spatial averages $\langle \phi \rangle$ and $\langle \psi \rangle$ is monitored to verify that the mean-field conditions remain stationary.  
All simulations are performed under fully periodic conditions, ensuring that pattern selection arises purely from intrinsic nonlinear dynamics and chemotactic coupling, without boundary effects.
This approach yields stable and accurate numerical integration for long time scales, enabling systematic exploration of the interplay between diffusion, reaction kinetics, and chemotactic coupling.

\section{Square geometry versus rectangular geometry}\label{geom}

\begin{figure}[!ht]
    \centering
    \includegraphics[width=\columnwidth]{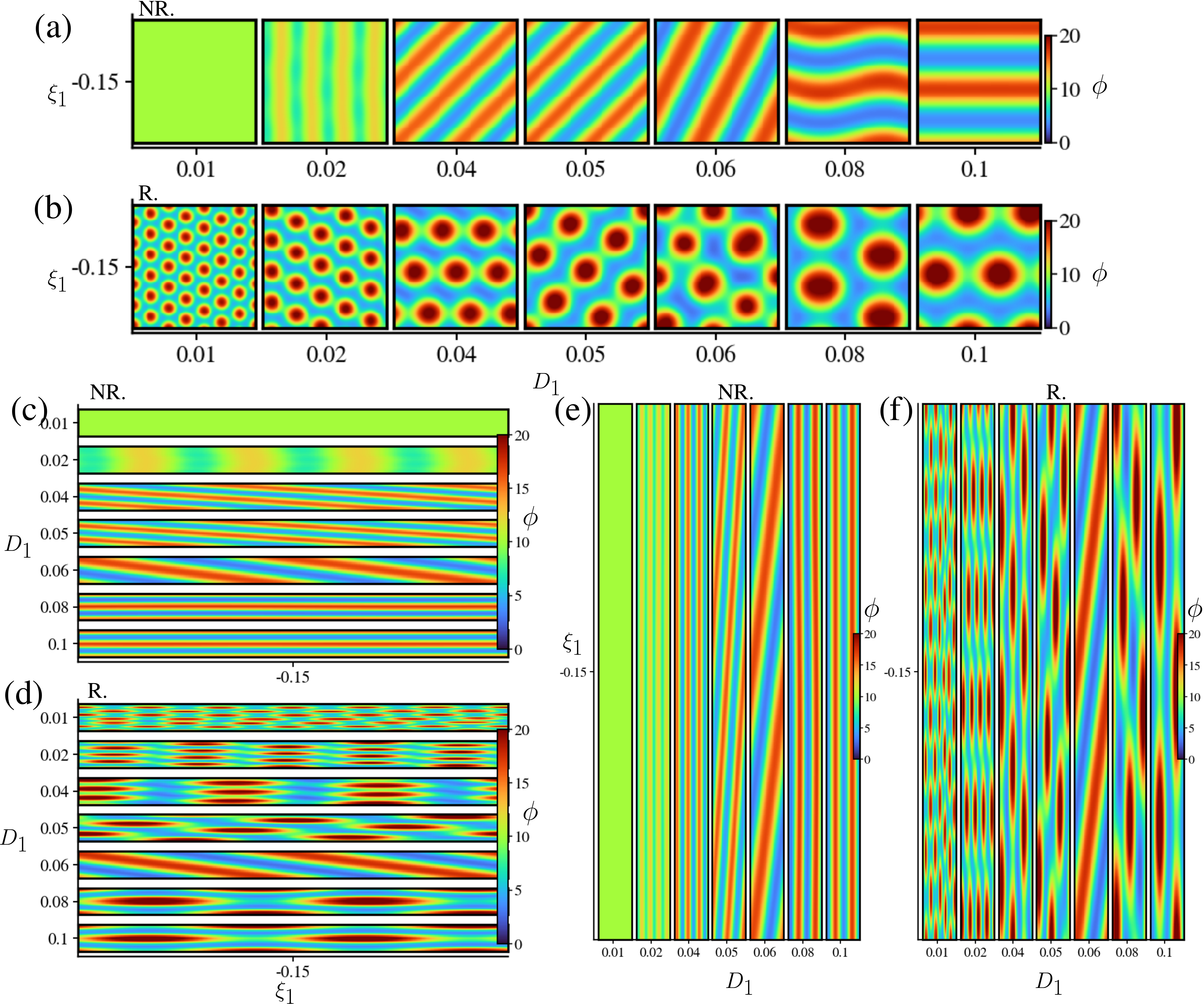}
    \caption{(Color online) Geometry-controlled pattern selection for fixed negative chemotactic coupling as the diffusion coefficient $D_1$ is varied.
    Panels (a,b) show the square system, $L_x=L_y=128$, for nonreciprocal (NR) and reciprocal (R) couplings, respectively, at fixed $\xi_1=-0.15$.
    Panels (c,d) show the corresponding space--parameter maps for the elongated geometry $L_x=256$, $L_y=32$, while panels (e,f) show the transposed geometry $L_x=32$, $L_y=256$. 
    }
    \label{fig:fig_R_NR_LxLy}
\end{figure}

We examine the role of the geometry by considering rectangular boxes of size $L_x\times L_y$ with $L_x=32$, $L_y=256$ together with the complementary case with $L_x=256$, $L_y=32$, and compare with the corresponding square geometry case having $L_x=L_y=128$. Quite intriguingly, we find that in case of striped patterns, the number of stripes is independent of the geometry: it is same in the rectangular geometries as in the square geometry. While our linear stability analysis suggests that the number of stripes $n$ should be $n\sim (L_x, L_y)/\lambda$, where $\lambda=2\pi/k_c$ with $k_c$ can be along any of the directions, our DNS studies indicate that $n$ is the more fundamental quantity, which remains geometry independent; see Fig.~\ref{fig:fig_R_NR_LxLy}. Thus a full nonlinear theory of the patterns should be in terms of $n$. Similar comments can be made about spots, where the number of spots plays an important role and appears to be independent of the geometry. Developing a nonlinear theory to account for this should be an important direction for future research.

\section{Initial condition dependence}\label{ini_cond}

\begin{figure}[!ht]
    \centering
    \includegraphics[width=\columnwidth]{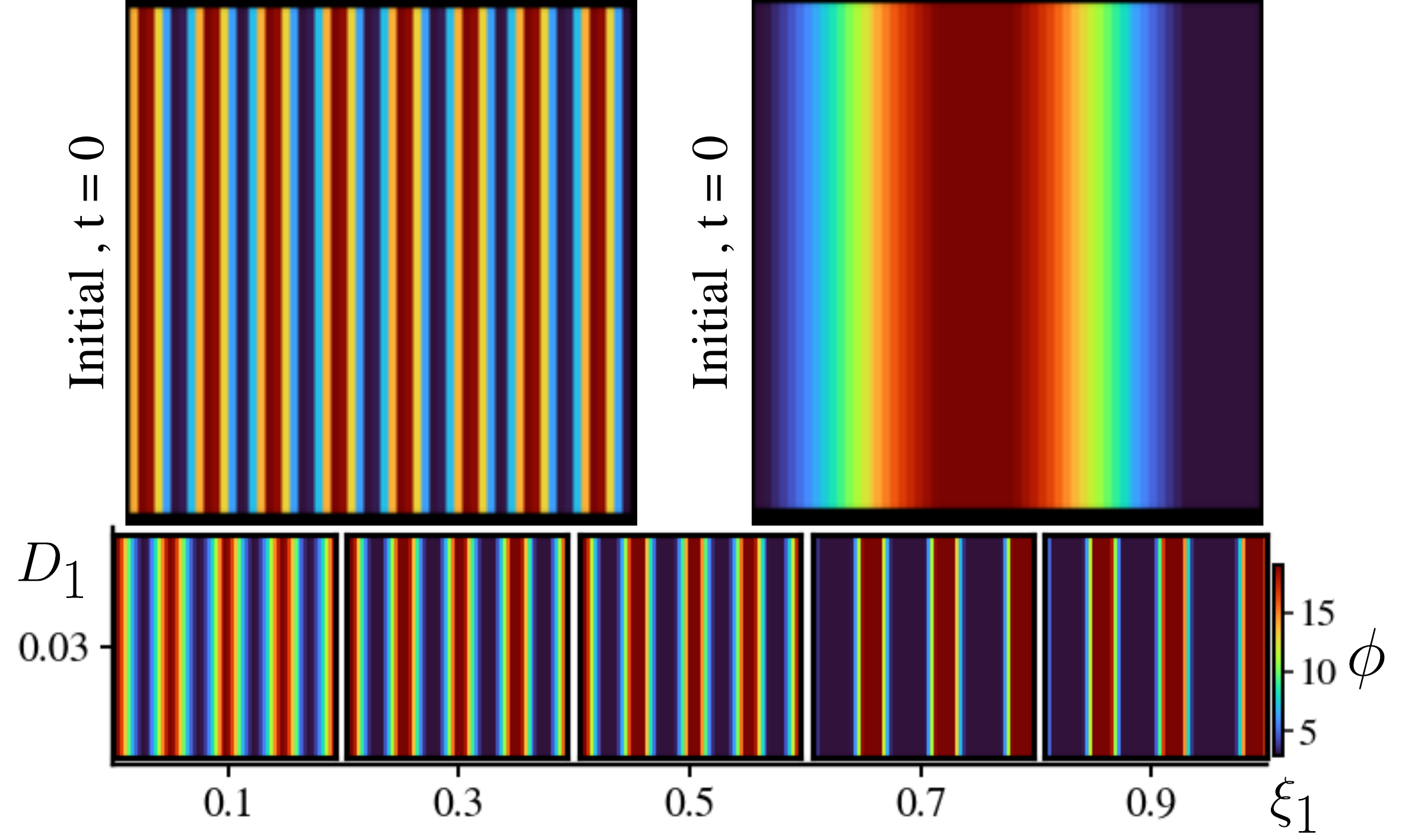}
    \caption{(Color online)
     Memory of the initial stripe number in the nonreciprocal positive-coupling regime. The two upper panels show stripe initial conditions for the density field $\phi(x,y)$ at $t=0$, prepared with different initial wavelengths.
    The lower row shows the corresponding steady-state profiles for increasing positive nonreciprocal coupling $\xi_1>0$, with $\xi_2=-\xi_1$, at fixed $D_1=0.03$.
    Despite starting from different stripe numbers, the final states converge to stripe patterns whose wavelength is selected primarily by the nonreciprocal coupling rather than by the imposed initial modulation.
    This demonstrates that, in the positive nonreciprocal regime, stripe formation is controlled by nonlinear mode selection and is largely insensitive to the initial stripe number.
    Parameters: $a=0.015$ and $b=10$.
    }
    \label{fig:fig_stripe_D}
\end{figure}

In this Section, we numerically explore the effects of non-random initial conditions on the eventual steady states. In particular, we focus on stripes as the initial condition and focus on their time evolution. We consider a square geometry and consider nonreciprocal coupling with $\xi_1=\xi=-\xi_2>0$. To explore this in detail, we have allowed the initial stripe conditions to contain an arbitrary number of stripes. We find that independent of the initial number of stripes, the system always achieves a unique value of $n_\text{stripe}$ in the steady states, controlled only by the various model parameters, but independent of the initial stripe numbers. Our results suggest that the time evolution of the stripe number $n$ should follow a phenomenological equation of the form
\begin{equation}
    \frac{\partial n}{\partial t} = - A_n (n- n_\text{stripe})^2,
\end{equation}
where $A_n>0$ is a phenomenological constant, allowing for $n=n_\text{stripe}$ as the stable steady state stripe number. See Fig.~\ref{fig:fig_stripe_D}.

\section{Nonlinear amplitude equations}\label{nonlin-ampli}
 A striking result from the present study is that by tuning the chemotaxis parameters $\xi_1,\xi_2$, it is possible to introduce a transition between patterned states with spots and stripes. This cannot be explained within the linear stability analysis, as we have seen above that the stability condition of a striped state is identical to that of a patterned state itself within the linear stability analysis. This then means stripes cannot be converted into spots by tuning $\xi_1,\xi_2$, which is contrary to the results from our DNS studies. We here argue heuristically by using perturbative reasoning that nonlinear effects should be responsible for such transitions. We study this in the present Section within an approximate, perturbative scheme.  For simplicity,  we retain only the nonlinear terms of chemotaxis origin and ignore all the other nonlinear terms. Our strategy is to calculate the perturbative corrections to the diffusivities $D_1, D_2$ and cross diffusivities $D_{\times 1}\equiv \xi_1 b, D_{\times 2}\equiv \xi_2 b/(a+b^2)$. 

 We start from (\ref{u-linear-amplitude}) and (\ref{v-linear-amplitude}), substitute these in the nonlinear equations  (\ref{mod1}) and (\ref{mod2}) and consider only the chemotaxis-generated nonlinearities. We then get
 \begin{eqnarray}
   && (-i\omega+D_1k^2) \tilde u_{{\bf k},\omega}+D_{\times 1}k^2 v_{{\bf k},\omega}=N_1({\bf k},\omega),\label{nonlinu1}\\
   && (-i\omega+D_2k^2) \tilde v_{{\bf k},\omega}+D_{\times 2}k^2 u_{{\bf k},\omega}=N_2({\bf k},\omega)\label{nonlinv1},
 \end{eqnarray}
where $\tilde u,\tilde v$ supplement the solutions obtained from the linearized dynamics, $N_1({\bf k},\omega)$ and $N_2({\bf k},\omega)$ are the spatio-temporal Fourier transforms of the chemotaxis-generated nonlinear terms in (\ref{mod1}) and (\ref{mod2}).
The solutions to (\ref{nonlinu1}) and (\ref{nonlinv1}) may be formally written as
\begin{equation}
    \left(\begin{array}{c}
    \tilde u_{{\bf k},\omega}\\
    \tilde v_{{\bf k},\omega}
    \end{array}\right) 
            = B((k,\omega)\left(\begin{array}{c}
                 N_1\\N_2
                \end{array}
\right),\label{nonlin-matrix}
\end{equation}
where 
\begin{equation}
    B(k,\omega)=\Delta_M(k,\omega)^{-1}\left(\begin{array}{cc}
    B_{11}(k,\omega) & B_{12}(k,\omega)\\
    B_{21}(k,\omega) & B_{22}(k,\omega)
    \end{array}
    \right),
\end{equation}
where
\begin{small}
\begin{eqnarray}
    &&\Delta_M(k,\omega)=(-i\omega+D_1k^2)(-i\omega + D_2 k^2) - D_{\times 1}D_{\times 2}K^4,\\
    && B_{11}(k,\omega)=\frac{-i\omega +D_2k^2}{\Delta_M},\\
    &&B_{12}(k,\omega)=\frac{-D_{\times 2}k^2}{\Delta_M},\\
    &&B_{21}(k,\omega)=\frac{-D_{\times 1}k^2}{\Delta_M},\\
    && B_{22}(k,\omega)=\frac{-i\omega + D_1k^2}{\Delta_M}.
\end{eqnarray}
\end{small}
We then obtain
\begin{small}
    \begin{eqnarray}
    &&\tilde  u({\bf k},\omega) = B_{11}(k,\omega) N_1 ({\bf k},\omega) + B_{12}(k,\omega) N_2 ({\bf k},\omega),\\
    &&\tilde v({\bf k},\omega) = B_{21}(k,\omega) N_1 ({\bf k},\omega) + B_{22}(k,\omega) N_2 ({\bf k},\omega).
\end{eqnarray}
\end{small}
The above equations are implicit equations for $\tilde u$ and $\tilde v$, since the latter appear on both lhs and rhs of the above equations. The full solutions for the fields $u$ and $v$ are then given by
\begin{eqnarray}
    &&u({\bf k},\omega)=u_L({\bf k},\omega) + \tilde u({\bf k},\omega),\label{u-full}\\
    &&v({\bf k},\omega)=v_L({\bf k},\omega) + \tilde v({\bf k},\omega),\label{v-full},
\end{eqnarray}
where $u_L,v_L$ refer to the Fourier transformed linear theory solutions in (\ref{u-linear-amplitude}) and (\ref{v-linear-amplitude}). Note that $N_1$ and $N_2$ are bilinear in the fields:  The explicit forms for $N_1,N_2$ are
\begin{eqnarray}
    &&N_1({\bf k},\omega)=-\xi_1\sum_{{\bf q},\Omega}{\bf k\cdot (k-q)}u_{{\bf q},\Omega}v_{{\bf k-q},\omega-\Omega},\\
    &&N_2({\bf k},\omega)=-\xi_2\sum_{{\bf q},\Omega}{\bf k\cdot (k-q)}u_{{\bf q},\Omega}v_{{\bf k-q},\omega-\Omega}.
\end{eqnarray}
Now expand $N_1,N_2$ in (\ref{u-full}) and (\ref{v-full}) above perturbatively and iteratively in $\xi_1,\xi_2$. in this spirit, $u({\bf k},\omega)=u_L({\bf k},\omega)$ and $ v({\bf k},\omega)=v_L({\bf k},\omega) $ at the zeroth order in $\xi_1,\xi_2$. Then in the iterative expansions of $N_1, N_2$, one obtains contributions trilinear in $u_L,v_L$, which in turn give contributions at the bilinear order in $\xi_1,\xi_2$ to the linear equations (\ref{mod1}) and (\ref{mod2}). These fluctuation-corrected linear equations may be used to calculate the effective amplitude equations for striped patterns, which generalize (\ref{amp-u1}) and (\ref{amp-v1}), producing effective diffusivities $D_1^\text{eff}, D_2^\text{eff} $ and cross diffusivities $D_{\times 1}^\text{eff}, D_{\times 2}^\text{eff}$, all of which are now nonlinear functions of $\xi_1,\xi_2$.  Using these ``renormalized'' diffusivities in the stability conditions~(\ref{band-stability}) for striped patterns, we can obtain the condition of stability of striped patterns, which should be quite different from the linear stability condition of a pattern itself. This points to the possibility of transitions between spots and stripes by tuning the chemotaxis parameters $\xi_1,\xi_2$.

\section{Effects of noises}\label{noises}

\begin{figure}[!ht]
 *   \centering
    \includegraphics[width=\columnwidth]{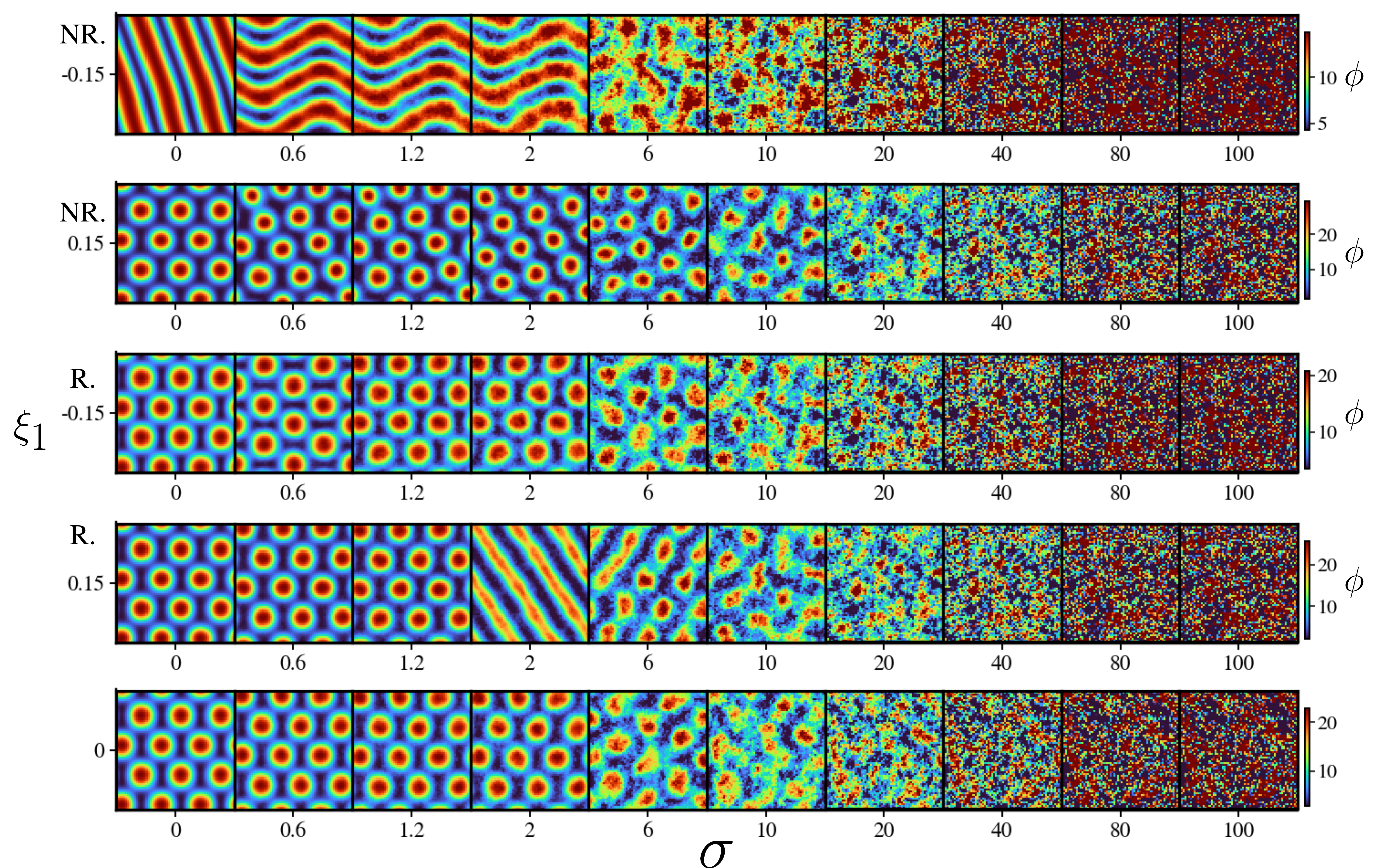}
    \caption{(Color online) { Snapshots of the density field $\phi$ for increasing noise strength $\sigma$
    at fixed chemotaxis parameters $\xi=0.15$ and $D_1=0.03$ for all four combinations. Bottom panel nonchemotactic result.}
    For small $\sigma$, the sharp pattern forms, while increasing noise leads to dissolution of spatially periodic patterns.
    The transition from a pattern state to disordered structures demonstrates the destabilizing role of noise in the chemotactic interaction dynamics. Parameters: $a = 0.015$ and $b = 10$.}
    \label{fig:fig_noise_fixedXi}
\end{figure}

\begin{figure}[!ht]
    \includegraphics[width=\columnwidth]{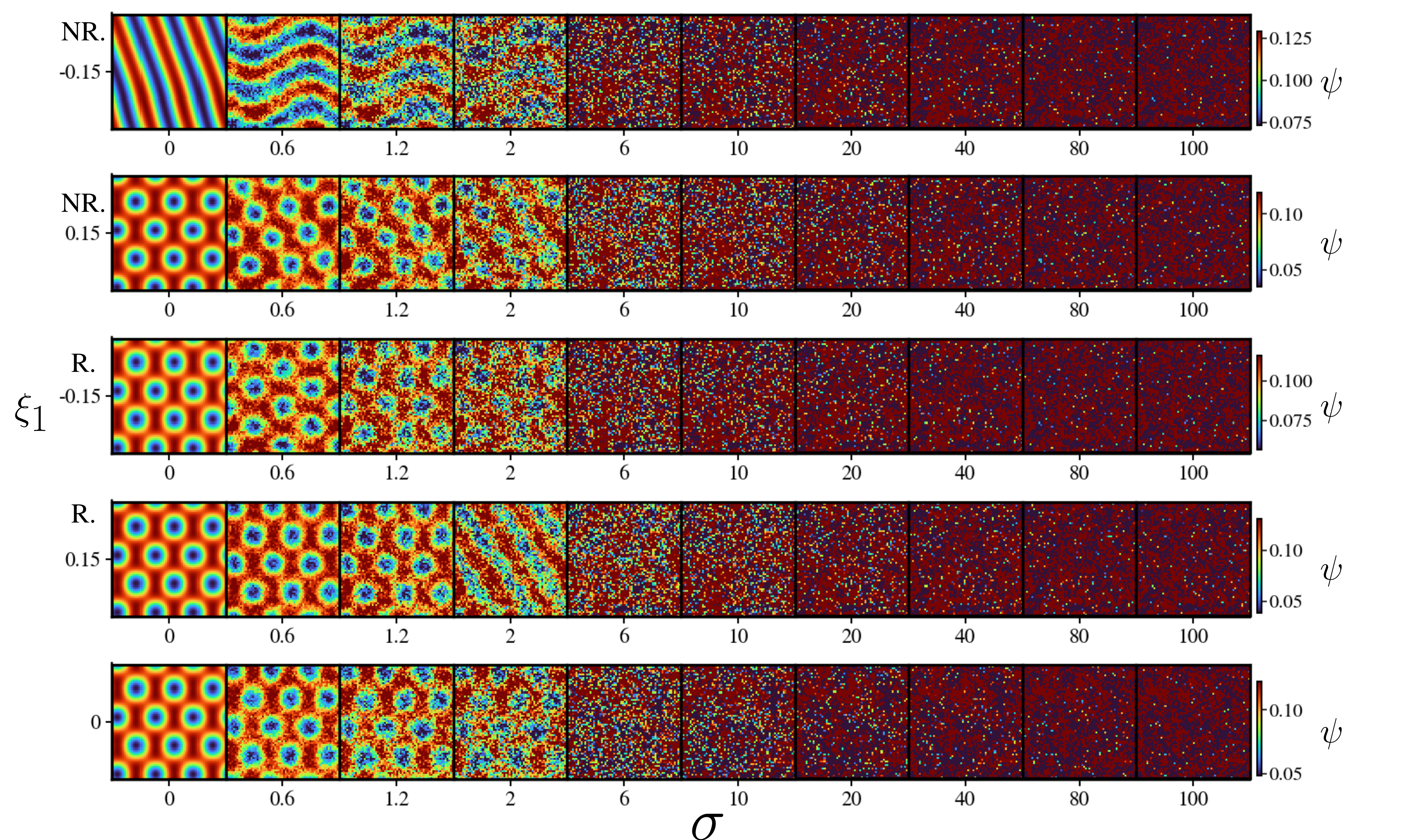}
    \caption{(Color online) { Snapshots of the density field $\psi$ for increasing noise strength $\sigma$
    at fixed chemotaxis parameters $\xi=0.15$ and $D_1=0.03$ for all four combinations. Bottom panel nonchemotactic result.}
    For small $\sigma$, the sharp pattern form, while increasing noise leads to dissolution of spatially periodic patterns.
    The transition from a pattern state to disordered structures demonstrates the destabilizing role of noise in the chemotactic interaction dynamics. Parameters: $a = 0.015$ and $b = 10$.}
    \label{fig:fig_noise_fixedXi_psi}
\end{figure}

Our linear stability analysis and the DNS results for patterns, as described above, ignore additive stochastic noise. However, real systems inevitably have noise in them, which are generalizations of the thermal noise in equilibrium systems. The effects of noise in pattern forming systems are a fascinating subject on their own. Some studies have indicated the possibility of transitions between patterns of different types induced by noise of specific distributions~\cite{noise1, noise2}. Here, we have confined ourselves to studying the robustness of the different patterns generated against additive noise in our DNS studies. To that end, we add the noises $\eta_1$ and $\eta_2$ in Eqs.~(\ref{mod1}) and (\ref{mod2}) and solve the resulting stochastic partial differential equations using pseudo-spectral methods. Noises $\eta_1$ and $\eta_2$ are assumed to be zero-mean and Gaussian-distributed with variances which are assumed to be equal. In equilibrium systems, these variances are proportional to the thermodynamic temperature $T$, and the proportionality factor is determined by the Fluctuation-Dissipation-Theorem (FDT)~\cite{chaikin}. In nonequilibrium systems as here, there is FDT and the noise variances just signify the strength of the stochastic noises. Our previous DNS results are the zero variance limit of the solutions of the noisy equations. As the variance is increased, we notice that the sharpness of the patterns formed at zero noise is decreased; eventually the patterns breaking up, giving rise to uniform states. Thus, for sufficiently high noises, steady patterns in the noiseless limit are no longer stable against perturbations by the noises. This is observed for both reciprocal and nonreciprocal cases; see Fig.~\ref{fig:fig_noise_fixedXi} and Fig.~\ref{fig:fig_noise_fixedXi_psi}. This is expected since higher or stronger noise usually leads to better mixing, thereby reducing segregation or patterns.

\section{Effects of diffusion on the steady patterns}\label{diff-eff}

We start from \eqref{qc} and look for its extremum with respect to variations in $D_1$, keeping all other parameters fixed. Setting $\partial k_c^2/\partial D_1=0$, we get
\begin{small}
\begin{eqnarray}
    D_1&=& \frac{1}{2d\left[-(a+b^2)+D\frac{b^2-a}{a+b^2}\right]}\nonumber \\&&\times\bigg[2Db\left( -\xi_2+\frac{2\xi_1b^2}{a+b^2}\right)\pm\bigg\{4D^2b^2\bigg(-\xi_2+\frac{2\xi_1b^2}{a+b^2}\bigg)^2 \nonumber \\&&+4D\frac{\xi_1\xi_2b^2}{a+b^2}\left[-(a+b^2)+D\frac{b^2-a}{a+b^2}\right]^2\bigg\}^{1/2}\bigg]\equiv D_{1\pm}.\label{d-eq}
\end{eqnarray}
\end{small}
The discriminant in \eqref{d-eq} is positive definite for $\xi_1\xi_2>0$, i.e., in the reciprocal chemotaxis case. Depending on the sign of $D(b^2-a)/(a+b^2)-(a+b^2)$, either $D_{1+}$ or $D_{1-}$ should be positive. Substituting that in \eqref{qc}, we get $k_c^2|_\text{max}$, the maximum of the selected wavevector. Thus, we get a nonmonotonic dependence on $D_1$. In the nonreciprocal case, $\xi_1\xi_2<0$. Thus, the discriminant may or may not be positive. Accordingly, $\partial k_c^2/\partial D_1=0$ may or may not have a solution.

\begin{figure*}[!ht]
    \centering
    \includegraphics[width=2\columnwidth]{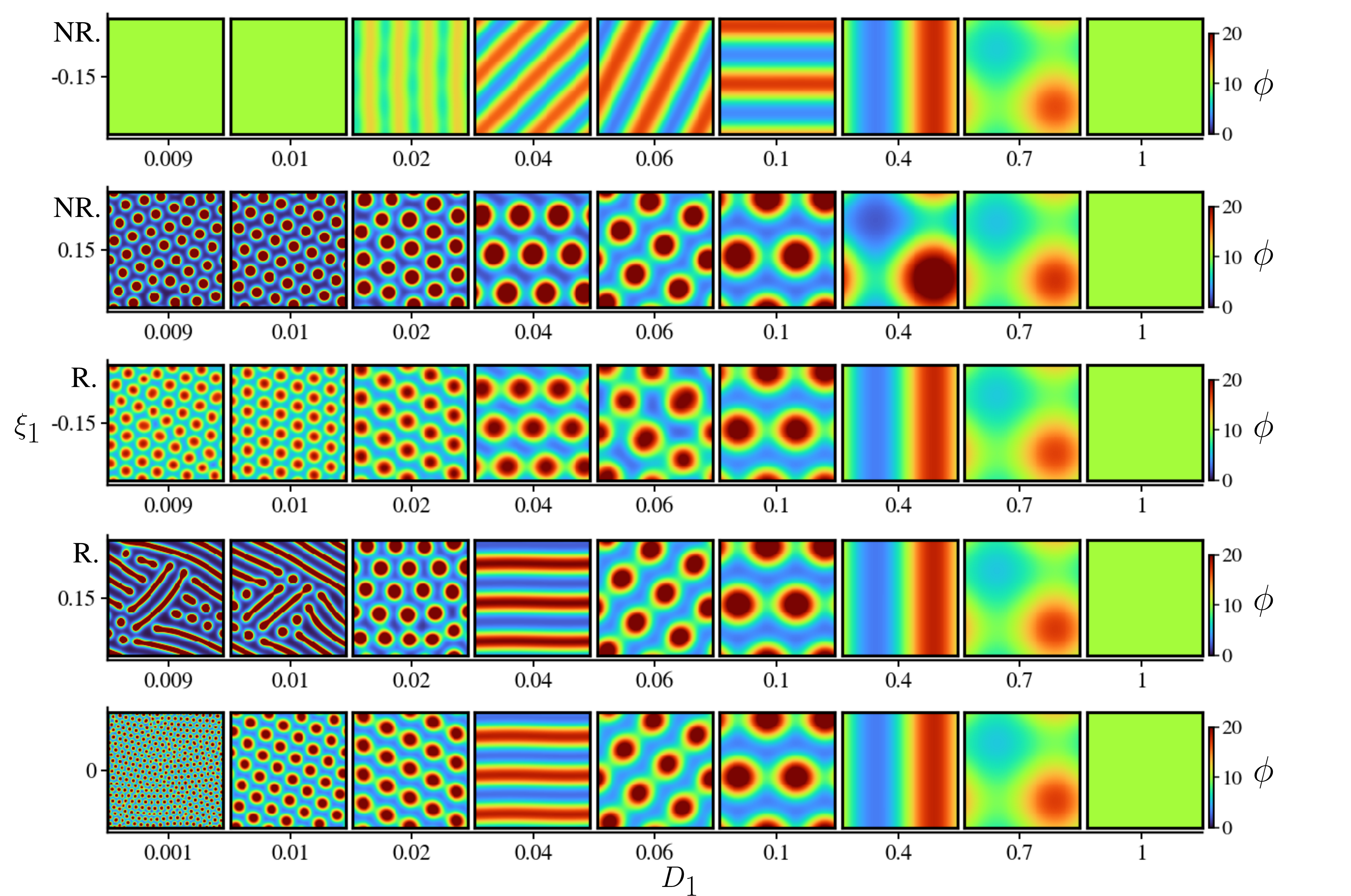}
    \caption{(Color online) { Snapshots of the density field $\phi$ for increasing diffusion coefficients $D_1$ and $D_2=1000\,D_1$, at fixed coupling parameters $\xi=0.15$ for all four combinations. Bottom panel nonchemotactic result.}
    For small $D_1$, the density remains nearly homogeneous, while increasing diffusion induces the spontaneous formation of spatially periodic patterns.
    The transition from a uniform state to disordered modulations and finally to regular stripe-like structures demonstrates the destabilizing role of cross diffusion in the chemotactic interaction dynamics. Parameters: $a = 0.015$ and $b = 10$. }
    \label{fig:fig_Diffusion_fixedXi}
\end{figure*}

To complement the above results from linear stability analysis, we use our DNS studies of (\ref{mod1}) and (\ref{mod2}) to explore the dependence of the patterns on $D_1$ for fixed $D$ and $\xi_1=\xi=\pm\xi_2$. We allow for $\xi$ to be positive or negative. Our results reveal complex dependence of the patterns formed on $D_1$  in the nonreciprocal and reciprocal cases. Results in the nonchemotactic limit ($\xi=0$) are also shown. Unsurprisingly for very large $D_1$, uniform states are found in all the cases. See Fig.~\ref{fig:fig_Diffusion_fixedXi}.

\begin{figure}
    \centering
    \includegraphics[width=0.9\linewidth]{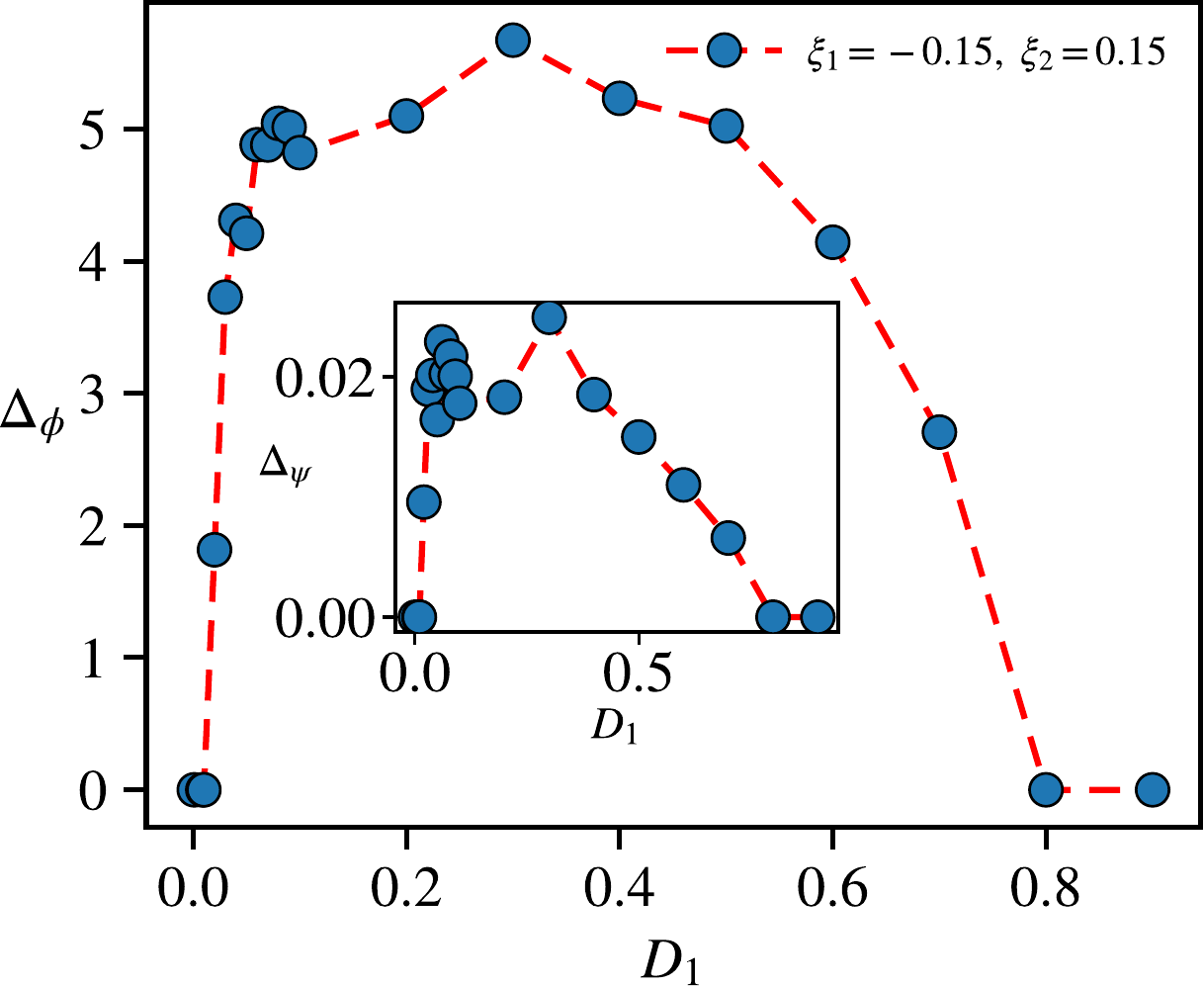}
    \caption{Variation of $\Delta_\phi$ (inset $\Delta_\psi$) as functions of $D_1$ for fixed $\xi_1=-0.15,\xi_2=0.15$. }
    \label{diffusion-nr-neg}
\end{figure}

\begin{figure}
    \centering
    \includegraphics[width=0.9\linewidth]{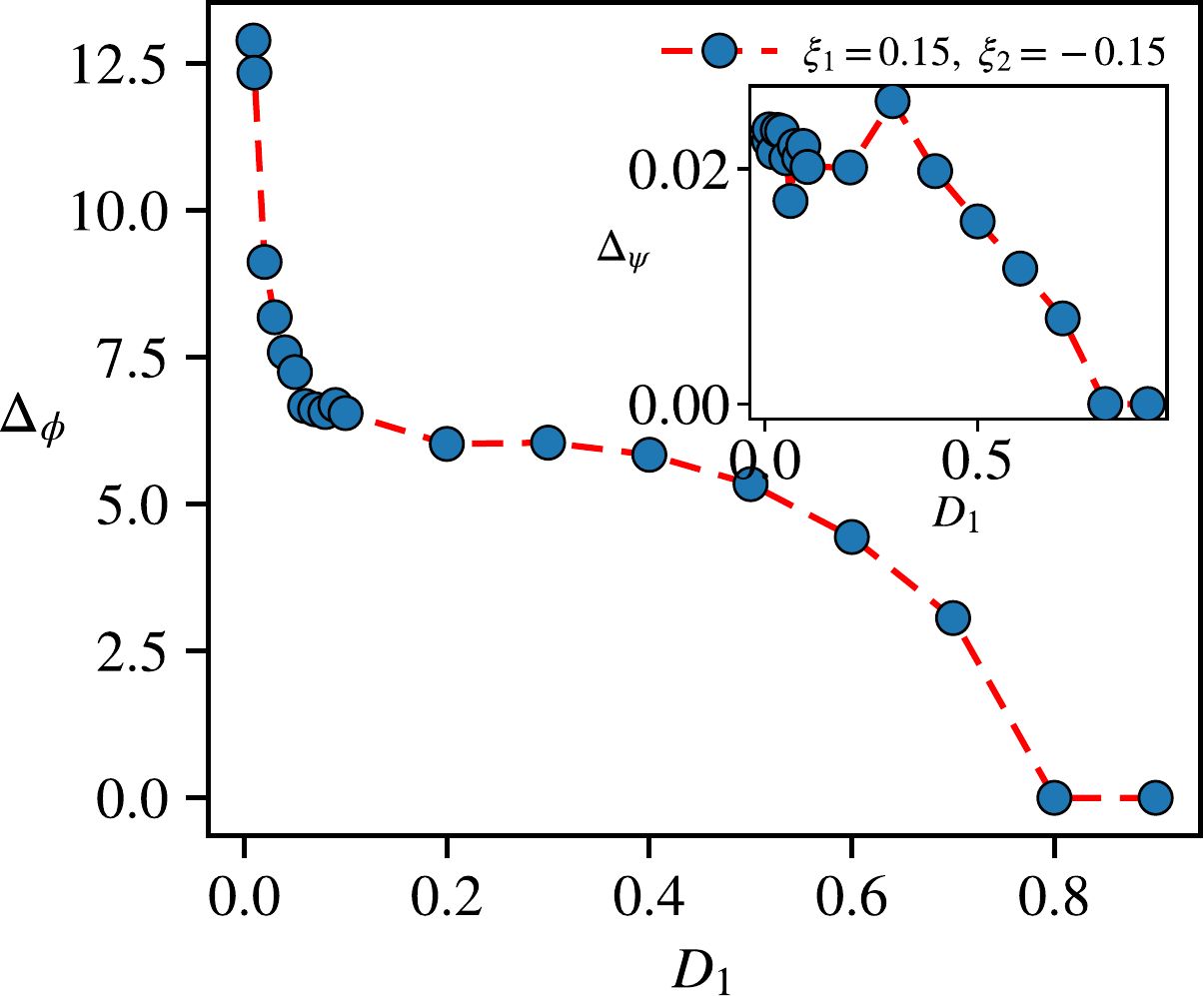}
    \caption{Variation of $\Delta_\phi$ (inset $\Delta_\psi$) as functions of $D_1$ for fixed $\xi_1=0.15,\xi_2=-0.15$}
    \label{diffusion-nr-pos}
\end{figure}

\begin{figure}
    \centering
    \includegraphics[width=0.9\linewidth]{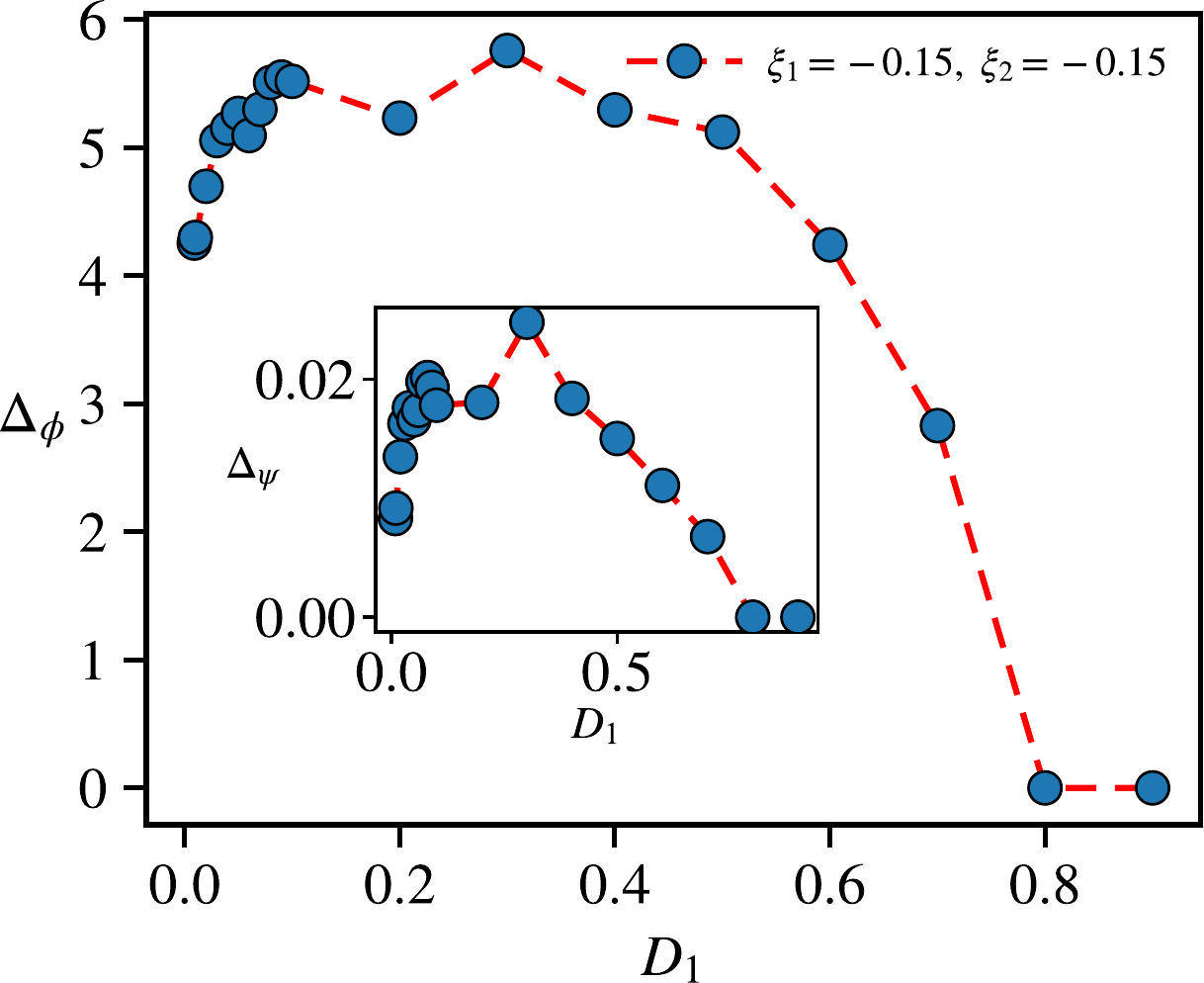}
    \caption{Variation of $\Delta_\phi$ (inset $\Delta_\psi$) as functions of $D_1$ for fixed $\xi_1=-0.15,\xi_2=-0.15$}
    \label{diffusion-r-neg}
\end{figure}

\begin{figure}
    \centering
    \includegraphics[width=0.9\linewidth]{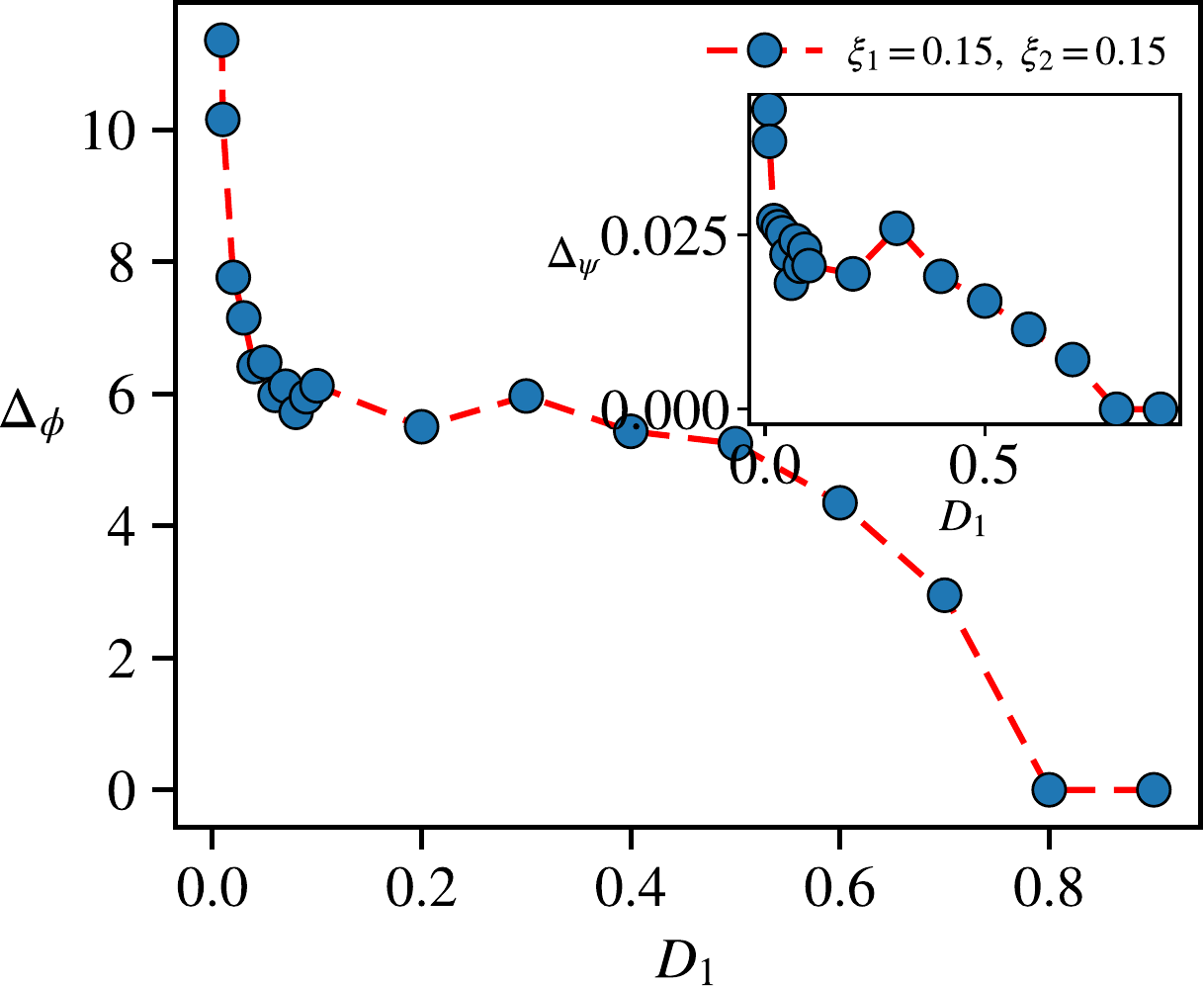}
    \caption{Variation of $\Delta_\phi$ (inset $\Delta_\psi$) as functions of $D_1$ for fixed $\xi_1=0.15,\xi_2=0.15$}
    \label{diffusion-r-pos}
\end{figure}

We quantify the snapshots in Fig.~\ref{fig:fig_Diffusion_fixedXi} by numerically calculating $\Delta_\phi,\Delta_\psi$, parametrized by $D_1$ in the different cases of extreme nonreciprocal chemotaxis and fully reciprocal chemotaxis.  The results are presented in Fig.~\ref{diffusion-nr-neg}, Fig.~\ref{diffusion-nr-pos}, Fig.~\ref{diffusion-r-neg} and Fig.~\ref{diffusion-r-pos}.

\section{Selected wavevector}\label{selected-kc}

We now revisit the selected wavevector $k_c$ in (\ref{qc}), and analyze its dependence on $\xi_1,\,\xi_2$. Given that the numeretor of (\ref{qc}) depends {\it linearly}) on $\xi_1,\xi_2$, whereas the denominator depends {\it bilinearly} on them, $k_c$ may have a nonmonotonic dependence on $\xi_1,\xi_2$. Notice also that at least in the nonreciprocal case $\xi_1\xi_2<0$, for which the denominator does not vanish for any magnitude of $\xi_1,\xi_2$, $k_c\rightarrow 0$ as $|\xi_1|,|\xi_2|$ diverges. This calls for finding the {\it maximum} of $k_c$ with respect to independent variations of $\xi_1,\xi_2$. Setting $\partial k_c^2/\partial\xi_1=0$, we get
\begin{eqnarray}
    &&2d\bigg[D_1D_2-\frac{\xi_1\xi_2}{a+b^2}\bigg]+\xi_2\bigg[-D_1(a+b^2)+D_2\frac{b^2-a}{a+b^2}\nonumber \\&&-\xi_2b+\frac{2\xi_1b^3}{a+b^2}\bigg]=0.\label{xi10}
\end{eqnarray}
Since $D_1D_2>\xi_1\xi_2/(a+b^2)$ for stability and  $-D_1(a+b^2)+D_2\frac{b^2-a}{a+b^2}-\xi_2 b+\frac{2\xi_1 b^3}{a+b^2}>0$ for $k_c^2>0$, (\ref{xi10}) can be satisfied {\it only if} $\xi_2<0$. Next, setting $\partial k_c^2/\partial\xi_2=0$ gives
\begin{eqnarray}
    &&-b(a+b^2)\bigg[D_1D_2-\frac{\xi_1\xi_2b^2}{a+b^2}\bigg]+\xi_1\bigg[-D_1(a+b^2)\nonumber \\&&+D_2\frac{b^2-a}{a+b^2}-\xi_2b+\frac{2\xi_1b^3}{a+b^2}\bigg]b^2=0.\label{xi20}
\end{eqnarray}
For reasons similar to those outlined for (\ref{xi10}), Eq.~(\ref{xi20}) has a solution {\it only if} $\xi_1>0$. Therefore, $\xi_1>0$ {\it and} $\xi_2<0$ are the necessary conditions for $k_c^2$ to have a maximum at some finite $\xi_1,\xi_2$. By using (\ref{xi10}) and (\ref{xi20}), we get
\begin{equation}
    2\xi_1 b^2 = - \xi_2 (a+b^2).\label{xi1xi2}
\end{equation}
Using (\ref{xi1xi2}) in (\ref{xi10}), we obtain
\begin{eqnarray}
    2b D_1 D_2 + \xi_2^2 b + \xi_2 \left[-D_1(a+b^2)+D_2\frac{b^2-a}{a+b^2}\right]=0,
\end{eqnarray}
solving, which we find
\begin{eqnarray}
    \xi_2&=& \frac{1}{2b}[D_1(a+b^2)-D_2\frac{b^2-a}{a+b^2}\pm \{[D_1(a+b^2)\nonumber \\&&-D_2\frac{b^2-a}{a+b^2}]^2 + 8b^2 D_1 D_2\}^{1/2}],\label{xi2sol}
\end{eqnarray}
where we have retained only the negative solution for $\xi_2$, since $\xi_2<0$.  Then $\xi_1$ can be obtained by using (\ref{xi1xi2}) and (\ref{xi2sol}). With these known values of $\xi_1,\xi_2$, $k_c^2|_{\text max}$ may be obtained from (\ref{qc}). These results mean that with $\xi_1>0,\xi_2<0$, as $\xi_1,|\xi_2|$  increase, $k_c^2$ will rise, which means that an increasingly dense pattern should be visible. This should continue until $k_c^2$ crosses  $k_c^2|_{\text max}$. If $\xi_1,|\xi_2|$  increase further, $k_c^2$ should decrease, eventually vanishing for very large $\xi_1,|\xi_2|$. In fact, for sufficiently small $\xi_1,|\xi_2|$ and for appropriate choices for $D_1, D_2$ for some $a,b$, there {\it may not} be any solution for $k_c^2$, giving rise to uniform states. Upon increasing $\xi_1,|\xi_2|$ a positive solution for $k_c^2$ is possible, resulting in a pattern, which becomes denser as $\xi_1,|\xi_2|$ increase until $k_c^2$ reaches $k_c^2|_{\text max}$, beyond which $k_c^2$ starts decreasing monotonically with rising $\xi_1,|\xi_2|$. This shows an intriguing {\it nonmonotonic} dependence of $k_c^2$ on $\xi_1,|\xi_2|$, possibly giving rise to a novel {\it chemotaxis-induced reentrant transition} between uniform states with an intervening patterned state, as $\xi_1,|\xi_2|$ increase from zero. For all other combinations of signs of $\xi_1,\xi_2$, there is no $k_c^2|_{\text max}$, with $k_c^2$ decreasing monotonically with increasing $|\xi_1|,|\xi_2|$.




\section{Simultaneous oscillations and patterns in the presence of chemotaxis}\label{osci-pattern}

\begin{figure}[!ht]
    \centering
\includegraphics[width=\columnwidth]{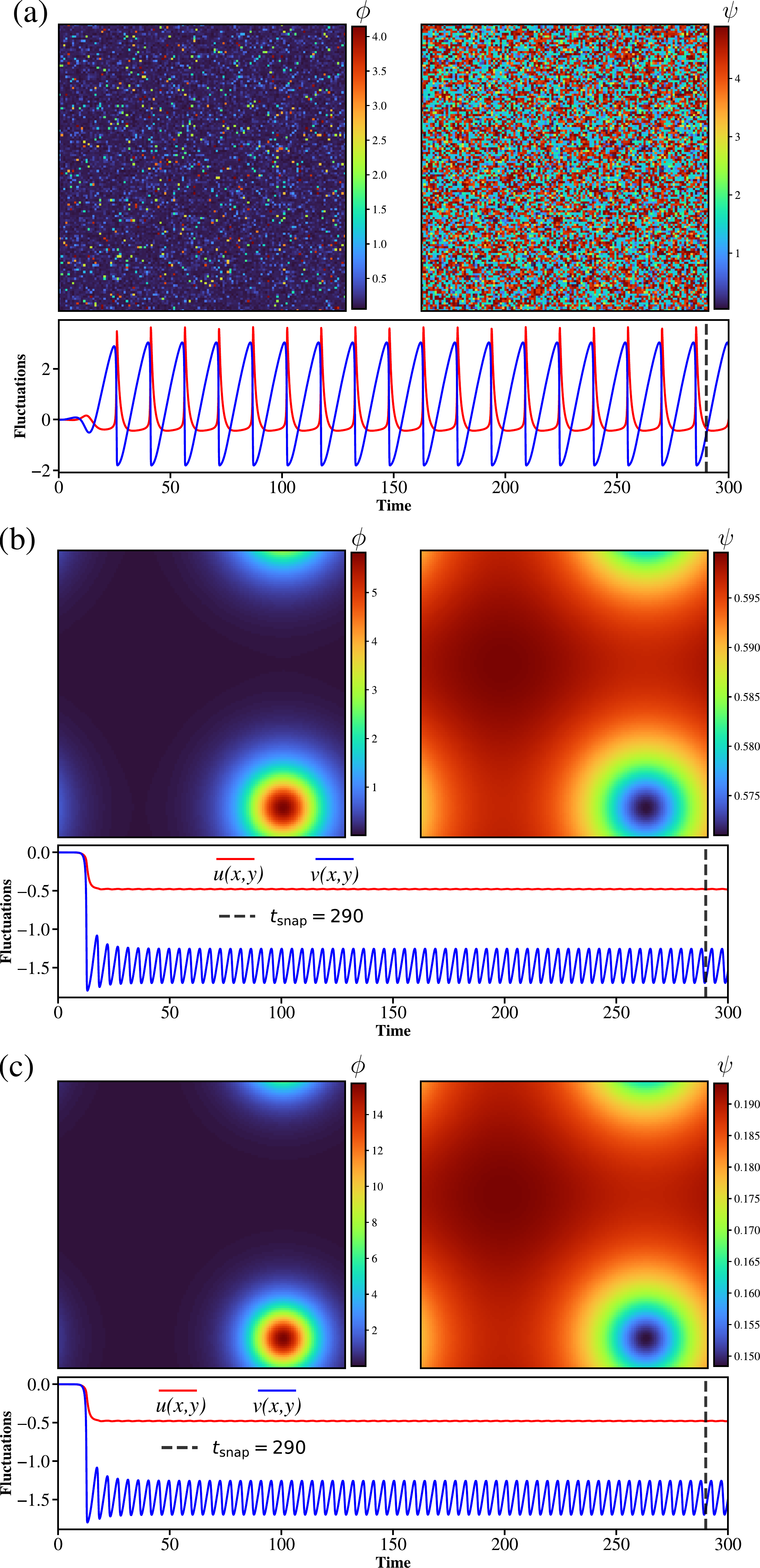}
    \caption{(Color online) Comparison of spatial density fields and temporal dynamics for different chemotactic couplings at fixed parameters $a=0.02$ and $b=0.5$.
    Panels (a)--(c) show snapshots of the densities $\phi$ and $\psi$ at $t_{\mathrm{snap}}=290$, together with the time evolution of local fluctuations $u(x,t)$ and $v(x,t)$ at a representative spatial point. (a)~In the absence of both diffusion and chemotactic coupling ($D_1=D_2=0$, $\xi_1=\xi_2=0$), the system exhibits sustained temporal oscillations in $\phi$ and $\psi$ without any spatial pattern formation.
    (b,c)~For finite diffusion ($D_1=0.2$, $D_2=200$) and chemotactic coupling ($\xi_1=\xi_2=0.10$ and $0.15$), spatial structure emerges and the temporal dynamics transitions to a damped oscillatory regime approaching steady inhomogeneous states. 
    }
\label{fig:fig_osci}
\end{figure}

Here we investigate the nature of the system for $a = 0.02, b = 0.5$. Without chemotaxis, this point lies in a region in the $(a,b)$, where both spatially uniform oscillations and steady patterns are possible in linear stability analysis. Our DNS studies reveal that without chemotaxis, the system undergoes sustained oscillations for both $\phi$ and $\psi$ without any pattern. However, in the presence of reciprocal chemotaxis with $\xi>0$, both $\phi(x,y)$ and $\psi(x,y)$ show steady patterns, but only $\psi(x,y)$ shows sustained oscillations also; see Fig.~\ref{fig:fig_osci}. More work is necessary for a systematic study of this, which is beyond the scope of the present work. 

\bibliography{ref_chemo}

\end{document}